\documentclass{aa}  

\usepackage{graphicx}
\usepackage{txfonts}
\usepackage{natbib}
\usepackage{booktabs}
\usepackage[font=small,labelfont=bf]{caption}

\begin{document} 

   \title{Space environment and magnetospheric Poynting fluxes of the exoplanet $\tau$ Boötis b}

   %\subtitle{I. Overviewing the $\kappa$-mechanism}

   \author{F. Elekes
          \inst{1}
          \and
          J. Saur\inst{1}\fnmsep\inst{2} %\thanks{Just to show the usage
          %of the elements in the author field}
          }

   \institute{Institute of Geophysics and Meteorology, University of Cologne,
              Pohligstr. 3, D-50969 Köln\\
              %
         %\and
            % Institute for Geophysics and Meteorology, University of Cologne,
            % Pohligstr. 3, D-50969 Köln\\
             \email{f.elekes@uni-koeln.de}\label{inst1}
             %\thanks{The university of heaven temporarily does not
            %         accept e-mails}
         \and 
         	 \email{jsaur@uni-koeln.de}\label{inst2}  
             }

   \date{Accepted: January 8, 2023}

% \abstract{}{}{}{}{} 
% 5 {} token are mandatory
 
  \abstract
   {The first tentative detection of a magnetic field on the Hot Jupiter type exoplanet $\tau$ Boötis b was recently reported by \citet{Turner2021}. The magnetic field was inferred from observations of circularly-polarized radio emission obtained with the LOFAR telescopes. The observed radio emission is possibly a consequence of the interaction of the surrounding stellar wind with the planet's magnetic field.}
   {We aim to better understand the near space environment of $\tau$ Boötis b and to shed light on the structure and energetics of its near-field interaction with the stellar wind. We are particularly interested in understanding the magnetospheric energy fluxes powered by the star-planet interaction and in localizing the source region of possible auroral radio emission.}
   {We perform magnetohydrodynamic simulations of the space environment around $\tau$ Boötis b and its interaction with the stellar wind using the PLUTO code. We investigate the magnetospheric energy fluxes and effects of different magnetic field orientations in order to understand the physical processes which cause energy fluxes that may lead to the observed radio emission given the proposed magnetic field strength in \citet{Turner2021}. Furthermore we study the effect of various stellar wind properties, such as density and pressure, on magnetospheric energy fluxes given the uncertainty of extrasolar stellar wind predictions.}
   {We find in our simulations that the interaction is most likely super--Alfvénic and energy fluxes generated by the stellar wind--planet interaction are consistent with the observed radio powers. Magnetospheric Poynting fluxes are of the order of 1--8 $\times10^{18}$ W for hypothetical open, semi-open and closed magnetospheres. These Poynting fluxes are energetically consistent with the radio powers in \citet{Turner2021} for a magnetospheric Poynting flux--to--radio efficiency $> 10^{-3}$ when the magnetic fields of the planet and star are aligned. In case of lower efficiency factors the magnetospheric radio emission scenario is according to the parameter space modeled in this study not powerful enough. A sub--Alfvénic interaction with decreased stellar wind density could channel Poynting fluxes on the order of $10^{18}$W towards the star. In case of a magnetic polarity reversal of the host star from an aligned to anti-aligned field configuration, expected radio powers in the magnetospheric emission scenario fall below the observable threshold. Furthermore we constrain the possible structure of the auroral oval to a narrow band near the open-closed field line boundary. Strongest emission is likely to originate from the night side of the planet. More generally, we find that stellar wind variability in terms of density and pressure does influence magnetospheric energy fluxes significantly for close-in magnetized exoplanets.}
   {}


   \maketitle

\section{Introduction}
Recently, tentative measurements of auroral radio emission from the hot Jupiter exoplanet  $\tau$ Boötis b were obtained with the \emph{Low Frequency Array} (LOFAR) \citep{Turner2021}. These observations might be considered the strongest evidence so far of an intrinsic magnetic field on a planet outside the solar system if the emission indeed originates from the planets vicinity. Then they also imply that $\tau$ Boötis b possess a magnetosphere which interacts with its surrounding stellar wind. The radio observations by \citet{Turner2021}, if confirmed, thus help pave the way for the field of extrasolar space physics. In this work we therefore use properties from the observed radio signals to derive new constraints on the space environment around  $\tau$ Boötis b.
\\
The massive Hot Jupiter $\tau$ Boötis b \citep{Butler1997} is a very good candidate to remotely observe a powerful interaction of a stellar wind with an exoplanet's magnetosphere for several reasons: With $\sim$16 pc, the $\tau$ Boötis system is relatively close to the solar system. The planet orbits its host star $\tau$ Boötis A at a short distance of 0.046 astronomical units \citep{Butler1997}. Additionally, its large mass ($> 5 M_{\text{Jupiter}}$) may cause its exobase to remain close to the planet, leading to a magnetosphere (MS) not completely filled with dense plasma and thus allowing for radio emission to be produced efficiently and to escape the planets vicinity \citep{Weber2018,Daley-Yates2018}.
\\
The tentative radio measurements obtained with LOFAR comprise two signals allegedly from the vicinity of $\tau$ Boötis b \citep{Turner2021}. The circularly polarized signals were detected in the 21 -- 30 MHz and 15 -- 21 MHz frequency bands, respectively. The emission possibly originates from gyrating, energetic electrons precipitating towards the planetary polar regions, emitting radio waves generated through the Electron Cyclotron Maser Instability (ECMI), which is expected to be the dominant mechanism for exoplanetary radio emission \citep{Zarka1998,Treumann2006}. From these signals the planetary magnetic field strength can be inferred directly since the emission frequency corresponds to the local electron gyro-frequency. The existing observations are consistent with the expectations on the emitted power from the radio-magnetic Bode's law \citep{Zarka2001,Zarka2007,Zarka2018}, on the polarization \citep[e.g. circular polarization;][]{Zarka1998,Griessmeier2005} and on the frequency \citep[i.e. slightly above Earth's ionospheric cutoff;][]{Griessmeier2007a,Griessmeier2011,Griessmeier2017}. The measured radio signal however needs multi-site follow up observations preferably at various radio wavelengths to confirm and to further constrain the magnetic field environment of $\tau$ Boötis b \citep{Turner2021}.

In addition to radio emission \citep[e.g.][]{Griessmeier2015,Farrell1999,Zarka2001,Zarka2007}
other indirect methods have been proposed to detect and constrain the magnetic
fields of exoplanets. These are, for example, optical signatures in the stellar chromosphere by means of Ca II H\&K line excess emission induced by star-planet interactions (SPI) \citep{Cuntz2000, Cauley2019,Shkolnik2003,Shkolnik2005,Shkolnik2008}, and asymmetries in near-ultraviolet
stellar light curves together with UV absorption signatures caused by transiting planetary bow shocks \citep{Vidotto2010,Vidotto2011,Llama2011}. The SPI and transit related observations can lead to false-positives \citep[e.g.][]{Turner2021,Kislyakova2016, Preusse2006, Lai2010,Kopp2011, Miller2012, Miller2015, Bisikalo2013, Alexander2016, Turner2016, Gurumath2018,Route2019} due to sets of model assumptions involved in the process.
Radio observations in contrast, can directly constrain the magnetic field
amplitude and are therefore less susceptible to false--positives \citep{Griessmeier2015}. The success of radio observations has been demonstrated in the past in the solar system. For example, Jupiter's magnetic field was discovered through radio observations
\citep{Franklin1958} before spacecraft confirmed it with in-situ magnetometer measurements.
\\
Since $\tau$ Boötis b may be the first exoplanet with a directly observed magnetic field it provides an unique opportunity to constrain the space environment around this exoplanet. However, various properties of $\tau$ Boötis b are unknowns such as radius, size and extend of its atmosphere above the 1 bar level as well as stellar wind parameters. $\tau$ Boötis A is a solar-like F7 IV-V star \citep{Gray2001}, coronal temperature and pressure might therefore be comparable to the sun. The coronal base density and consequently the stellar wind mass loss rate poses the most uncertain free parameter of previous studies of the stellar wind from $\tau$ Boötis A \citep{Vidotto2012, Nicholson2016}. Recently new constraints on stellar winds of M dwarf stars were reported utilizing astrospherical absorption signatures induced by the interaction of the stellar wind with the interstellar medium \citep{Wood2021}. The question naturally arises if stellar wind -- planet interactions might as well produce observable signatures capable to provide constrains on stellar wind properties such as density (i.e. mass loss rate) or pressure (i.e. temperature), which will be addressed in this paper.
\\
The proximity of Hot Jupiters to their host stars can potentially cause sub--Alfvénic star -- planet interactions, which is expected to produce observable signatures in the stellar \citep[e.g. chromospheric emission Ca II H \& K line emission][]{Shkolnik2003,Shkolnik2008, Cauley2019} or the planetary atmosphere \citep[e.g. auroral radio emission][]{Cohen2018, Turnpenney2018,Bastian2022, Kavanagh2021, Kavanagh2022}. Such magnetic SPI in exoplanetary systems were excessively studied by means of magnetohydrodynamic simulations \citep[e.g.][]{Preusse2006,Preusse2007,Zhilkin2020,Varela2018,Varela2022}, partly with a focus on the far--field interaction incorporating self--consistent stellar wind models \citep[e.g.][]{Strugarek2014,Strugarek2019a,Strugarek2019b, Cohen2011, Cohen2014,Vidotto2015,Vidotto2017}. Modeling of super-Alfvénic stellar wind -- magnetosphere interactions was sparsely done to our knowledge although a large number of close-in exoplanets might be exposed to, at least temporarily, super-Alfvénic wind conditions \citep{Zhilkin2019} (i.e. orbiting outside the Alfvén surface).
In case of super-Alfvénic or, more precisely, super-fast magnetosonic stellar wind flows a bow shock forms upstream of the planetary magnetosphere because of the flow being faster than the fastest magnetohydrodynamic wave mode. In this case the planet does not interact with the star since waves are not able to propagate upstream. This might be the case for $\tau$ Boötis b, as the planet is likely exposed to a super--fast stellar wind according to \cite{Nicholson2016}.
\\
The generation of radio emission from exoplanets, its properties and dependence on stellar wind and planetary parameters was studied intensively using numerical simulations \citep{Nichols2016,Varela2016,Varela2018,Turnpenney2020,Daley-Yates2018,Kavanagh2020} for other or generic exoplanets. However, little to no emphasis was given to studying the detailed spatial structure and energetics of magnetospheric Poynting fluxes that ultimately deliver the available electromagnetic energy capable of driving planetary auroral emissions at various wavelengths. 
\\
In order to better understand the space environment around $\tau$ Boötis b we perform magnetohydrodynamic (MHD) simulations of the near space environment of $\tau$ Boötis b and its magnetic field interacting with the surrounding stellar wind plasma using the PLUTO code. The stellar wind model is based on wind simulations \citep{Vidotto2012,Nicholson2016} driven by magnetic surface maps derived from magnetic measurements of $\tau$ Boötis A \citep{Marsden2014,Mengel2016, Jeffers2018}. The magnetic field estimate of the planet's intrinsic field, based on the tentative magnetic field strengths derived by \citet{Turner2021}, is used to model the planetary magnetosphere. We specifically aim to better understand the magnetospheric energy fluxes around $\tau$ Boötis b and of Hot Jupiter type exoplanets in general that are exposed to similar stellar wind conditions. We also address the question on how stellar wind variability in the time-independent case affects magnetospheric Poynting fluxes and therefore possible radio powers generated by the interaction.
\\
\\
The paper is structured in the following way: An overview of the physical model to describe the plasma interaction $\tau$ Boötis b with the surrounding stellar wind is given in Sect. \ref{section:numerical_simulation}. The numerical setup is summarized in Sect. \ref{section:method}, details about the stellar wind model can be found in Sect. \ref{section:stellar_wind}. The $\tau$ Boötis b model is described in Sect. \ref{section:TauBoötis_model}. In the subsequent Sect. \ref{section:results} we show our results, starting with a general description of the interaction in Sect. \ref{section:structure_of_interaction}, followed by a study of the spatial structure of Poynting fluxes in Sect. \ref{section:poynting_spatial_structure}. Then we discuss the energetics of the interaction in Sect. \ref{section:energetics}, where we also compare possible radio emission output with the observations by \citet{Turner2021}. The results are followed by a discussion about the role and importance of the stellar wind to power the energy fluxes in the magnetosphere of the exoplanet in Sect. \ref{section:discussion_stellarwind}. At last, we discuss possible auroral radio emission and its detectability in scope of stellar wind variability (Sect. \ref{section:auroralRadioEmission}).
%--------------------------------------------------------------------
\section{Numerical simulation} \label{section:numerical_simulation}
In this section we introduce our physical model and the numerics to describe the interaction of $\tau$ Boötis b and its intrinsic magnetic field with its surrounding stellar wind. The MHD model together with the numerical model and coordinate system are presented in Sect. \ref{section:method}. We introduce the stellar wind that is included as boundary condition for the plasma variables in Sect. \ref{section:stellar_wind} followed by the description of parametrizations of physical processes introduced by the planet and its atmosphere in Sect. \ref{section:TauBoötis_model}.
\subsection{Method}\label{section:method}
We performed single-fluid ideal, non-resistive and non-viscous MHD simulations using the open-source code PLUTO (v. 4.4) in spherical coordinates \citep{Mignone2007}. The MHD equations to solve are
\begin{eqnarray}
\frac{\partial \rho}{\partial t} + \nabla \cdot\left[\rho \vec{v}\right] &=& Pm_n - L m_p \label{continuity-equation} \\
\frac{\partial \rho \vec{v}}{\partial t} + \nabla \cdot \left[ \rho \vec{v} \vec{v}  + p- \vec{B} \vec{B} + \frac{1}{2}  B^{2} \right] &=& -(L m_p + \nu_n \rho)\vec{v} \label{momentum-equation} \\
\frac{\partial E_t}{\partial t} + \nabla \cdot \left[(E_t + p_t)\vec{v} - \vec{B}(\vec{v}\cdot\vec{B}) \right] &=& -\frac{1}{2} (L m_p + \nu_n \rho)v^{2} \nonumber 
\\ & & -\frac{3}{2} (L m_p + \nu_n \rho) \frac{p}{\rho} \nonumber 
\\ & & +\frac{3}{2} (P m_n + \nu_n \rho) \frac{k_B T_n}{m_n} \label{energy-equation}\\
\frac{\partial \vec{B}}{\partial t}- \nabla \times \left[\vec{v} \times \vec{B}\right] &=&0\; , \label{induction-equation}
\end{eqnarray}
where $\rho \vec{v}$ is the momentum density, $\vec{v}$ the velocity, $\rho$ the mass density, $p_t$ the total pressure (e.g. magnetic and thermal) and $p$ the thermal pressure. $\vec{B}$ is the magnetic flux density, $-\vec{v}\times \vec{B}$ in Eq. \ref{induction-equation} is the electric field in the ideal limit with infinite electrical conductivity.
$E_t$ is the total energy density, $E_t = \rho e + \rho v^2 /2+ B^2 / 2\mu_0$, and $e$ the specific internal energy.
The system is closed by the equation of state in the form $p = \rho e (\gamma - 1)$, where $\gamma$ is the ratio of specific heats for the adiabatic case.\\
As for magnetic diffusion we do not include a diffusion term in the induction equation (Eq. \ref{induction-equation}) but point out that numerical diffusion, especially for coarse grids such as in our simulation, introduce numerical diffusion sufficient to allow for reconnection \citep[see][]{Varela2018}. To justify this assumption we performed test simulations incorporating magnetic diffusion and found it to not influence the results of this paper significantly (see appendix \ref{section:diffusion} for a detailed discussion on this topic).\\
We include plasma production, $P$, and loss terms, $L$, (Eqs. \ref{continuity-equation}-\ref{energy-equation}) to account for photo-ionization, dissociative recombination together with associated momentum and internal energy transfer between neutral atmospheric and magnetospheric plasma particles as well as ion-neutral collisions. We note that the neutral species is not simulated and altered by the interaction with the ion species. Details on how plasma production and loss are modeled can be found in Sect. \ref{section:TauBoötis_model}. The mass of plasma particles is denoted by $m_p$ and $m_n$ describes the mass of neutral particles. We assume the plasma to completely consist of ionized hydrogen atoms, $m_p = m_{H^+}$. The atmosphere only consists of neutral molecular hydrogen,  $m_n = m_{H_2}$.
\\
The conservative form of Eqs. \ref{continuity-equation} -- \ref{induction-equation} are integrated using a approximate hll-Riemann solver (Harten, Lax, Van Leer) with the diffusive \emph{minmod} limiter function. The $\nabla \cdot \vec{B} = 0$ condition was ensured by the mixed hyperbolic--parabolic divergence cleaning technique \citep{Dedner2002,Mignone2010}.
\\
The spherical grid consists of 256 non--equidistant radial, 64 and 128 equally spaced angular grid cells in $\theta$ and $\phi$ dimension respectively. 
The radial grid is divided into three regions. From 1 to 1.2 planetary radii ($R_p$) the grid contains 10 uniform cells. After that from 1.2 to 12 $R_p$ the next 150 cells increase in size with a factor of $\sim$ 1.01 per cell. The last 96 cells from 12 $R_p$ towards the outer boundary at 70 $R_p$ increase gradually with a factor of $\sim$ 1.02.
The positive x axis points parallel to the relative velocity $v_0$ of the stellar wind in the frame of the planet. The stellar wind magnetic field is assumed to be perpendicular to $v_0$ and is anti-parallel to the z axis. The y axis completes the right handed coordinate system. Co-latitude $\theta$ is measured from the positive z-axis, longitudes $\Phi$ are measured from the positive y axis within the xy plane. The origin is located at the planetary center. We run all simulations for approximately 3.6 h physical time until a quasi steady-state is reached in the vicinity of the planet ($r<30$). Small fluctuations cannot be avoided although larger scale structure and dynamics within the MS remain already almost constant after approximately 2 hours physical time.
\subsection{Stellar wind model}\label{section:stellar_wind}
\begin{table*}
	\caption{Physical simulation parameters. Details of the stellar wind and planet model are discussed in Sects. \ref{section:stellar_wind} and \ref{section:TauBoötis_model}.}      
	\label{table:parameters}      
	\centering                                   
	\begin{tabular}{c c c c c}          
		\hline\hline                       
		 &  Symbol & Value & Source & Note \\    
		\hline\hline
		\textbf{$\tau$ Boötis b}   &&&&\\                                
		Planet radius 	& $R_p$ & 72875 km &  \citet{Wang2011} & Theoretical\\      
		Orbital period & $P_{orb}$ & $3.31$d & \citet{Butler1997,Wang2011} & \\
		Semi-major axis & $a$ & $0.046$ au & \citet{Butler1997} & \\
		Atm. surface density 	& $n_{n,0}$ & 8$\times10^{12}$ m$^{-3}$ & --&  \\
		Atm. scale height & H & $0.06\,R_p$ & -- & 3 radial grid cells \\
		Magn. flux density (eq) & $B_p$ & $455\mu$T & \citet{Turner2021} & Average of observations 1,2 \\
		\hline\hline 
		\textbf{Basic Stellar wind model} &&&&\\
		Therm. pressure & $p_{sw}$ & $2.5 \times 10^{-5}$ Pa& \citet{Nicholson2016} & \\
		Ion density & $n_{sw}$ & $1.4\times 10^{12}$ m$^{-3}$ & \citet{Nicholson2016} & \\
		Velocity & $v_{sw}$ & $224.5$ km s$^{-1}$ & \citet{Nicholson2016} & Stellar frame of reference\\
		Relative velocity & $v_{0} $ & $270.98$ km s$^{-1}$ & -- & $v_{0}=\sqrt{v_{sw}^2 + 4\pi^2\times a^2/P_{orb}^2}$ \\
		Magnetic flux density & $B_{sw}$ & $2.715 \; \mu$T & \citet{Nicholson2016} &\\  
		\hline
		Alfvén Mach number & $M_A$ & $5.36$ &  &  \\  
		Fast mode Mach number & $M_f$ & $1.9$ &  & \\ 
		Plasma beta & $\beta$ & $8.31$ &  &\\  
		\hline\hline                    
	\end{tabular}
\end{table*}
The derived stellar wind parameters from \citet{Nicholson2016} resemble those of the sun, such as the polytropic index, $\gamma = 1.1$ \citep{Doorsselaere2011}, and the stellar coronal base temperature, which is not well constrained by observations, is set to $2\times10^{6}$ K as typical value for the solar coronae \citep{Nicholson2016,Vidotto2012,Doorsselaere2011}. The magnetic field of $\tau$ Boötis A was studied excessively during several epochs and magnetic surface maps as well as several magnetic polarity reversals were observed \citep{Donati2008,Fares2009,Fares2013}.
The coronal base density remains an educated guess based on a comparison of emission measure (EM) values obtained from X-Ray spectra of $\tau$ Boötis A \citep{Vidotto2012, Maggio2011}. Due to the uncertainty of the base density estimate, different stellar wind densities will be investigated separately in the scope of magnetospheric Poynting fluxes and possible radio powers in Sects. \ref{section:results:stellar-wind-variability} and \ref{section:discussion_stellarwind}.
\\ 
The stellar wind is applied through constant in--flow boundary conditions at the upstream hemisphere ($\Phi = 0$ to $180^\circ$). The magnetic field is assumed to be perpendicular to the relative velocity $v_0$ of the wind (i.e. parallel to the negative z axis). The in--flow velocity of the plasma, which we call the relative velocity $v_0$, is parallel to the x--axis and is composed of the radial velocity of the wind $v_{sw}$ and the orbital velocity of the planet. The adopted plasma parameters of the wind are summarized in Table \ref{table:parameters} which were averaged over the several epochs studied by \citet{Nicholson2016}. 

\subsection{$\tau$ Boötis b model}\label{section:TauBoötis_model}

\noindent
We assumed a radially symmetric neutral atmosphere with a scale height of $H=4373$ km. Thus, the scale height extends over three radial grid cells and consequently the neutral atmosphere is sufficiently resolved within the numerical grid. We assume an atmosphere consisting of molecular hydrogen as it is, followed by helium, the most abundant constituent of the Jovian atmosphere \citep{Atreya2003}. The collisional cross-section is assumed to be $\sigma_{in} = 2\times10^{-19}\, \text{m}^2$ for H$^+$ -- H$_2$ collisions with momentum transfer for low--eV relative velocities between the colliding particles \citep{Tabata2000}. In our simulations the collision frequency is $\nu_{in} \approx 0.5$ s$^{-1}$, so that $\nu_{in} = \bar{v}\sigma_{in}n(r)$, where $\bar{v} \approx v_0$ denotes a typical velocity in the system and $n_n(r)$ is the atmosphere number density as function of radial distance from the center,
\begin{equation}\label{eq:atmosphere}
	n_n(r) = n_{n,0} \exp\left(\frac{R_p - r}{H}\right) \;,
\end{equation}
where $n_{n,0} = 8\times 10^{12}$ m$^{-3}$ is the surface number density. Based on test studies, we found that for $n_{n,0} \approx 8\times 10^{12}$ m$^{-3}$ the ion-neutral collisions nearly completely bring the incoming plasma flow to a halt in the atmosphere. This results in plasma pile up in form of a shell around the planet. Increasing the density would thus not produce a larger interaction.
\\
We use a simplified description of photo-ionization. We neglect the shadow zone exerted by the planet's body and parameterize plasma production through photo-ionization using only the radial dependence of the neutral atmosphere density,
\begin{equation}\label{eq:ionisation}
	P(r) = \nu_{ion} n_n(r) \; .
\end{equation} 
 The radial symmetric ionization partially mimics some night side ionization through electron impact ionization. For the photo-ionization frequency of hydrogen exposed to a solar-like UV radiation environment at a distance of approximately 0.046 AU from the star we take the value from \citet{Kislyakova2014}, $\nu_{ion} = 6\times 10^{-5}$ s$^{-1}$.
\\
Plasma loss is introduced through recombination of hydrogen ions. The loss term therefore depends on the plasma density, 
\begin{equation}\label{eq:recombination}
	L(\vec{r},t) = \alpha n(\vec{r},t) (n(\vec{r},t) - n_{sw}) \;.
\end{equation}
Plasma loss is switched off if the plasma density falls below the background density (i.e. $n(\vec{r},t) \leq n_{sw}$) as stellar wind ions and electrons recombine significantly slower due to the higher electron temperatures in the stellar wind.
Given an electron temperature of roughly $T_e \approx 7500$ K for a Hot Jupiter exoplanet's ionosphere with semi-major axis of 0.046 AU around a sun-like star derived by \citet{Koskinen2010} and using the formula of \citet{Storey1995},
\begin{equation}
	\alpha = 4 \times 10^{-12} \left(\frac{300 \text{K}}{T_e}\right)^{0.64} \text{cm}^3\, \text{s}^{-1}\; ,
\end{equation}
 we find the hydrogen ion recombination rate, $\alpha$, to be $5.1\times10^{-19}$ m$^3$ s$^{-1}$.
Further discussion about the underlying assumption about our atmosphere model can be found in appendix \ref{section:appendix:atmosphere}.\\
Recent tentative auroral radio measurements from $\tau$ Boötis b give a first observational constraint on its magnetic field strength. \citet{Turner2021} found the polar surface magnetic flux density $B_p$ to lie between 7.5 and 10.7 G for two right-handed circularly polarized signals. We assume a dipole field and adopt the average value of both Stokes V$^+$ signals \citep{Turner2021}, $B_{p} = 9.1$ G, for our simulations. Furthermore we study the effect of dipole orientation on the stellar wind -- planet interaction through simulating an open ($0^\circ$ tilt), semi-open ($90^\circ$ tilt) and closed MS ($180^\circ$ tilt), where the tilt is measured with respect to the negative z axis. The various tilts are realized by rotating the stellar background magnetic field accordingly so that the planetary dipole axis is always parallel to the z-axis. Given the strong magnetic variability of $\tau$ Boötis A \citep[e.g. several magnetic polarity reversals were observed as well as a chromospheric activity cycle in terms of S--indices of roughly 240 days][]{Donati2008,Fares2009,Fares2013,Mengel2016,Mittag2017, Jeffers2018} we are also able to study the effect of the host star's magnetic field topology on the stellar wind--planet interaction and associated magnetospheric energy fluxes.
\\
The magnetic field is implemented using the insulating--boundary method by \citet{Duling2014} which ensures that no radial electric currents exist within the insulating boundary, which we assume to be the planet's neutral atmosphere below its ionosphere.

\section{Results} \label{section:results}
\begin{figure}
	\centering  
	\includegraphics[width=0.999\linewidth]{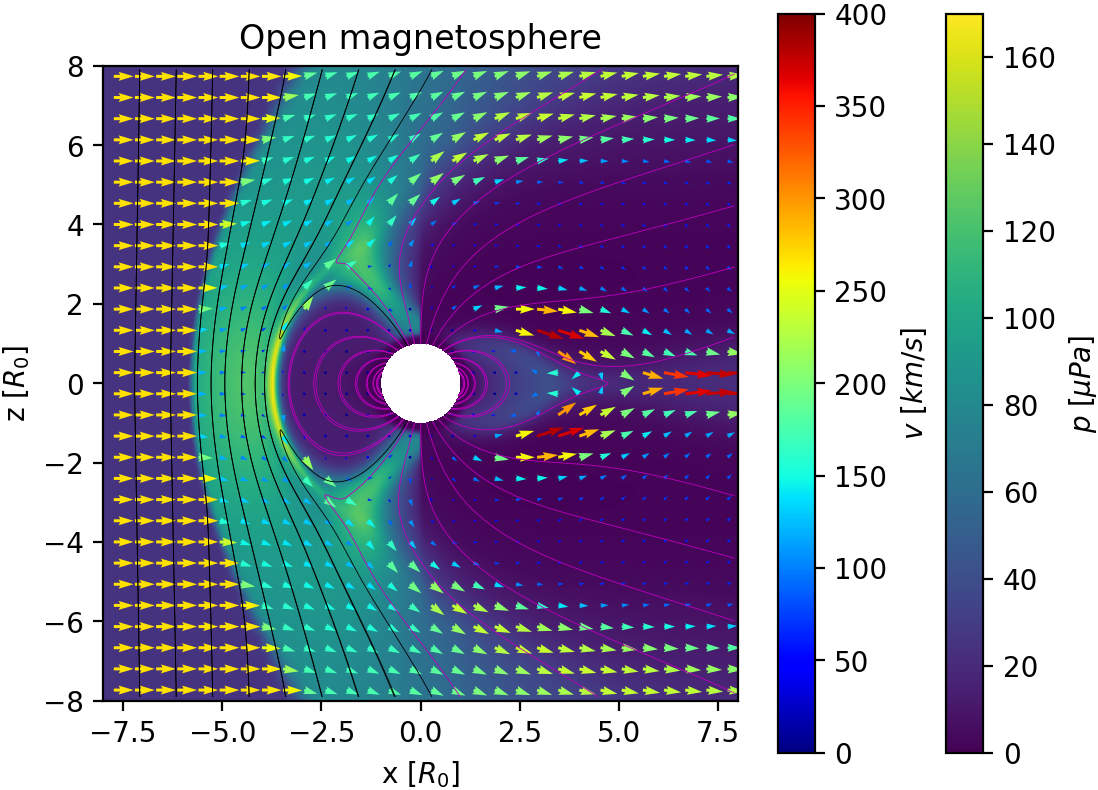}\\
	\includegraphics[width=0.999\linewidth]{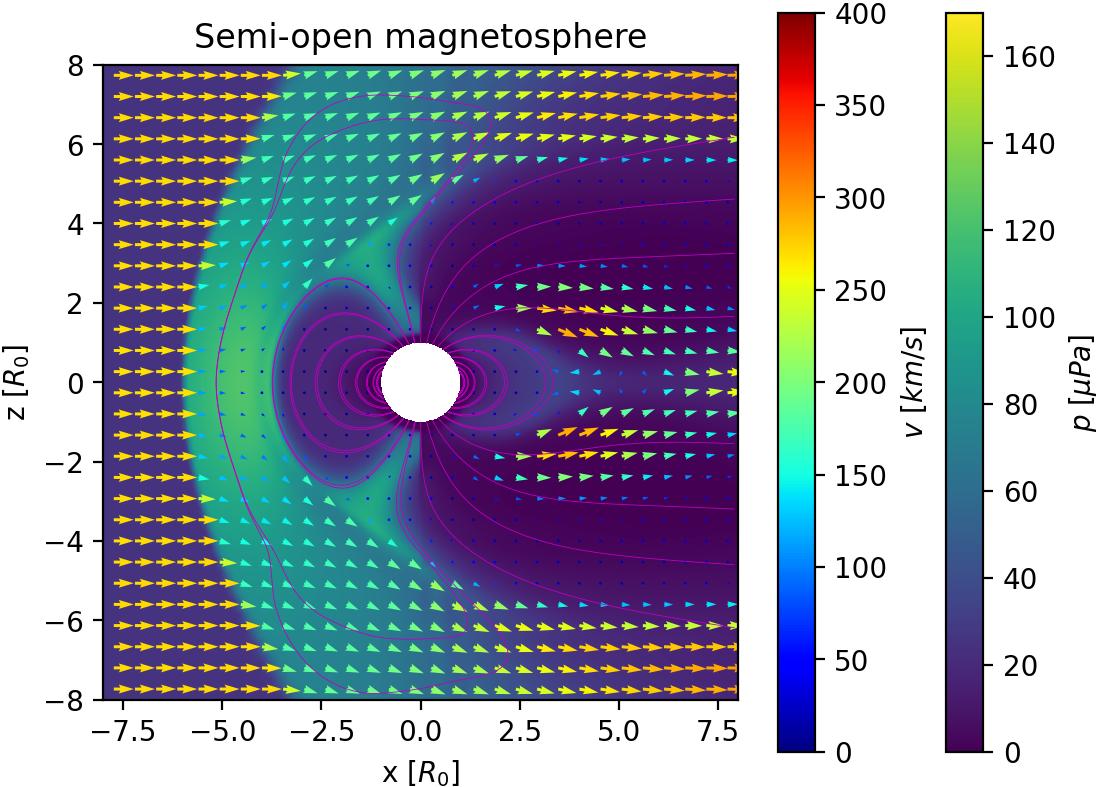}\\
	\includegraphics[width=0.999\linewidth]{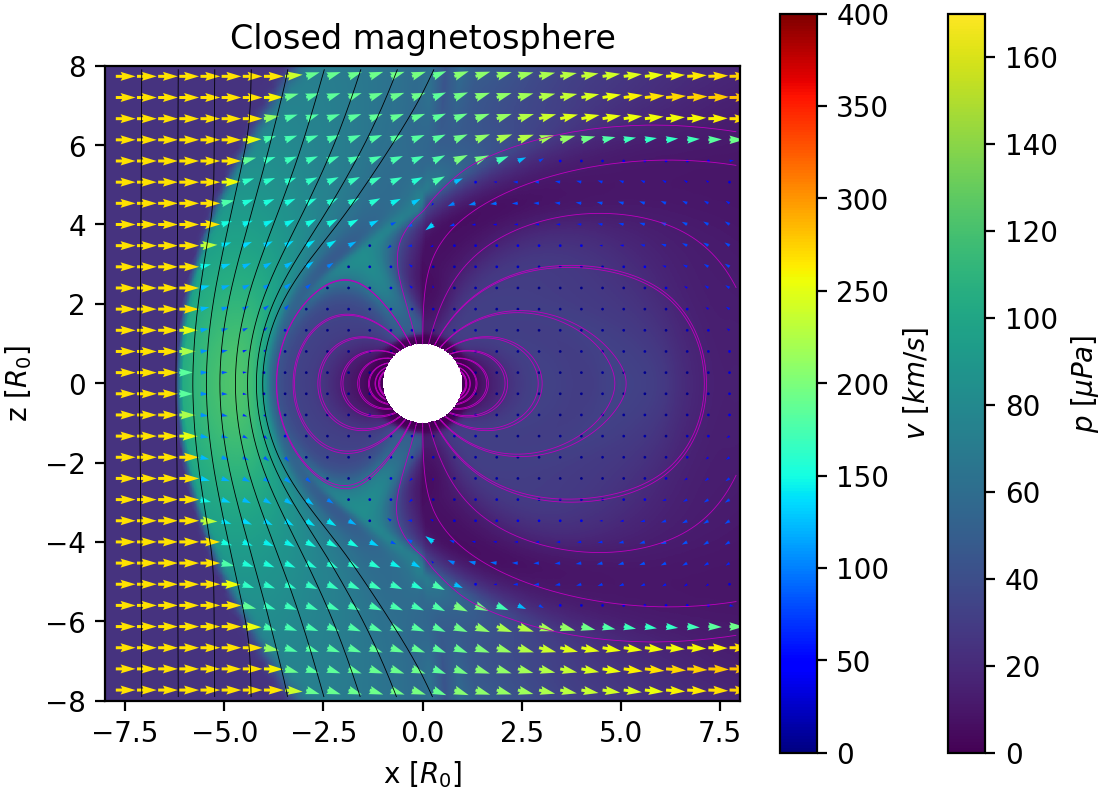}
	\caption{Velocity fields (colored arrows, left colorbars) and plasma pressure (color contours, right colorbars) in the xz-plane for the open MS ($\theta_B=0^\circ$, top), semi-open ($\theta_B=90^\circ$, middle) and closed ($\theta_B=180^\circ$, bottom) MS case. Projected stellar wind magnetic field lines are indicated as black solid lines within the xz-plane parallel to the ambient magnetic field. Closed and open magnetospheric field lines are colored in magenta.}
	\label{fig:vel_plots}
\end{figure}
In this section we first present results of our modeling which provides an overview of the space plasma environment of $\tau$ Boötis b (Sect. \ref{section:structure_of_interaction}). Then we study in detail the magnetospheric Poynting fluxes in Sect. \ref{section:poynting_fluxes}. 
\subsection{Structure of the interaction} \label{section:structure_of_interaction}
The simulated plasma velocities and pressures according to the basic model (Table \ref{table:parameters}) are displayed in Fig. \ref{fig:vel_plots} for the open ($\theta_B = 0^\circ$), semi-open MS ($\theta_B = 90^\circ$) and closed MS ($\theta_B = 180^\circ$) case. The magnetic field tilt $\theta_B$ is the angle between the external field (parallel to the z axis) and the planet's magnetic moment. We note that, due to the symmetries chosen in our model studies, the stellar wind and intrinsic magnetic field are not inclined with respect to the z-axis, therefore we also show projected field lines (black solid lines) in the xz-plane. Color contours denote plasma pressure in $\mu$Pa (right colorbar). Arrows represent velocity components, their magnitudes are color coded (left color bar). The length of arrows indicate the magnitudes of the shown components. Spatial dimensions are given in units of planetary radii. 
\\
The intrinsic magnetic field and its corresponding MS poses an obstacle to the stellar wind flow coming from negative x -- direction. The flow outside the MS is super-Alfvénic ($M_A=5.36$) and super-fast magnetosonic ($M_f = 1.6$) (see Table \ref{table:parameters}), where $M_f = v_0/(v_A^2 + c_s^2)^{1/2}$, with the sound speed $c_s = \sqrt{\gamma p_{sw}/\rho_{sw}}$, polytropic index $\gamma = 1.1$ \citep{Nicholson2016} and Alfvén velocity $v_A = B_{sw}/\sqrt{\mu_0 \rho_{sw}}$. The super-fast interaction enforces a bow shock to be formed roughly $5 R_p$ in front of the planet followed by a fairly thick magnetosheath. Since no wave is able to propagate upstream, the stellar wind plasma is unperturbed until the bow shock. The structure of the MS strongly depends on the internal field orientation as visible in Fig. \ref{fig:vel_plots} with an increase of overall MS size towards higher magnetic axis tilts. For the open and semi-open MS (Fig. \ref{fig:vel_plots} top and middle plot respectively) two magnetic lobes form, separated by a thin plasma sheet, where open magnetic field lines connect to the stellar wind field several planetary radii downstream (not shown in the plots). The day side magnetopause, defined by the location of the last closed field line, lies between 3 and 3.5 $R_p$, while the night side magnetopause is located at roughly 5 $R_p$ for the open and semi-open MS respectively. The downstream side magnetopause is very narrow in the z -- direction as expected due to the magnetic field lines convected downstream together with the stellar wind flow and due to the magnetic stresses stretching the magnetic field. The closed MS case (Fig. \ref{fig:vel_plots}, bottom plot) has a night side magnetopause lying several planetary radii ($\sim 17R_p$) downstream (not shown in the plots). While the day side magnetopause is controlled by the stellar wind thermal and magnetic pressure balanced with those exerted by the planet's surroundings, the night side MS is influenced by reconnection (i.e. the merging of  planetary with stellar wind field lines). Magnetic reconnection is most efficient for a magnetic moment parallel to the ambient field (here the z-axis), therefore the fraction of open planetary field lines connected to the star decreases significantly with an intrinsic field moment being directed anti-parallel to the stellar field. As the stellar wind plasma primarily penetrates the MS along magnetic field lines, the amount of plasma and thermal pressure decreases as well with increasing magnetic axis tilt. \\ 
We note that, as visible in Fig. \ref{fig:vel_plots} (bottom), the magnetosphere is completely closed. This is due to the perfect anti--parallel alignment of the planetary and stellar wind magnetic field.
\\
Within the MS the flow velocity is strongly reduced and has weak upstream components in the negative x direction due to magnetic tension exerted on planetary field lines. Magnetic reconnection takes place at the upstream and downstream side where velocities, both within and outside the MS, are strongly enhanced due to acceleration through released magnetic energy. Velocities are slightly larger at the flanks of the MS compared to the upstream side and exceed the initial stellar wind velocity at the downstream side where stellar wind as well as planetary field lines merge together again and accelerate the plasma.
\\
Thermal pressures are strongly enhanced within the magnetosheath, where stellar wind plasma is decelerated abruptly and compressed, so that kinetic energy is converted into heat. Plasma may penetrate the MS along open magnetic field lines in the polar cusps where pressure is enhanced as well. The cusps act as channels for plasma transport into the MS. There is a trend towards lower pressures in the cusps for increasing magnetic axis tilt. This is directly connected to the amount of stellar wind plasma advected towards the planet as the amount of injected plasma is related to the ability of magnetic field lines to merge with the ambient field. This becomes increasingly difficult for planetary magnetic moments having components anti-parallel to the ambient field, therefore the area fraction of open magnetic field lines and thus the size of the plasma injection channel is maximal for a completely open MS. Here pressures up to 160 $\mu$Pa can be reached while the closed MS case shows pressures up to roughly 90 $\mu$Pa.

\begin{figure*}
	\centering
	\includegraphics[width=0.48\linewidth]{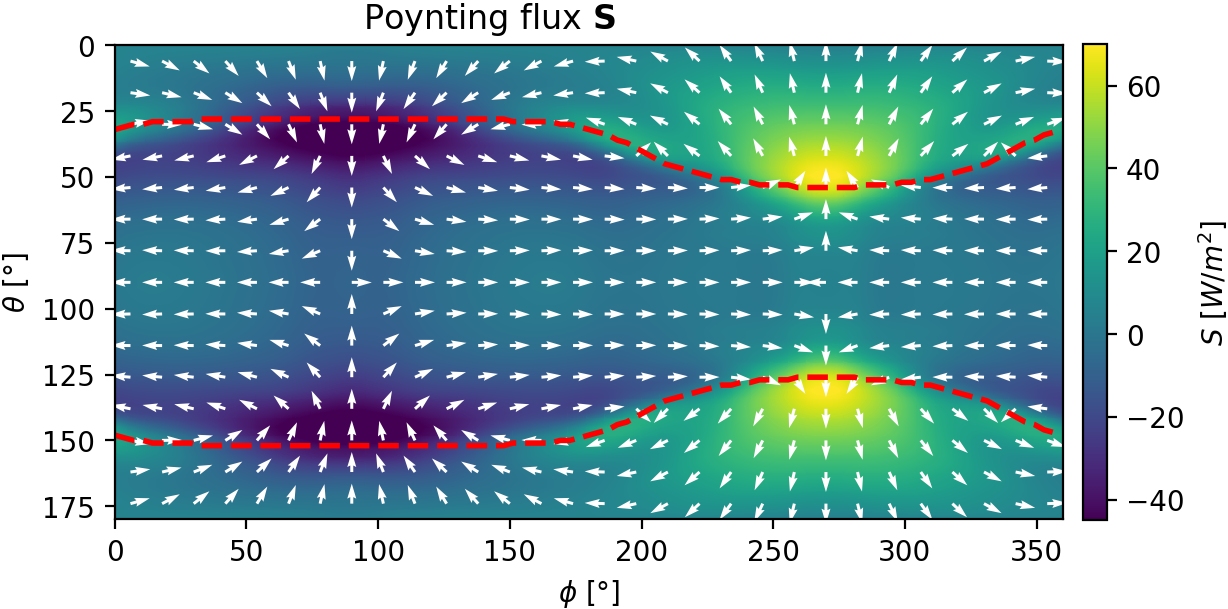}
	\includegraphics[width=0.48\linewidth]{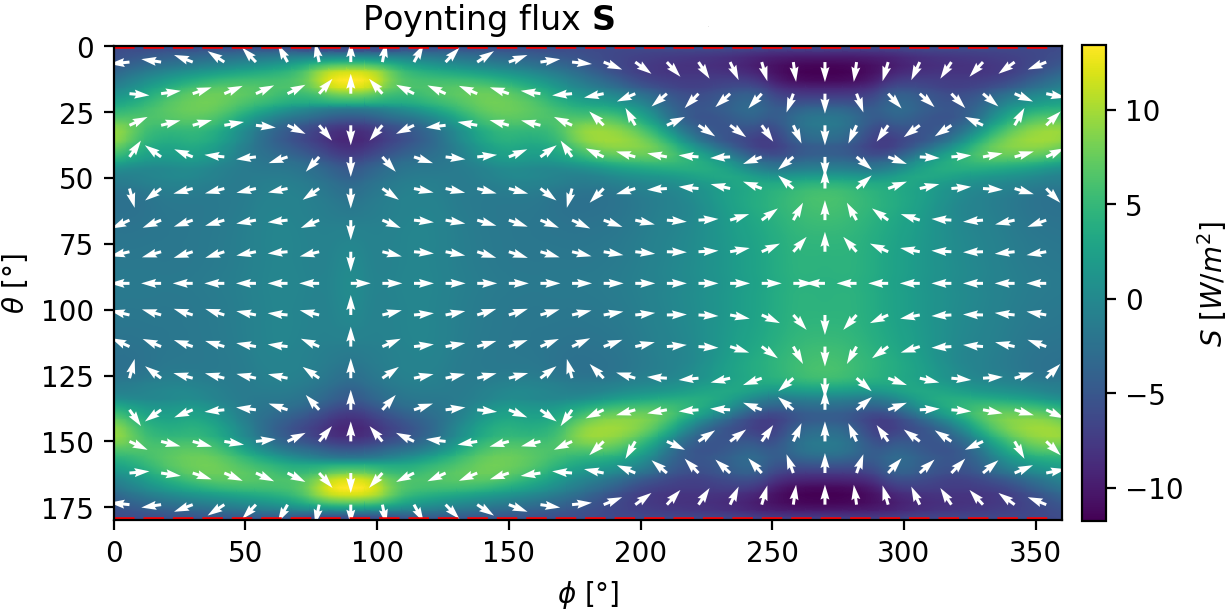}\\
	\includegraphics[width=0.48\linewidth]{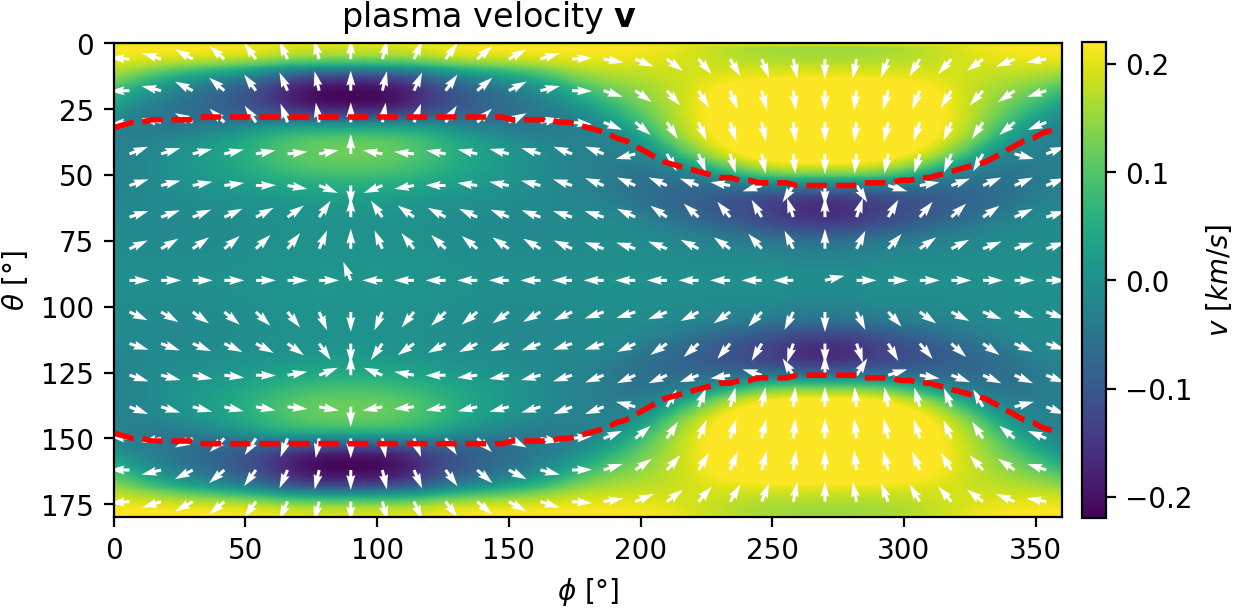}
	\includegraphics[width=0.48\linewidth]{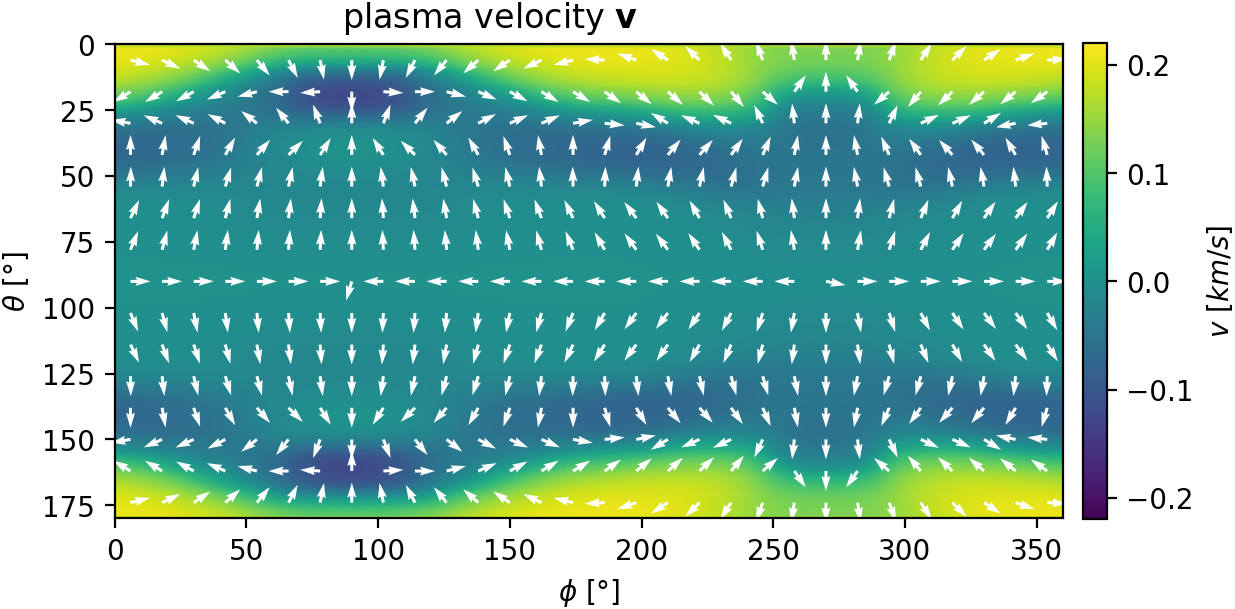}\\
	\includegraphics[width=0.48\linewidth]{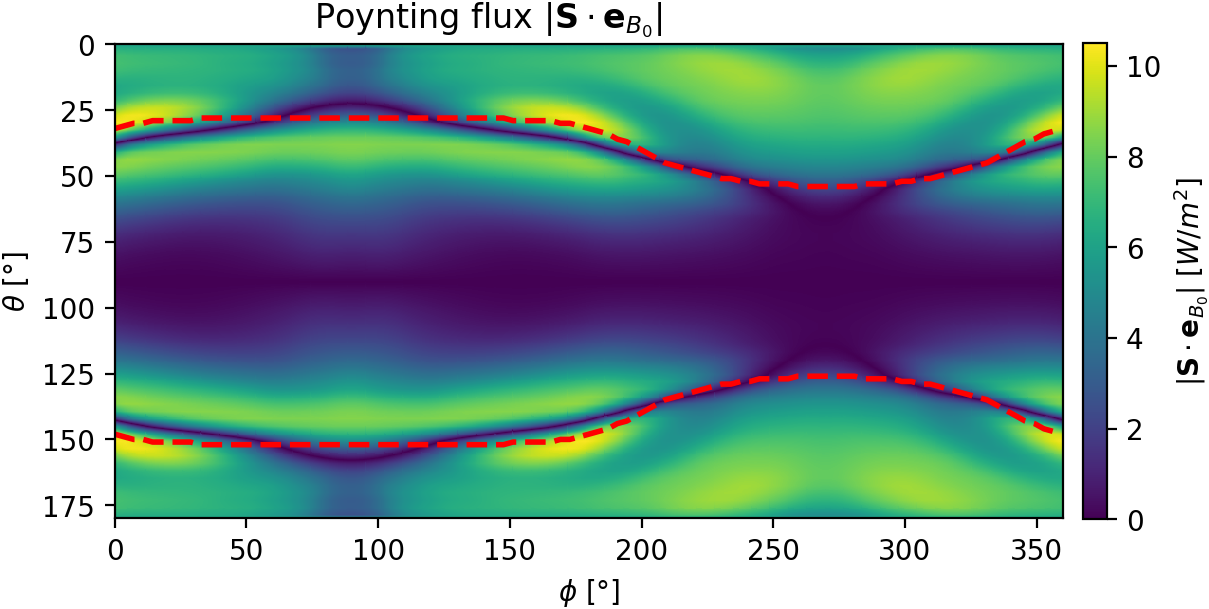}
	\includegraphics[width=0.48\linewidth]{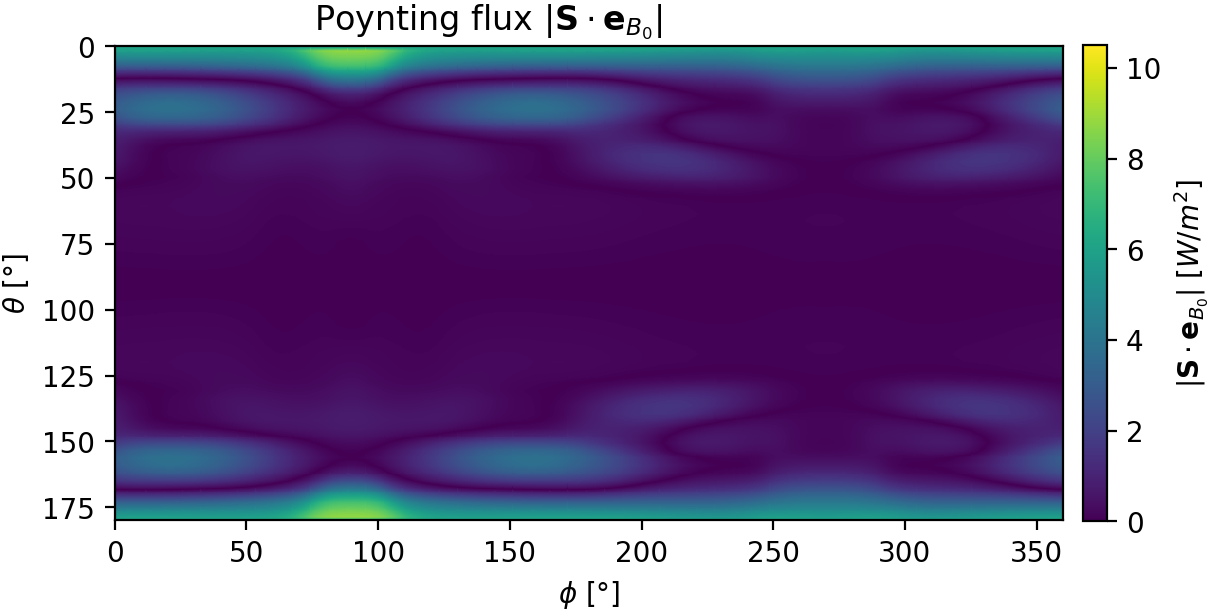}\\
	\caption{Mercator projections of the Poynting flux (upper row), plasma velocity (middle row) and absolute values of Poynting flux components parallel to the unperturbed planetary field plus small background field from stellar wind (bottom row). The results are shown at an altitude of one planetary radius above the surface. The left column displays maps for the open MS case ($\theta_B = 0^\circ$), the right column for the closed MS case ($\theta_B = 180^\circ$). Arrows indicate normalized angular components, color contours denote radial components. Red dashed lines indicate the location of the OCFB.}
	\label{fig:Poynting_vector}
\end{figure*}
\subsection{Poynting fluxes and aurorae} \label{section:poynting_fluxes}
We are interested in understanding the electromagnetic coupling of the stellar wind with the magnetosphere of the exoplanet, its atmosphere and ionosphere. The energy fluxes associated with these electromagnetic coupling processes provide the energetics for the auroral emission from the exoplanet's magnetosphere at radio and other wavelengths. Therefore we study the Poynting flux to calculate the maximum available electromagnetic energy fluxes. 
We describe the spatial structure of magnetospheric Poynting fluxes in  Sect. \ref{section:poynting_spatial_structure}. Then we study the energetics of the interaction and effects of stellar wind variability on magnetospheric energetics in the subsequent sections \ref{section:energetics} and \ref{section:results:stellar-wind-variability}.
\subsubsection{Spatial structure}\label{section:poynting_spatial_structure}
We first describe the spatial structure of the Poynting fluxes and plasma velocities within the MS as displayed in the top and middle row of Fig. \ref{fig:Poynting_vector} respectively. The plots show Mercator projections of the angular vector components over a spherical shell with radius $2 R_p$. The angle $\theta$ denotes the co-latitude, and $\phi$ the longitude. Arrows indicate the orientation of vectorial quantities and color contours their magnitudes. Positive values indicate radial components pointing away from the planet. Red dashed lines represent the open -- closed planetary field line boundaries (OCFB). Magnetic field lines with both foot points on the planetary surface are closed field lines. Each field line having only one foot point on the planet is an open field line. The OCFB separates areas with open from areas with closed field lines. Thus, the OCFB also represents the magnetopause at that specific radial location. The open and closed MS cases are shown left and right, respectively.
\\
The Poynting vector $\vec{S}$ can be rewritten in the ideal MHD case using the convective electrical field \citep[e.g.][]{Saur2013}
\begin{equation}\label{eq:PoyntingFlux}
	\vec{S} = \frac{\vec{E}\times \vec{B}}{\mu_0} = \frac{-\vec{v} \times \vec{B}\times \vec{B}}{\mu_0} = \frac{ B^2}{\mu_0} \vec{v}_\perp \;,
\end{equation}
which is bodily carried by the plasma flow perpendicular to magnetic field lines, denoted by the perpendicular velocity $\vec{v}_\perp$. The Poynting flux describes the transport of magnetic enthalpy, which is a factor of two larger than the magnetic energy density $B^2/2\mu_0$ \citep[e.g.][]{Saur2013}. In the remainder of this work we mostly present Poynting fluxes, but need to consider the factor of two when we compare magnetic energy densities with thermal (i.e. internal energy) or kinetic energy densities based on their flows.\\
For both, the open and closed MS case, flow velocities are strongly reduced at the upstream ($\phi = 0$--$180$ degrees) and downstream ($\phi = 180$ -- $360$ degrees) side down to speeds below 10 km s$^{-1}$. This occurs due to interaction with the intrinsic magnetic field and momentum transfer with the neutral atmosphere. The OCFB is located at roughly $\theta \approx 30^\circ$ and $150^\circ$ in the open MS case. Due to the perfectly anti--parallel configuration of the stellar wind and planetary magnetic field no open planetary field lines exist in the closed MS model. This has also been observed in sub-Alfvénic simulations using this field topology \citep{Ip2004,Strugarek2015}.
\\
\\
\textbf{Open Magnetosphere model (Fig. \ref{fig:Poynting_vector} left):}
The very narrow vertical extend of the downstream closed field line region of the open MS is caused by magnetic tension due to the magnetized stellar wind. Highest velocities are found within the open field line region mainly at the downstream side where plasma is accelerated downstream through magnetic tension on open lines.
\\
Strong Poynting fluxes occur where plasma velocities have strong components perpendicular to the magnetic field. They are found within the open field line region mainly at the downstream side with outward directed Poynting fluxes. Comparatively strong Poynting fluxes, but directed toward the planet, are located on the upstream side near the magnetopause. Within the closed field line region and especially near the equator Poynting fluxes mostly vanish.
\\
\\
\textbf{Closed Magnetosphere model (Fig. \ref{fig:Poynting_vector} right):}
In the closed MS model highest velocities can be found near the planetary poles confined to an area below 25 degrees co--latitude and similar in the south. This high velocities are caused by tension on high latitude closed field lines that are strongly stretched towards the downstream side by the stellar wind and reach up to 17 planetary radii.
\\
Poynting fluxes oriented away from the planet are confined to narrow bands encircling the high latitude polar regions between 40 and 80 degrees co--latitude and similar in the south. At the upstream side Poynting fluxes vanish near the equatorial regions due to plasma flow being mainly aligned with planetary field lines. Inward oriented Poynting fluxes occur near the polar axis slightly shifted towards the downstream side.
\\
\\
We now study the Poynting fluxes parallel to the unperturbed background magnetic field because in the solar system magnetospheres they are considered the root energy fluxes from which a small fraction can be converted into auroral radio emission. Poynting fluxes provide the energy from which wave--particle interaction can draw energy to accelerate electrons \citep[e.g. for Jupiter][]{Hill2001,Saur2021}. The resulting energetic electrons then can be subject to the electron maser instability \citep{Treumann2006,Zarka2007}.
The interaction of the stellar wind with $\tau$ Boötis b's magnetic field perturbs the magnetic and electric field, which causes the Poynting fluxes. To quantitatively assess the associated Poynting flux, we therefore use the unperturbed magnetic background field $\vec{B}_{0}=\vec{B}_{p,0} + \vec{B}_{sw}$ (i.e. the initial dipole and stellar wind field) to calculate the Poynting flux on this field, $\vec{S}\cdot \vec{e}_{B_{0}}$. The unit vector $\vec{e}_{B_{0}}$ points in the direction of unperturbed magnetic field lines. These projections give insight on where electromagnetic energy is transported either through propagating magnetic disturbances (i.e. Alfvén waves) or convection. The bottom row of Fig. \ref{fig:Poynting_vector} shows $|\vec{S}\cdot \vec{e}_{B_0}|$ for the open MS (left) and closed MS (right). We note that only absolute values are shown in the plots in order to clearly identify zero or near-zero power densities. 
\\
Strongest energy transport along unperturbed field lines occurs over narrow bands encircling the polar open field line regions at the flanks of the planet where velocities are nearly perpendicular to the magnetic field as seen in Fig. \ref{fig:Poynting_vector}. Moreover, the spatial structure of Poynting fluxes along $B_0$ is strictly symmetric with respect to the equator (at $\theta = 90^\circ$). A significant amount of energy is transported parallel to the unperturbed field within the polar open field line regions in the open MS case. Parallel energy fluxes reach values up to 10 Wm$^{-2}$ at the flanks of the planet just outside the closed field line regions. Poynting fluxes up to 9 Wm$^{-2}$ are found at the downstream side, above the OCFB. For both, open and closed MS model, strongest convected energy can be found extensively in high latitude regions due to high velocities perpendicular to the magnetic field. Here the planetary field lines are most mobile in a sense that they are bent over towards the downstream side by the stellar wind. For the closed MS parallel Poynting fluxes up to roughly 9 Wm$^{-2}$ can be found directly at the planetary poles slightly shifted towards the upstream side. At lower latitudes parallel Poynting fluxes up to 6 Wm$^{-2}$ are confined to narrow bands at the flanks of the planet. Auroral emission is expected to be strong where Poynting fluxes are large, hence near the OCFB (e.g. mostly confined to the $L=3$--$3.5$ shell at the upstream side) and in the polar regions for both MS models. They vanish completely along the equator. Generally said Poynting fluxes are significantly weaker and confined to the small polar regions for the closed magnetosphere model compared to the open MS case. In the open MS model strong parallel Poynting fluxes cover the whole open field line area with their maximum at the flanks of the planet in contrast to the closed MS where the regions of strongest parallel Poynting fluxes are partitioned into smaller areas around the planetary poles.

\subsubsection{Energetics of the interaction}\label{section:energetics}
%--------------------------------------------------- One column table
%----------------------------------------------------------------- 
\begin{table}
	\caption{\label{table:results} Integrated magnetospheric Poynting fluxes for different magnetic field topologies}
	\centering
	\begin{tabular}{lccc}
		\hline\hline
		\textbf{Model}&Dipole tilt&$P_a$ [W]\tablefootmark{a}&$P_a || \vec{B}_0$ [W]\tablefootmark{b}\\
		\hline
		Open MS & $0^{\circ}$ & 3.46e+18 & 8.73e+17\\
		Semi-open MS & $90^{\circ}$ & 1.77e+18 & 4.88e+17\\
		Closed MS & $180^{\circ}$ & 6.91e+17 & 1.09e+17\\
		\hline
	\end{tabular}
	\tablefoot{
		Integrated Poynting fluxes over a sphere with radius $r=2R_p$.\\
		\tablefoottext{a}{Magnetospheric Poynting flux (Eq. \ref{eq:S-integral-mag})}\\
		\tablefoottext{b}{Magnetospheric Poynting flux parallel to $\vec{B}_0$ (Eq. \ref{eq:S-integral-B0})}
	}
\end{table}
To estimate the total available Poynting flux, which serves as the root energy flux, we assume for simplicity that the radio emission is generated in a shell 1 $R_p$ above the surface of the exoplanet. This particular choice is inspired by the fact that radio emission around Jupiter and other solar system planets arises from altitudes about 1$R_p$ (or larger) above the planet's surface \citep[e.g.][]{Zarka1998,Hess2011} where strong electron acceleration takes place \citep[e.g. for Jupiter][]{Mauk2020}. Poynting fluxes within the magnetosphere of $\tau$ Boötis b only vary little as function of distance from the planet (see appendix \ref{section:appendix:S_vs_r} for a discussion on the choice of $r$).
\\
Available electromagnetic power for possible conversion into electron acceleration and radio emission is given by the divergence of the Poynting flux in this shell with Volume V,
\begin{equation}\label{eq:S-integral}
	P_a = \int_{V} \nabla \cdot \vec{S} dV = \int_{A_{shell}} \vec{S} \cdot \hat{\vec{n}} \,dA_{shell} \;,
\end{equation}
where $A_{shell}$ is the surface area of the shell and $\hat{\vec{n}}$ the surface normal vector. To investigate the maximal Poynting flux which can be dissipated in the shell we assume that the Poynting flux entering the shell from above or below is dissipated within the shell. For mathematical simplicity we further let the thickness of the shell grow infinitesimally small such that
\begin{equation} \label{eq:S-integral-mag}
	P_a = \int_{A_{sphere}} |\vec{S}_r| \,dA_{sphere} \;,
\end{equation}
with $A_{sphere}$ the area of the sphere located at 2 $R_P$ from the center.
In physical terms it means that the possible dissipation in the shell can be supplied with energy fluxes from below the shell (i.e. coming from the planet's ionosphere) or from above the shell (i.e. coming from the magnetosphere or stellar wind). Ultimately, the energy flux is coming from the stellar wind, but the energy flux can be reflected or converted in the ionosphere and can be redirected away from the planet again. This integrated Poynting flux serves as a proxy for maximum available electromagnetic energy dissipated within an auroral acceleration region. \\
Alternatively, we integrate the components of the Poynting flux parallel to the unperturbed magnetic field, $\vec{B}_0 = \vec{B} - \delta\vec{B}$, where $\delta\vec{B}$ denotes the magnetic field perturbation generated by the interaction. These Poynting fluxes take into account the energy flux of which a fraction can directly contribute to particle acceleration and powering the ECMI driven emission,
\begin{equation}\label{eq:S-integral-B0}
	P_{a,||} = \int_{A} |\vec{S}\cdot \vec{e}_{B_0}|\, dA\;,
\end{equation}
where $\vec{e}_{B_0}$ is the unit vector pointing in direction of $\vec{B}_0$. We refer to this Poynting flux component as the auroral Poynting flux. As opposed to Eq. \ref{eq:S-integral-mag}, $P_{a,||}$ (Eq. \ref{eq:S-integral-B0}) serves as a more realistic estimator for calculating auroral energy dissipation since Eq. \ref{eq:S-integral-mag} includes significant contribution of convected energy which is likely not converted into particle acceleration.
Table \ref{table:results} summarizes integrated Poynting fluxes according to Eq. \ref{eq:S-integral-mag} (third column) for all three intrinsic magnetic field orientations. The 4th column shows integrated Poynting fluxes along the unperturbed field (Eq. \ref{eq:S-integral-B0}). Integrated Poynting fluxes range from $3.5\times 10^{18}$ down to $6.9\times10^{17}$ W for the open towards the closed MS model. Poynting fluxes along the unperturbed field (Eq. \ref{eq:S-integral-B0}) amount to $\sim 9\times10^{17}$ and $10^{17}$ W for the open and closed MS respectively. The effect of magnetic topology on convected energy within the magnetosphere is therefore significant as the powers differ by almost one order of magnitude. Magnetic stress due to the stellar wind interaction can work on the magnetospheres less strongly if the magnetosphere is closed, thus giving rise to weaker flows and therefore weaker convected Poynting fluxes. The trend is similar for Poynting fluxes along $\vec{B}_0$, $P_{a,||}$, but here the powers are reduced by almost an order of magnitude below the integrated total Poynting fluxes $P_a$.

\subsubsection{Influence of stellar wind variability on magnetospheric energetics}\label{section:results:stellar-wind-variability}
\begin{figure}
	\centering
	\includegraphics[width=0.499\textwidth]{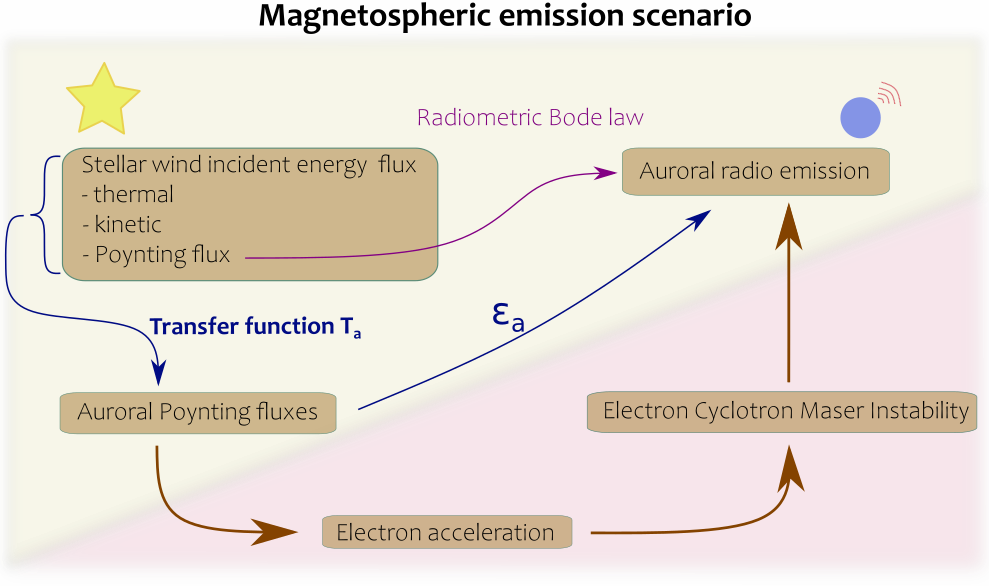}
	\caption{Schematic illustrating the several steps from incident stellar wind energy flux towards auroral radio emission. The transfer function $T_a$ (see Sect. \ref{section:discussion_stellarwind_variability}) describes the conversion from incident stellar energy to auroral Poynting fluxes. The conversion efficiency from auroral Poynting fluxes (Eq. \ref{eq:S-integral-B0}) to radio emission, $\epsilon_a$, implicitly includes the efficiency of electron acceleration and the ECMI mechanism. The steps within the pink shaded area are not included in our model. Brown arrows indicate physical processes, blue arrows denote model parameters quantifying energy conversion and the magenta arrow the radiometric scaling law.}
	\label{fig:EfficienciesSchematic}
\end{figure}
For modeling the space environment of $\tau$ Boötis b, the properties of its surrounding stellar wind carry very large uncertainties, in particular the stellar wind density. In \citet{Nicholson2016} and \citet{Vidotto2012}, the coronal base density was estimated by choosing the electron density so that it can reproduce electron measure (EM) observations of $\tau$ Boötis A. The energy fluxes within the MS are powered by and limited by the maximum incident power of the stellar wind flow transferring onto the magnetospheric obstacle. \citet{Zarka2007} found that the observed radio power of solar system planets is nearly a constant fraction of the incident kinetic and magnetic energy convected through the obstacle's cross section, $\pi R^{2}_{mp}$, where $R_{mp}$ is the magnetospheric stand-off distance of the magnetized planet. This energy is utilized in perturbing the topology of the planets magnetic field which in turn results in currents induced by changes in magnetic flux. Therefore the incident power controls the energetics within the MS. 
%----------------------------------------------------------------- 
\begin{figure}
	\centering
	\includegraphics[width=0.49\textwidth]{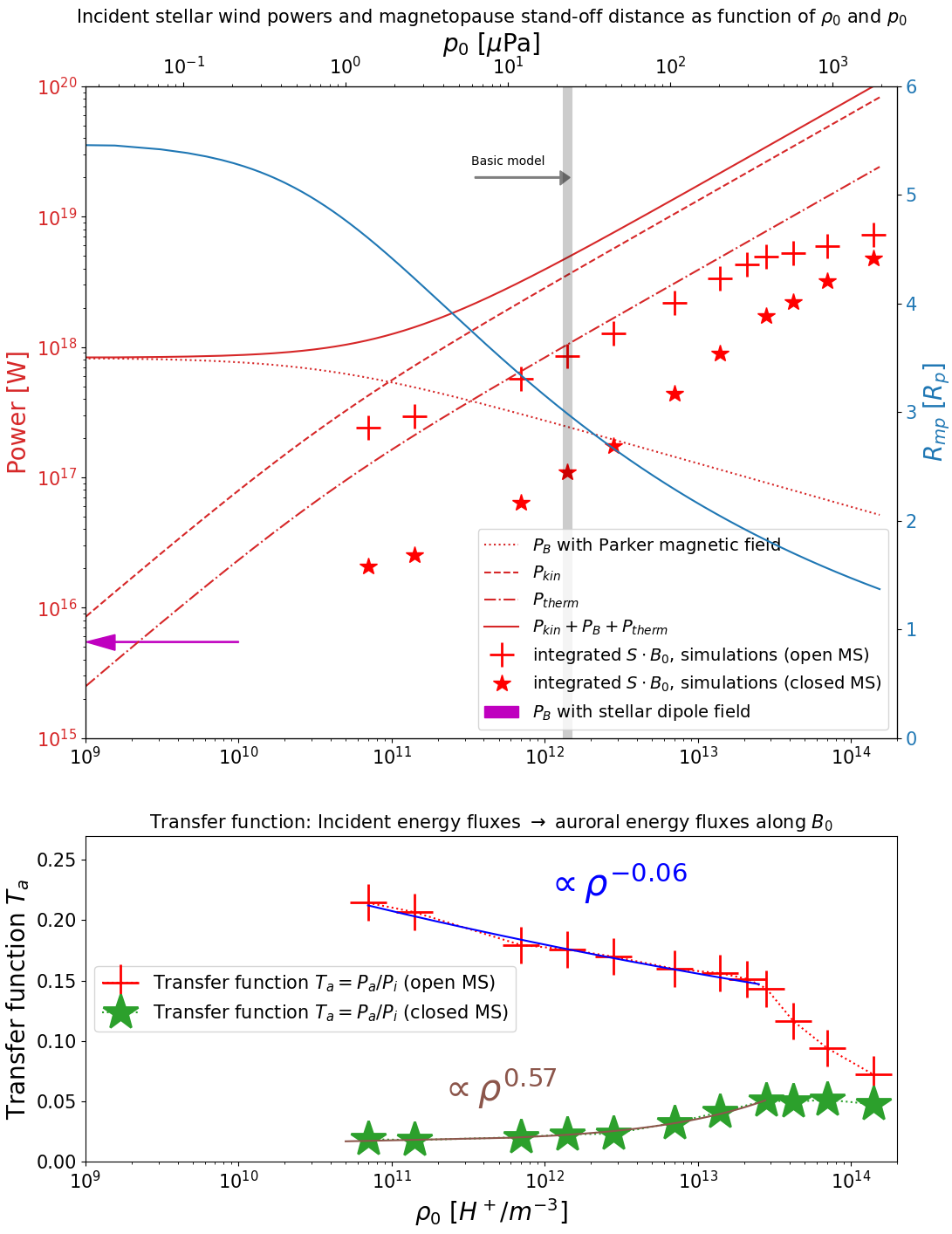}
	\caption{\textbf{Top:} Analytically calculated incident kinetic (red dashed line), Poynting (red dotted line) and thermal (red dashed--dotted line) energy fluxes convected through the magnetospheric cross section $\pi R_{mp}^2$ as function of stellar wind density at the position of $\tau$ Boötis b. The pressure varies accordingly through $p_0 \propto \rho_0$ (see Table \ref{table:parameters} for the basic stellar wind model). The powers are calculated using Eqs. \ref{eq:Pkin} - \ref{eq:Ptherm} and \ref{eq:incident-power_total}. The red solid line represents the sum of the three energy components. The Magnetosphere stand-off distance $R_{mp}$ (blue solid line) is calculated with the same set of $\rho_0$ and $p_0$ using Eq. \ref{eq:Rmp}. The stand-off distance can be obtained from the right y-axis, given in terms of planetary radii $R_p$. The magenta arrow indicates the convected power in the limiting case where the plasma density in the astrosphere approaches zero and the planet is only exposed to the stellar dipole field (see Sect. \ref{section:beyondMHD} for discussion of this case). Red crosses show simulated Poynting fluxes along the unperturbed magnetic field over a spherical shell with radius $2R_p$ for the closed MS (Eq. \ref{eq:S-integral-B0}). Red stars indicate simulated Poynting fluxes for the closed MS.\\
		\textbf{Bottom:} Transfer function $T_a=P_a/P_i$ as a function of $\rho_0$ and $p_0$ (i.e. the conversion from incident energy to auroral Poynting fluxes). The scaling behavior according to power laws of $T_a$ is indicated by spectral indices above the lines.}
	\label{fig:S_vs_rho}
\end{figure}
%-----------------------------------------------------------------
The magnetic Poynting flux, $P_{B}$, and the kinetic energy flux, $P_{kin}$, convected through the obstacle's cross section can be calculated as follows
 \begin{eqnarray}\label{eq:Pkin}
P_{kin} &=& \frac{1}{2} \rho_0  v_0^2 \cdot \pi R_{mp}^2 v_0 \\ \label{eq:PB}
P_B     &=& \frac{B_{sw}^2}{\mu_0} \cdot \pi R_{mp}^2 v_0  \; .
\end{eqnarray}
Additionally, the thermal energy flux should be considered as well as it cannot be neglected for close-in orbits where stellar wind temperature, $T$, pressure and density are high,
\begin{equation} \label{eq:Ptherm}
	P_{therm} = \frac{3}{2} n_{sw} k_B T \cdot \pi R_{mp}^2 v_0 \; ,
\end{equation}
with $n_{sw}$ being the stellar wind particle density and $v_0$ denoting the incident stellar wind velocity.
\\
The magnetopause distance $R_{mp}$ can be obtained from an equilibrium between stellar wind and planetary ram ($p_{ram}=\rho_{sw}v_0^2$), magnetic ($p_B = B^2/2\mu_0$) and thermal pressure. Both, the magnetospheric thermal and ram pressures are considered negligible, thus $p_{ram,sw} + p_{B,sw} + p_{therm,sw} = p_{B,pl}$, where the subscript \emph{sw} stands for stellar wind and \emph{pl} for planet. The magnetopause distance (or magnetospheric stand-off distance) can then be calculated from
\begin{equation}\label{eq:Rmp}
	R_{mp} = R_p  B_{p}^{1/3} \left[ 2\mu_0 \left( \frac{1}{2} \rho_{sw} v^2_{0} + p_{sw}\right)  + B^2_{sw} \right]^{-1/6} \; .
\end{equation}
All parameters can be found in Table \ref{table:parameters}. The parameter $v_{0}$ refers to the relative velocity between the stellar wind and planet and $B_{p}$ to the planetary surface magnetic field at the equator.\\
A certain fraction of the total incident power, 
\begin{equation} \label{eq:incident-power_total}
	P_{i} = P_{kin} + P_B + P_{therm} \; ,
\end{equation}
intersecting with the magnetopause can eventually be converted for the generation of radio emission within the magnetosphere. The fraction of total incident energy, $\epsilon$, that may result in radio emission is expected to range from $10^{-5}$ to $\sim 3\times 10^{-3}$ (i.e. $P_{radio} = \epsilon P_{i}$) in the radiometric Bode law (see Fig. \ref{fig:EfficienciesSchematic}) \citep{Zarka2007}.
We point out that various efficiencies for converting incident energy flux into electromagnetic radiation are discussed in the literature. For example, the efficiency of conversion from magnetospheric, auroral Poynting fluxes to radio emission, which accounts for the efficiency of electron acceleration through wave-particle interaction and the efficiency of the electron-cyclotron maser, should be separated from the generic efficiency factor obtained from the radiometric Bode law \citep{Zarka2007} (see Fig. \ref{fig:EfficienciesSchematic}).
For Jupiter's radio emission the efficiency for conversion from magnetospheric, auroral Ponyting fluxes to radio emission is roughly 0.3 -- 3$\times10^{-4}$ \citep{Saur2021}. We denote this efficiency by $\epsilon_a$ (Fig. \ref{fig:EfficienciesSchematic}).
\\
As the stellar wind density is the most uncertain parameter we performed simulations with densities ranging from $0.05\cdot\rho_0$ to $100\cdot \rho_0$ (see Table \ref{table:parameters} for the basic model). To get an understanding on how stellar wind variability affects the structure of the magnetosphere we show xz--plane slices similar to Fig. \ref{fig:vel_plots} for the two extreme cases ($0.05\cdot\rho_0$ and $100\cdot \rho_0$) in the appendix \ref{section:appendix:structure_of_interaction} (Fig. \ref{fig:vel_plots_extremeCases}). We do not solve a self consistent stellar wind model but instead follow the solar wind solution of \citet{Parker1958} where the solution of the solar wind velocity $v(r)$ is independent of the coronal base density $n_{c,0}$. In this solution, stellar mass and base temperature control $v(r)$ and $T(r)$, where $r$ is the distance from the sun. For simplicity of the parameter study of this subsection, we  choose an isothermal approach and change the density together with the pressure $p_0$ (and therefore $T$) according to $p_0 \propto \rho_0$ (see equation \ref{eq:Ptherm}). We therefore keep the temperature constant and consequently the velocity does not change according to \citet{Parker1958}. Given the average stellar mass loss rate of $\tau$ Boötis A of $\dot{M}\approx 2.3\times 10^{-12}\,M_{\sun}$ yr$^{-1}$ estimated by \citet{Nicholson2016} (see also our basic model, Table \ref{table:parameters}) the parameter range of stellar wind densities considered in this parameter study translates to mass loss rates between $1.15\times 10^{-13}\,M_{\sun}$ yr$^{-1}$ and $2.3\times 10^{-10}\,M_{\sun}$ yr$^{-1}$ since $\dot{M}\propto\rho_0$.
\\
We integrate the Poynting flux along the unperturbed field over a spherical shell with radius $2R_p$ (e.g. Eq. \ref{eq:S-integral-B0}) in order to obtain an understanding of how much incident energy flux is eventually converted to auroral Poynting fluxes. Resulting powers are shown as red crosses and stars (open and closed MS respectively) in Fig. \ref{fig:S_vs_rho} as a function of $\rho_0$ (and $p_0$). The simulated convected energy fluxes follow the trend of incident energy flux estimates (red solid line in Fig. \ref{fig:S_vs_rho}) but are reduced to fractions of the total incident energy flux, $P_i$, between 15 and 20 \% for the open MS and between 1 and 5 \% for the closed MS. 
Changes to the stellar wind density $\rho_0$ (and in the same manner $p_0$) affect the incident power inflicted on $\pi R_{mp}^2$ but also influence the magnetospheric cross section in an opposite manner, as it can be seen in Fig. \ref{fig:S_vs_rho} (blue solid line). The magnetospheric stand-off distance scales according to $R_{mp}\propto (\rho_{sw} v_0^2 + p_{sw})^{-1/6}$ and the incident energy flux with $P_i \propto \rho_{sw}+p_{sw}$, therefore the incident energy flux increase dominates over the effect of a shrinking MS due to increasing thermal and kinetic pressure. This is also validated by our simulation results (Fig. \ref{fig:S_vs_rho}), implying a approximately linear scaling of auroral Poynting fluxes with $\rho_0$ and $p_0$ at least in the regime between $3\times10^{11}$ and $3\times10^{13} H^+$ m$^{-3}$. Below the point where stellar wind magnetic energy dominates over thermal and kinetic energy near $10^{11} H^+$ m$^{-3}$, auroral Poynting fluxes seem to saturate near 2--3 $\times 10^{17}$ W (open MS) and near 1--2$\times10^{16}$ W (closed MS). Above $3\times10^{13} H^+$ m$^{-3}$ the increase of auroral Poynting fluxes with $\rho_0$ (and $p_0$) deviates further from the course of incident flux, implying a saturation towards $10^{19}$ W (open MS). This, however, has to be validated further through future simulations.

%-----------------------------------------------------------------
\section{Discussion}\label{section:discussion}
 In this section we discuss the importance of the stellar wind on magnetospheric energetics (Sect. \ref{section:discussion_stellarwind}) and on possible auroral radio emission (Sect. \ref{section:auroralRadioEmission}).
 \subsection{Importance of the stellar wind on magnetospheric energetics}\label{section:discussion_stellarwind}
 In the following sections we study the conversion of incident to dissipated power within the magnetosphere (Sect. \ref{section:discussion_stellarwind_variability}) as a function of stellar wind density and pressure. We also discuss the limiting case of an absent stellar wind (Sect. \ref{section:beyondMHD}).
 \subsubsection{Stellar wind variability, its effect on magnetospheric energetics and scaling behavior of auroral Poynting fluxes}\label{section:discussion_stellarwind_variability}
 We separate the considered stellar wind density and pressure range introduced in Sect. \ref{section:results:stellar-wind-variability}, Fig. \ref{fig:S_vs_rho} in two regimes: \\
 Regime 1 ranges from a vanishing stellar wind up to a density at roughly $10^{11} H^+$ m$^{-3}$ where kinetic and thermal energy fluxes fall below the persistent magnetic energy flux which dominates the flow (compare red curves in Fig. \ref{fig:S_vs_rho}). Above roughly $10^{11} H^+$ m$^{-3}$ the flow is super-Alfvénic ($M_A \approx2$) and super-fast ($M_f\approx 1$). The interaction is super-Alfvénic for the whole parameter space used in our simulations and sub-fast only for the lowest simulated density ($\rho_{sw}=7\times10^{10}$  $H^+$ m$^{-3}$, $M_A\approx 1.2$). The incident energy nearly stagnates below $\rho_0 =10^{10}\;H^+$ m$^{-3}$  (red dotted line). Below this point the incident energy flux asymptotically approaches its minimum at $8\times10^{17}$ W as we assume that only the plasma density decrease but the incident magnetic field is kept constant.
 In this regime it can be expected that the stellar wind magnetic field solution transitions from the Parker solution (e.g. $B \propto r^{-2}$) to a pure stellar multipole (here dipole) solution (e.g. $B_{sw} = B_{star} \propto r^{-3}$) with decreasing stellar wind density. Eventually, when the stellar wind density hypothetically approaches zero, only the dipolar stellar magnetic field interacts with the planetary magnetic field. This limiting case will be separately discussed in Sect. \ref{section:beyondMHD}.
 \\
 Regime 2 ranges from roughly $10^{11}H^+$ m$^{-3}$ up to arbitrarily high stellar wind densities. Here kinetic and thermal energy fluxes dominate the flow. We will now focus on this regime.
 Considering the total energy flux convected through the magnetospheric cross section $\pi R_{mp}^2$, $P_{total}$ (red solid line in Fig. \ref{fig:S_vs_rho}), we observe a nearly constant efficiency of conversion from incident stellar wind energy towards magnetospheric Poynting fluxes at auroral altitudes (we assumed $r\approx 2R_p$) with increasing density and pressure. We calculate the transfer function $T_a$ as the conversion ratio from total incident energy flux $P_i$ to the simulated auroral Poynting fluxes parallel to the unperturbed field (Eq. \ref{eq:S-integral-B0}), $P_a$ (red crosses and stars in Fig. \ref{fig:S_vs_rho}), within the  MS, such that $T_a = P_a/P_{i}$ (see Fig. \ref{fig:EfficienciesSchematic} for a schematic illustrating the role of $T_a$). The transfer function also contains information on the magnetic topology and thus the efficiency of reconnection. The transfer function is displayed in the lower panel of Fig. \ref{fig:S_vs_rho}. For the open MS, auroral Poynting fluxes decrease as a function of $\rho_0$ and $p_0$ according to an approximately constant ratio up to $\sim 3\times10^{13} H^+$ m$^{-3}$. For higher densities and pressures, the transfer function scales with an exponent of $\sim -0.4$, indicating a decrease of efficiency for conversion from incident to auroral energy fluxes. As the MS is increasingly compressed due to stronger ram and thermal pressures, the magnetopause eventually crosses the spherical shell with $r=2R_p$ after a critical density of $\sim 3\times10^{13} H^+$ m$^{-3}$ and pressure of $3\times10^{-4}$ Pa. 
 \\
 For the closed MS, $T_a$ behaves differently. The transfer function instead increases slightly from $7\times10^{10}$ to $\sim 3\times10^{13} H^+$ m$^{-3}$ following a power law with exponent $\sim 0.6$. The opposite behavior compared to the open MS transfer function might be a consequence of a geometry where less reconnection occurs. The stronger ram and thermal pressure exert stronger tension on planetary field lines which in turn release stronger energy fluxes during reconnection.

 \subsubsection{Beyond the MHD limit}\label{section:beyondMHD}
 The more the stellar wind density decreases, the emptier the heliosphere of $\tau$ Boötis A becomes. In analogy with the solar wind, the stellar wind density upstream of the magnetosphere of $\tau$ Boötis b may change by orders of magnitude. The solar wind density is observed to vary by more than two orders of magnitude \citep[see e.g.][]{Chane2012} such that Earth's bow shock can disappear and the Earth develops Alfvén wings. In the hypothetical limit when the density approaches zero, $\tau$ Boötis b will still be exposed to the stellar magnetic field $B_\tau(r)$ (which then decreases according to $B_\tau(r) \propto r^{-3}$ instead of $B_{sw} \propto r^{-2}$) and will propagate through it. Therefore the incoming Poynting flux of the star does not, in contrast to kinetic and thermal energy flux, vanish and is maintained by the relative motion between the stellar and planetary magnetic field. In case of an empty heliosphere, the interaction around $\tau$ Boötis b is not magnetohydrodynamic any more, but turns electromagnetic. Then the movement of $\tau$ Boötis b within the external magnetic field of the star is a unipolar or homopolar interaction (i.e. a moving conductive object with external field similar to a current generator in classical electromagnetism).
 In the case of a stellar magnetic field rotating at the same speed as the planet orbits around its host star (i.e. in case of total orbital and spin synchronization), non-existent changes in magnetic flux lead to a system where no work can be done by the magnetic fields. Therefore no currents are induced and the magnetosphere remains energetically silent. Although it is expected that close-in exoplanets are in nearly tidally locked rotation, \citet{Murray2000} suggest in their textbook that completely synchronous rotation might not be possible if the planet has no permanent magnetic quadrupole moment and its eccentricity is not zero, which is the case for $\tau$ Boötis b \citep{Wang2011}. Taking the rotation period of $\tau$ Boötis A, $P_{\tau} = 3.1 \pm 0.1$ d \citep{Brown2021, Mengel2016}, and the sidereal rotation period of $\tau$ Boötis b, $P_{orb}= 3.31$ d \citep{Butler1997,Wang2011}, we can calculate the relative velocity between the stellar and planetary magnetic field $v_0 = 2\pi a\times (1/P_{\tau} - 1/P_{orb}) \approx 10.4$ km s$^{-1}$, where $a=0.046$ AU is the semi-major axis of $\tau$ Boötis b. We use an average surface magnetic flux density of $\tau$ Boötis A obtained by \citet{Marsden2014,Mengel2016,Jeffers2018}, $B_{\tau,0} = 2.4$ G, and calculate the flux density at $0.046$ AU using the dipole formula, $B_{\tau}(r=a) = B_{\tau,0} \cdot (a/R_{\tau})^{-3} \approx 0.73$ G, where $R_{\tau} = 1.43R_{\odot}$ is the stellar radius \citep{Bonfanti2016}. The magnetospheric stand-off distance in this case is defined purely by the balance of stellar and planetary magnetic pressure, i.e. 
 \begin{equation}
 	{R}_{mp} = R_p \cdot ( B_{\tau}(a)/B_{p} )^{-1/6} \approx 8.6 \;R_p \;.
 \end{equation}
 Now we can calculate the stellar magnetic power convected on the MS using 
 \begin{equation}
 	P_{B} = B_{\tau}(a)^2/\mu_0 \cdot v_0 \pi R_{mp}^2 \approx 8\times10^{15}\;\text{W}\;.
 \end{equation}
 This value is included in Fig. \ref{fig:S_vs_rho} as magenta arrow. We can conclude the following:
 \\ 
 (\textbf{a}) The maximum possible magnetospheric stand-off distance is reached in complete absence of a stellar wind and if the stellar and planetary magnetic fields are anti-parallel. If the stellar and planetary fields are parallel (i.e. open MS) the planetary field lines are all connected to the stellar field and no magnetopause can be determined. In the closed MS case the magnetopause lies at roughly 8.6 planetary radii in the upstream direction. 
 \\
 (\textbf{b}) Even if there is no stellar wind, the magnetic interaction between the stellar and planetary magnetic field still has the potential to drive an interaction with an available power limit of roughly $8\times10^{15}$ W due to the relative motion of $\tau$ Boötis b in the stellar magnetic field. Radio emission would still be possible although very weak; corresponding radio fluxes at Earth's position would be far below today's telescope sensitivity limit. Considering Poynting flux--to--radio power efficiencies between $10^{-4}$ and $10^{-2}$, radio powers can reach values between $10^{11}$ and $10^{14}$ W. These emitted powers exceed the strongest radio sources within the solar system by several orders of magnitude, with Jupiter's aurora being the strongest radio emitter \citep[$P_{radio} \approx 10^{10} - 10^{11}$ W][]{Zarka2007}, although Jupiter's emission is, in contrast to close-in exoplanets, powered by internal, rotationally driven mechanisms.
 The transition from a magnetosphere interaction with a stellar wind field \citep[according to the Parker solution][]{Parker1958} to an interaction with a pure dipolar stellar magnetic field goes with an energetic transition followed by a decrease of maximum emitted radio power. This might pose a possible opportunity for constraining stellar wind densities in the future. Solving a self consistent stellar wind model and comparing auroral Poynting fluxes for different stellar wind base densities could reveal the critical density range where the transition from a stellar magnetic field dominated electrodynamic interaction to a stellar wind dominated magnethydrodynamic interaction takes place. Comparing the magnetospheric Poynting fluxes and corresponding radio powers with possible future observations could reveal if the stellar wind density lies below or above the critical density.
 
 \subsection{Magnetospheric Poynting fluxes and auroral radio emission}\label{section:auroralRadioEmission}
 In this section we discuss possible radio emission scenarios (Sect. \ref{section:radioEmissionScenarios}), study how magnetic topology of the interaction as well as stellar wind variability affects auroral radio emission output (Sect. \ref{section:tilt_aurora}). We also discuss the possibility of a sub--Alfvénic emission scenario (Sect. \ref{section:subAlfvenic_aurora}) as well as a rotation driven magnetosphere of $\tau$ Boötis b in Sect. \ref{section:rotation_aurora}, followed by a discussion on possible source regions and radio frequencies of auroral emission in Sect. \ref{section:properties-aurora}.
 \subsubsection{On the different radio emission scenarios}\label{section:radioEmissionScenarios}
 There are several scenarios capable of generating observable radio emission that must be distinguished from each other. 
 \\
 \textbf{(1)} If the stellar wind is sub--Alfvénic, Alfvén waves are able to propagate upstream towards the stellar atmosphere along Alfvén wings and possibly drive electron acceleration and radio emission in the stellar vicinity. Local radio emission within the magnetosphere can also be generated in this scenario. We will refer to this scenario as sub--Alfvénic emission scenario. This scenario is discussed briefly in Sect. \ref{section:subAlfvenic_aurora}.
 \\
 \textbf{(2)} If the stellar wind is super--Alfvénic, no MHD wave is able to propagate upstream. The stellar wind--magnetosphere interaction however drives Poynting fluxes within the magnetosphere which may to some extent generate auroral radio emission (see Fig. \ref{fig:EfficienciesSchematic}). We refer to this scenario as the magnetospheric emission scenario. Due to the stellar wind being super--Alfvénic for all simulations we will focus on the magnetospheric emission (Sect. \ref{section:tilt_aurora}).
 \\
 \textbf{(3)} In a rotation dominated magnetosphere scenario the rotating planet and its magnetic field causes co--rotation of magnetospheric plasma that, at some point, breaks-down due to conservation of angular momentum if radial mass transfer takes place. This co--rotation breakdown exerts magnetic stresses on the field lines that are the root cause of auroral Poynting fluxes which in turn drive auroral radio emission. This scenario is discussed in Sect. \ref{section:rotation_aurora}.
 
 \subsubsection{Effects of magnetic field tilt and stellar wind variability on auroral radio emission -- Magnetospheric emission scenario}\label{section:tilt_aurora}
\begin{figure}
 	\centering
 	\includegraphics[width=0.99\linewidth]{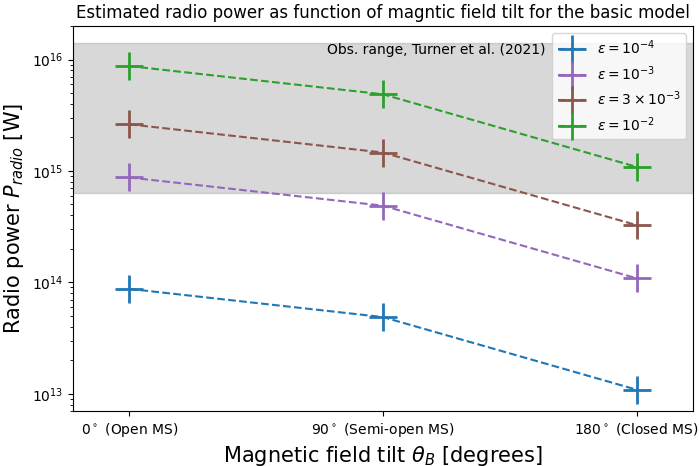}
 	\caption{Expected radio powers as function of planetary magnetic axis tilt using auroral-to-radio power conversion efficiencies between $10^{-4}$ and $10^{-2}$. The auroral Poynting fluxes $\vec{S}\cdot \vec{B_0}$ are integrated over a spherical shell with radius $2R_p$. The gray shaded area represents observational limits given by \citet{Turner2021}.}
 	\label{fig:power_vs_tilt}
 \end{figure}

 Figure \ref{fig:power_vs_tilt} shows radio powers as function of magnetic axis tilt. Radio powers are obtained from multiplying integrated auroral Poynting fluxes (i.e. Eq. \ref{eq:S-integral-B0}), which serve as a proxy for maximum available electromagnetic energy that is transported along magnetic field lines, with efficiency factors for converting magnetospheric Poynting fluxes to radio--power, $\epsilon$, ranging from $10^{-4}$ to $10^{-2}$. This range covers proposed \citep{Zarka2007} and observed efficiency factors \citep[e.g. $\epsilon \approx 10^{-4}$ for Jupiter][]{Saur2021}. The modeled magnetic field tilt can also be interpreted as stellar magnetic field orientation within this work, allowing us to study the effect of varying stellar magnetic field polarity on magnetospheric Poynting fluxes and limits for associated radio emission. Radio powers within the limits inferred from observations by \citet{Turner2021} lie within the gray shaded area. It is visible that efficiency factors in the range of $\epsilon \approx (0.3-1)\times10^{-2}$ deliver radio powers most consistent with observations if the MS is open or at least semi open given the basic model (Table \ref{table:parameters}). This indicates that the efficiency of auroral Poynting fluxes driving electron acceleration and the electron cyclotron maser emission may be higher in the magnetosphere of $\tau$ Boötis b than in the Jovian magnetosphere \citep{Saur2021}. Electric fields generated by reconnection between stellar wind and planetary magnetic field lines are expected to contribute significantly to powering electron acceleration and therefore the ECMI \citep{Jardine2008}.
 In our studies we find reconnection to indirectly play an important role (Fig. \ref{fig:power_vs_tilt}) because auroral Poynting fluxes and consequently radio powers drop by nearly an order of magnitude from an open to a closed MS. This is due to magnetic stress exerted by the stellar wind interaction being less strong for closed magnetospheres.
 The polarity of $\tau$ Boötis A's magnetic field switches every approximate 360 days \citep{Fares2013}. Shorter cycles in magnetic activity levels (by means of S-indices) were also observed \citep{Mengel2016}. A difference of half an order of magnitude to almost an order of magnitude can therefore be caused by a polarity reversal of $\tau$ Boötis A's magnetic field. This results in radio emission whose observability is expected to fluctuated periodically in a nearly 1-year cycle. We note that the stellar wind magnetic field strength was kept constant in our parameter study, although in reality the field strength may vary strongly and influence produced radio emission significantly \citep{See2015}. 
 \\
 
 \begin{figure}
 	\centering
 	\includegraphics[width=0.99\linewidth]{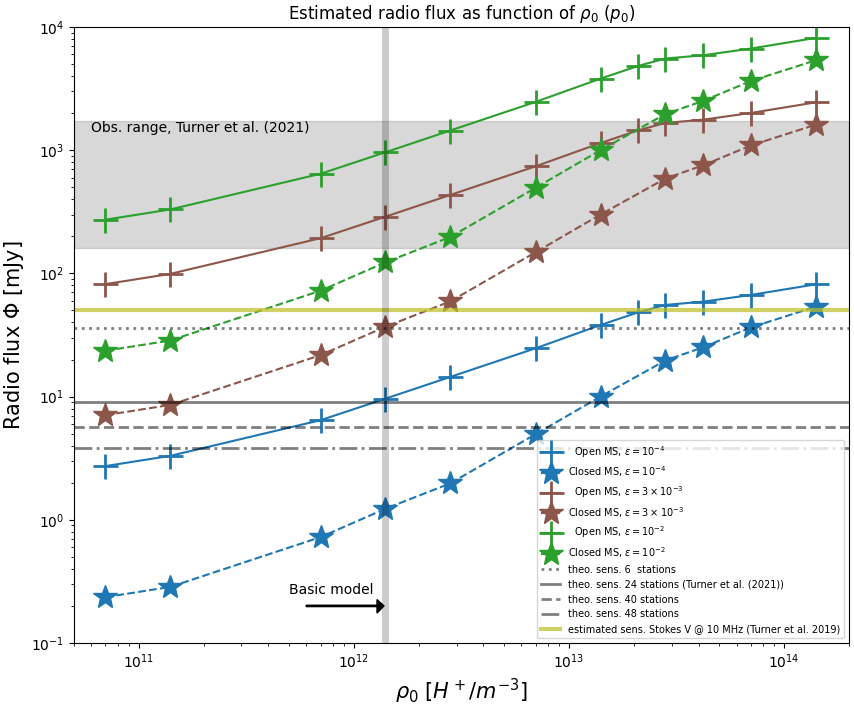}
 	\caption{Radio flux (Eq. \ref{eq:radio-flux}) as function of stellar wind density and pressure for different efficiency factors. Colored solid and dashed lines represent fluxes for the open and closed MS model. Observational limits \citep{Turner2021} are indicated by the gray shaded area. Horizontal gray lines display theoretical sensitivity limits of the LOFAR telescope. The real sensitivity for Stokes V signals obtained from \citet{Turner2019} is plotted as yellow line. The vertical gray line marks the basic model (Table \ref{table:parameters}).}
 	\label{fig:radioflux_vs_rho}
 \end{figure}
 The emitted radio flux observed at Earth's position can be calculated with \citep{Griessmeier2005,Griessmeier2007a}
 \begin{equation}\label{eq:radio-flux}
 	\Phi = \frac{P_{radio}}{\Omega\, \delta\nu\, d^2 }\;,
 \end{equation}
 where $\Omega$ is the solid angle of the beam and $\delta\nu$ the emission bandwidth that is approximately equal to the maximum gyro-frequency \citep{Griessmeier2007a}, $\nu_{g,e} \approx 24$ MHz. We assume a solid angle of $\Omega=1.6$ sr similar to Jupiter's decametric radio emission \citep{Zarka2004}. The distance to the $\tau$ Boötis system is 15.6 pc. We calculate the radio flux for both, the open and closed MS model, as a function of $\rho_0$ and $p_0$ with radio efficiencies $\epsilon_a = 10^{-4}$--$10^{-2}$. The results are displayed in Fig. \ref{fig:radioflux_vs_rho}. Solid and dashed colored lines represent radio fluxes originating from open and closed MS's, respectively. The gray shaded area again denotes the range of observed radio fluxes form \citet{Turner2021}. Horizontal gray lines indicate theoretical sensitivity limits of the LOFAR telescopes for 20MHz. As stated by \citet{Turner2019}, the realistic sensitivity might be slightly lower for circularly polarized (Stokes V) signals. We therefore include the expected sensitivity calculated by \citet{Turner2019} as yellow line. The results in Fig. \ref{fig:radioflux_vs_rho} indicate that radio efficiencies between $\sim 3\times 10^{-3}$ and $\sim 10^{-2}$ are most consistent with the tentative observations \citep{Turner2021}. The efficiency accounts for several steps from conversion of auroral Poynting fluxes to radio emission (e.g. wave-particle interaction, electron acceleration and ECMI), therefore an efficiency of the order of 1--10\% might be unrealistic. The efficiency for Jupiter's auroral emission is roughly $\epsilon_a = 0.3$--$3\times10^{-4}$ \citep{Saur2021}, therefore $\epsilon_a = 10^{-2}$ might be too high. Moreover high plasma densities within the magnetosphere injected by the dense stellar wind and due to strong irradiation which results in high ionization rates and inflated atmospheres \citep[e.g. for $\nu$ And b see][]{Erkaev2022}, may further decrease the ECMI efficiency or even prevent it \citep{Weber2017,Weber2018,Daley-Yates2018}. Assuming the radio efficiency to lie near $10^{-3}$, the radio flux from a closed MS falls below the detection threshold (yellow line). Therefore, in case of a polarity reversal of $\tau$ Boöts A's magnetic field (i.e. from aligned with the planetary field to anti-aligned), the radio signal would not be observable anymore in case of radio efficiency equal or below $\sim 3\times10^{-3}$. In case of $\epsilon_a \approx 10^{-4}$ all radio fluxes for the basic model fall below the sensitivity limit. The observability increases, however, if stellar wind density and pressure rises, rendering $\epsilon_a = 10^{-3}$--$10^{-2}$ to possible efficiencies to observe emission from open and closed MS's. Additionally, the ECMI efficiency \citep{Treumann2006,Weber2017} as well as efficiency of electron acceleration through wave-particle interaction decreases dramatically with increasing plasma density \citep{Saur2018}, making the higher density and pressure regime a less likely scenario to explain the tentative observations. 
 As the pressure rises, the magnetopause is getting closer to the planet, reducing the space of magnetospheric diluted plasma regions between the magnetopause and atmosphere where radio emission might occur. We therefore conclude that the basic model (vertical gray line) and slightly different configurations represent the most likely scenarios if the emission is indeed generated by stellar wind driven auroral Poynting fluxes. In this case radio emission is only observable, if the stellar wind and planetary magnetic fields are aligned (i.e. the magnetosphere is open). Given the high efficiencies ($\epsilon > 10^{-3}$) needed by our model in order to generate radio emission which is consistent with the tentative observations, the magnetospheric emission scenario might not be energetic enough to explain the observations.
 
 \subsubsection{Sub--Alfvénic emission scenario}\label{section:subAlfvenic_aurora}
 Although there is no sub-Alfvénic interaction within the parameter space we considered, the possibility of such an interaction and its consequences on possible radio emission should not be neglected. By choosing a stellar wind density of $\rho_{sw} = 0.03$  $\rho_0$ we find an Alfvénic Mach number of $M_A\approx 0.9$. In this case Alfvén waves may propagate back to the star through Alfvén wings connecting the planetary magnetic field with the star. The electromagnetic energy channeled through this flux tube can be calculated using the model of \citet{Saur2013},
 \begin{equation}
 	P_S = 2\pi R_{mp}^2 \frac{\left(\bar{\alpha} B_{sw} \cos\theta\right)^2}{\mu_0} M_A v_0 \; ,
 \end{equation}
 where $\theta = 0^\circ$ is the angle which describes the deviation of the flow from being perpendicular to the stellar wind magnetic field, $R_{mp} \approx 5 R_p$ the magnetospheric stand--of distance and $\bar{\alpha}$ the interaction strength. Due to the planet presumably possessing an ionosphere which favors a strong plasma interaction we choose $\bar{\alpha}\approx 1$. Using $v_0$ and $B_{sw}$ from our basic model (Table \ref{table:parameters}), the energy flux channeled through Alfvén wings parallel to magnetic field lines amounts to $P_S = 1.2\times 10^{18}$ W. This energy flux may contribute to electron acceleration and consequently to ECMI driven radio emission in the space environment near the star. Taking the same range of Poynting flux to radio efficiencies ($10^{-4}$--$10^{-2}$) the expected radio power ranges from roughly $10^{14}$ to $10^{16}$ W which partially overlaps with the range of radio power deduced by \citet{Turner2021}. The emission frequency would, however, be much lower compared to the signals observed by \citet{Turner2021} due to the low magnetic field strength of $\tau$ Boötis A \citep{Turner2021}. Additionally \citet{Turner2021} state that stellar emission would be less strongly circularly polarized compared to planetary emission. Nevertheless the root energy flux able to possibly drive radio emission near the stellar atmosphere from such a sub--Alfvénic interaction exceeds the auroral Poynting fluxes in the magnetospheric emission scenario by roughly half an order of magnitude. Consequently this scenario might provide energetically a possible explanation for the observed radio emission if the interaction is indeed sub--Alfvénic. This scenario requires the stellar wind density to be reduced by at least a factor of $\sim$ 30 compared to our basic model. This, however, could well be within the realm of possible densities given the observed density variations of the solar wind even tough the intervals of variability are short \citep{Chane2012}.

 \subsubsection{Is auroral radio emission from $\tau$ Boötis b rotationally driven?}\label{section:rotation_aurora}
 One of the possible generators for auroral emissions is radial mass transport within the magnetosphere as it is the case for Jupiter \citep[e.g.][]{Hill2001,Zarka2018,Zarka2021}. Plasma is moving radially outwards due to centrifugal forces within an exoplanet's MS. For small distances from the planet (or stellar host) the plasma co-rotates with the host. After a certain distance $\hat{L}$ in units of planetary radii, $\hat{L}\times R_p$, the co-rotation breaks down due to conservation of angular momentum (i.e. the plasma orbits its host with a smaller angular velocity than those from the rotation of the host). The relative velocity exerts magnetic stresses on the frozen-in field lines, the tension accelerates plasma along the field lines back to the host. The distance of co-rotation breakdown can be estimated using the co-latitude of the region, where auroral emission occurs, $\theta_{a}$, following \citet{Hill2001}, $\sin(\theta_{a})=\hat{L}^{-1/2}$. \citet{Saur2021} derived a so called auroral power potential for magnetized rotation dominated hosts,
 \begin{equation}\label{eq:auroral_power_potential}
 	S_{pot} = B^2_{host} \Omega^2_{host} R^2_{host}\;,
 \end{equation}
 where the subscript \emph{host} refers in our case to the planet $\tau$ Boötis b. The hosts magnetic flux density is denoted by $B_{host}$, $\Omega_{host}\approx 2\pi/P_{orb}= 2.2\times10^{-5}$ rad s$^{-1}$ is its angular velocity and $R_{host}=R_p$ the planet's radius. We note that the discussion in this subsection is only based on theoretical considerations since we neglected planetary rotation in our MHD model. Following \citet{Saur2021} we can calculate the energy flux due to mass transport, $P_{mag}$, by using
 \begin{equation}\label{eq:auroral_magnetic_power}
 	P_{mag} = S_{pot} \left( \pi \frac{R_p^2}{\hat{L}^2} \right) \Sigma_p \;,
 \end{equation}
 where $\Sigma_p$ denotes the Pedersen conductance. We assume a value of 1 S and 10 S that are of the same order of magnitude as observed conductances near Jupiter \citep[e.g.][]{Hill2001,Strobel1983,Hinson1998}. For a hot Jupiter exoplanet the ionospheric Pedersen conductivity might be larger due to the larger ionization \citep{Koskinen2010}. However, also the height of the resulting ionospheric layer and thus the mobility of the electrons and ions within the atmosphere also plays an important role for the values of its conductivity. To our knowledge no dedicated study for the conductance of $\tau$ Boötis b is available. We also note that in Eq. \ref{eq:auroral_power_potential} the distance of co--rotation breakdown $\hat{L}$ is proportional to $\Sigma^{1/4}$ \citep{Hill2001}. Thus $P_{mag}$ in Eq. \ref{eq:auroral_magnetic_power} depends effectively weaker on $\Sigma_{p}$ (i.e. $P_{mag}$ is proportional to $\Sigma_p^{1/2}$). In our simulations we integrate the ion component of the Pedersen conductivity along the z--axis at the poles from the surface ($r=R_p$) to $r=1.1$ $R_p$ where the plasma density peaks and find the height integrated Pedersen conductance to be of the order of 10 S. We find, with aurora occurring at field lines with L-parameter smaller than $\sim 4$ that have to lie within the MS (e.g. $\hat{L} \approx 3.5$), the auroral power potential to be $S_{pot} \approx 0.5$ W m$^{-2}$ S$^{-1}$ and the magnetic power to lie between $P_{mag} \approx 10^{14}\;W$ and $10^{15}\;W$. The auroral power potential of Jupiter, for comparison, is roughly $600$ W m$^{-2}$ S$^{-1}$. The resulting energy flux between the magnetosphere and ionosphere lies near the Jovian value  \citep[$3.1\times10^{14}$ W][]{Saur2021}. This result is several orders of magnitude below the power of magnetospheric Poynting fluxes powered by the stellar wind -- planet interaction in our model (e.g. $10^{17}$ -- $10^{18}$ W, Table. \ref{table:results}). Therefore we can safely assume that the MS of $\tau$ Boötis b is not rotationally dominated which is in agreement with the expected slow rotation nearly synchronized with its orbit around the star.

 \subsubsection{Properties and source regions of possible auroral radio emission}\label{section:properties-aurora}
 Auroral radio emission is mainly characterized by signal strength and frequency. The peak frequency is defined by the local electron gyrofrequency,
 \begin{equation}\label{eq:gyrofrequency}
 	\nu_{g,e} = \frac{|e|B}{2\pi m_e}\;,
 \end{equation}
 where $e$ is the electron charge, $m_e$ its mass and $B$ the local magnetic flux density \citep{Farrell1999,Griessmeier2007b,Griessmeier2007a}. We already discussed possible radio powers from $\tau$ Boötis b in Sects. \ref{section:energetics} and \ref{section:tilt_aurora}, now we will focus on possible signal frequencies. \citet{Turner2021} observed signals within the 21 -- 30 MHz range.
 The magnetic flux density inferred from the maximal signal frequency by the authors is almost certainly underestimated since ECMI driven auroral radio emission typically arises from altitudes above the polar regions of the planet's atmosphere \citep{Zarka1998, Hess2011} due to three reasons: (\textbf{a}) High ionospheric electron densities cause the electron plasma frequency $\nu_{p,e}$ to exceed the local gyro frequency $\nu_{g,e}$, thus prohibiting the transmission of emitted waves; (\textbf{b}) radio wave amplification through the ECMI mechanism works most efficiently in regions where $\nu_{g,e} >> \nu_{p,e}$ \citep{Treumann2006,Weber2017} and (\textbf{c}) the electrons are usually accelerated in regions where the plasma density along a field line is the smallest (or $B/n$ is maximum) \citep[e.g.][]{Saur2018}. There wave particle interaction to accelerate electrons is maximum efficient.\\
 The electron plasma frequency is given by
 \begin{equation}\label{eq:plasmafrequency}
 	\nu_{p,e} = \sqrt{\frac{n_e e^2}{\epsilon_0 m_e}}\frac{1}{2\pi}\;,
 \end{equation}
 where $n_e = n_i$ is the particle density in a quasi-neutral plasma and $\epsilon_0$ the vacuum permittivity.
 The source regions of Jupiter's radio emission lie several planetary radii above the surface \citep[e.g.][]{Zarka1998} where strongest electron acceleration is expected \citep[e.g.][]{Mauk2017}. The maximum gyro frequency in our model (directly above the magnetic poles) is $\sim 24$ MHz. Auroral emission, however, typically arises from near the OCFB. In our simulations we indeed see that strongest Poynting fluxes are confined to narrow bands encircling the OCFB (or magnetopause) near L=3 and L=3.5 shell field lines (Fig. \ref{fig:Poynting_vector}). Since source regions of ECMI induced emission typically lie at higher altitudes the magnetic flux density and thus emission frequency is even further reduced.
 The conditions (\textbf{a}--\textbf{c}) for efficient ECMI driven radio emission might particularly be fulfilled for regions with low plasma density at the night side of the planet where the shadow exerted by the planet prevents photo-ionization. Since $\tau$ Boötis b is likely in tidally locked rotation around its host star, the night side might exhibit relatively low temperatures, favoring recombination of electrons and hydrogen ions as well as further electron capture processes. In the shadow zone (i.e. permanent night side of the planet), ECMI induced emission might be more likely and might as well occur in lower altitudes, where the magnetic field is stronger (and thus emission frequency is higher).

%-----------------------------------------------------------------
\section{Conclusions}\label{section:conclusion}
In this study we modeled the magnetosphere of the Hot Jupiter exoplanet $\tau$ Boötis b and its interaction with the stellar wind. In order to study magnetospheric energetics by means of Poynting fluxes we performed MHD simulations of the near space environment of $\tau$ Boötis b with the stellar wind being injected into the simulation domain. The magnetic field of $\tau$ Boötis b is constrained by tentative observations of auroral radio emission \citep{Turner2021}. The stellar wind model is based on simulations where observed magnetic surface maps of $\tau$ Boötis A were utilized \citep{Nicholson2016,Vidotto2012}. We investigated magnetospheric energetics, available electromagnetic power and limits of possible radio emission originating from above the planet's polar regions. We also explored the possibility of a rotation dominated magnetospheric emission scenario as well as a sub--Alfvénic (stellar) emission scenario. The effect of stellar wind variability in terms of density, pressure and magnetic field orientation on magnetospheric energetics was additionally investigated. 
\\
\\
We find that the stellar wind--magnetosphere interaction is super-fast and super-Alfvénic for the majority of stellar wind densities and pressures considered in this study (from $1.4\times10^{11}$ H$^+$ m$^{-3}$ and $2.5\times10^{-6}$ Pa to arbitrarily high stellar wind densities and pressures). For the minimum density and pressure considered in this work ($\rho_{sw} = 7\times10^{10}$ H$^+$ m$^{-3}$ and $p_{sw} = 1.3\times10^{-5}$ Pa) the interaction is super--Alfvénic and sub--fast. Maximum available electromagnetic energy convected within the magnetosphere amounts to 3.5$\times10^{18}$ W, 1.8$\times10^{18}$ W and 7$\times10^{17}$ W for an open, semi-open and closed magnetosphere configuration. Auroral Poynting fluxes at altitudes of 1$R_p$ above the planet reach powers of 8.7$\times10^{17}$ W down to 1.1$\times10^{17}$ W for the open and closed magnetosphere, indicating a strong decrease of stellar wind energy converted to auroral Poynting fluxes as a function of magnetic field tilt (either stellar or planetary). 
\\
We present Poynting flux maps of the planet's near space environment. Strongest Poynting fluxes are confined to narrow bands encircling the open-closed field line boundaries (i.e. the magnetopause) that is displaced towards the planet's downstream side due to magnetic stresses exerted by the stellar wind and its high plasma $\beta$. Electromagnetic energy transported parallel to unperturbed field lines are as well confined to bands near the magnetopause and the polar open-field line regions.
\\
Considering the stellar wind density as free parameter, we choose values between 0.05 and 100 times the basic density, $\rho_{sw} = 1.4\times10^{12}$ H$^+$ m$^{-3}$ (Table \ref{table:parameters}), and change the stellar wind pressure in the same manner in order to keep the wind isothermal. Comparing the incident total energy flux convected through the magnetospheric cross section, we find that the transfer function $T_a$ (i.e. the amount of stellar wind incident energy flux converted to magnetospheric Poynting fluxes parallel to unperturbed field lines) amounts to a fraction of roughly 20 \% for the open and 2 \% for the closed MS. We find that $T_a\propto\rho_0^{-0.06}$ and $T_a\propto\rho_0^{0.57}$ for the open and closed MS respectively in the density range 7$\times10^{11}$ -- 2$\times10^{13}$ H$^+$ m$^{-3}$. Increasing stellar wind ram and thermal pressure and thus incident energy flux causes the shrinking of the magnetosphere. The reduced capability of the MS for receiving stellar wind energy is outperformed by the increasing pressure and density. The scaling of $T_a$ and thus energy output of the magnetosphere nearly stagnates from that point on where the magnetopause turns so small that the planetary atmosphere starts to interact with pre-bow shock and magnetosheath flow. 
\\
We investigated different radio emission scenarios. A rotation dominated magnetospheric emission scenario could be ruled out due to the expected long rotation period of $\tau$ Boötis b. The estimated magnetic power generated by the planet's rotation is on the order of $P_{mag} \approx 10^{14-15}$ W. As this would require a very high Poynting flux to radio efficiency of $10^{-2}$, we expect that this scenario is not powerful enough to drive the tentatively observed radio emission.\\
We considered several efficiencies for conversion of auroral Poynting fluxes to radio emission and compared the results with observations together with theoretical limits based on the tentative observations by \citet{Turner2021}. We find that generation of radio emission near $\tau$ Boötis b requires relatively high efficiencies ($\epsilon_a\approx 1$--$5\times10^{-3}$) compared to estimates from Jupiter's rotationally driven aurora (up to$\sim 3\times10^{-4}$) \citep{Saur2021} in order to explain the observed emission if we assume the emission to originate from the magnetosphere of $\tau$ Boötis b. Furthermore we find a strong dependence of radio emission observability on magnetosphere topology. In case of a polarity reversal of $\tau$ Boötis A's magnetic field from an aligned to anti-aligned (with respect to the planetary field) configuration, the estimated radio power falls below the observational limit. We therefore expect an on--off nature of detectable radio emission in phase with $\tau$ Boötis A's magnetic cycle.
\\
We also considered the case of a sub--Alfvénic interaction which requires the stellar wind density to be at least a factor of 30 lower compared to our basic model. The Poynting flux channeled towards the host star due to this interaction amounts to $\sim 10^{18}$ W which slightly exceeds the power converted by the magnetospheric emission scenario.
\begin{acknowledgements}
This project has received funding from the European Research Council (ERC) under the European Union’s Horizon 2020 research and innovation programme (grant agreement No. 884711).
\\
We furthermore thank the Regional Computing Center of the University of Cologne (RRZK) for providing computing time on the DFG-funded (Funding number: INST 216/512/1FUGG) High Performance Computing (HPC) system CHEOPS as well as support.
\\
We also thank Stefan Duling for providing several basic model implementations for the Pluto code used in this study as well as for helpful discussions.
\end{acknowledgements}

\bibliographystyle{aa}
\bibliography{article}

\begin{thebibliography}{104}
\expandafter\ifx\csname natexlab\endcsname\relax\def\natexlab#1{#1}\fi

\bibitem[{{Alexander} {et~al.}(2016){Alexander}, {Wynn}, {Mohammed}, {Nichols},
  \& {Ercolano}}]{Alexander2016}
{Alexander}, R.~D., {Wynn}, G.~A., {Mohammed}, H., {Nichols}, J.~D., \&
  {Ercolano}, B. 2016, \mnras, 456, 2766

\bibitem[{{Atreya} {et~al.}(2003){Atreya}, {Mahaffy}, {Niemann}, {Wong}, \&
  {Owen}}]{Atreya2003}
{Atreya}, S.~K., {Mahaffy}, P.~R., {Niemann}, H.~B., {Wong}, M.~H., \& {Owen},
  T.~C. 2003, \planss, 51, 105

\bibitem[{{Bastian} {et~al.}(2022){Bastian}, {Cotton}, \&
  {Hallinan}}]{Bastian2022}
{Bastian}, T., {Cotton}, B., \& {Hallinan}, G. 2022, arXiv e-prints,
  arXiv:2206.14099

\bibitem[{{Bisikalo} {et~al.}(2013){Bisikalo}, {Kaigorodov}, {Ionov}, \&
  {Shematovich}}]{Bisikalo2013}
{Bisikalo}, D.~V., {Kaigorodov}, P.~V., {Ionov}, D.~E., \& {Shematovich}, V.~I.
  2013, Astronomy Reports, 57, 715

\bibitem[{{Bonfanti} {et~al.}(2016){Bonfanti}, {Ortolani}, \&
  {Nascimbeni}}]{Bonfanti2016}
{Bonfanti}, A., {Ortolani}, S., \& {Nascimbeni}, V. 2016, \aap, 585, A5

\bibitem[{{Brown} {et~al.}(2021){Brown}, {Marsden}, {Mengel}, {Jeffers},
  {Millburn}, {Mittag}, {Petit}, {Vidotto}, {Morin}, {See}, {Jardine},
  {Gonz{\'a}lez-P{\'e}rez}, {Gonz{\'a}lez-P{\'e}rez}, \& {BCool
  Collaboration}}]{Brown2021}
{Brown}, E.~L., {Marsden}, S.~C., {Mengel}, M.~W., {et~al.} 2021, \mnras, 501,
  3981

\bibitem[{{Butler} {et~al.}(1997){Butler}, {Marcy}, {Williams}, {Hauser}, \&
  {Shirts}}]{Butler1997}
{Butler}, R.~P., {Marcy}, G.~W., {Williams}, E., {Hauser}, H., \& {Shirts}, P.
  1997, \apjl, 474, L115

\bibitem[{{Cauley} {et~al.}(2019){Cauley}, {Shkolnik}, {Llama}, \&
  {Lanza}}]{Cauley2019}
{Cauley}, P.~W., {Shkolnik}, E.~L., {Llama}, J., \& {Lanza}, A.~F. 2019, Nature
  Astronomy, 3, 1128

\bibitem[{{Chan{\'e}} {et~al.}(2012){Chan{\'e}}, {Saur}, {Neubauer}, {Raeder},
  \& {Poedts}}]{Chane2012}
{Chan{\'e}}, E., {Saur}, J., {Neubauer}, F.~M., {Raeder}, J., \& {Poedts}, S.
  2012, Journal of Geophysical Research (Space Physics), 117, A09217

\bibitem[{Cohen {et~al.}(2014)Cohen, Drake, Glocer, Garraffo, Poppenhaeger,
  Bell, Ridley, \& Gombosi}]{Cohen2014}
Cohen, O., Drake, J.~J., Glocer, A., {et~al.} 2014, The Astrophysical Journal,
  790, 57 (13pp)

\bibitem[{{Cohen} {et~al.}(2011){Cohen}, {Kashyap}, {Drake}, {Sokolov},
  {Garraffo}, \& {Gombosi}}]{Cohen2011}
{Cohen}, O., {Kashyap}, V.~L., {Drake}, J.~J., {et~al.} 2011, \apj, 733, 67

\bibitem[{{Cohen} {et~al.}(2018){Cohen}, {Moschou}, {Glocer}, {Sokolov},
  {Mazeh}, {Drake}, {Garraffo}, \& {Alvarado-G{\'o}mez}}]{Cohen2018}
{Cohen}, O., {Moschou}, S.-P., {Glocer}, A., {et~al.} 2018, \aj, 156, 202

\bibitem[{{Cuntz} {et~al.}(2000){Cuntz}, {Saar}, \& {Musielak}}]{Cuntz2000}
{Cuntz}, M., {Saar}, S.~H., \& {Musielak}, Z.~E. 2000, \apjl, 533, L151

\bibitem[{{Daley-Yates} \& {Stevens}(2018)}]{Daley-Yates2018}
{Daley-Yates}, S. \& {Stevens}, I.~R. 2018, \mnras, 479, 1194

\bibitem[{{Dedner} {et~al.}(2002){Dedner}, {Kemm}, {Kr{\"o}ner}, {Munz},
  {Schnitzer}, \& {Wesenberg}}]{Dedner2002}
{Dedner}, A., {Kemm}, F., {Kr{\"o}ner}, D., {et~al.} 2002, Journal of
  Computational Physics, 175, 645

\bibitem[{{Donati} {et~al.}(2008){Donati}, {Moutou}, {Far{\`e}s}, {Bohlender},
  {Catala}, {Deleuil}, {Shkolnik}, {Collier Cameron}, {Jardine}, \&
  {Walker}}]{Donati2008}
{Donati}, J.~F., {Moutou}, C., {Far{\`e}s}, R., {et~al.} 2008, \mnras, 385,
  1179

\bibitem[{{Duling} {et~al.}(2014){Duling}, {Saur}, \& {Wicht}}]{Duling2014}
{Duling}, S., {Saur}, J., \& {Wicht}, J. 2014, Journal of Geophysical Research
  (Space Physics), 119, 4412

\bibitem[{{Erkaev} {et~al.}(2022){Erkaev}, {Weber}, {Grie{\ss}meier}, {Lammer},
  {Ivanov}, \& {Odert}}]{Erkaev2022}
{Erkaev}, N.~V., {Weber}, C., {Grie{\ss}meier}, J.~M., {et~al.} 2022, \mnras,
  512, 4869

\bibitem[{{Fares} {et~al.}(2009){Fares}, {Donati}, {Moutou}, {Bohlender},
  {Catala}, {Deleuil}, {Shkolnik}, {Collier Cameron}, {Jardine}, \&
  {Walker}}]{Fares2009}
{Fares}, R., {Donati}, J.~F., {Moutou}, C., {et~al.} 2009, \mnras, 398, 1383

\bibitem[{{Fares} {et~al.}(2013){Fares}, {Moutou}, {Donati}, {Catala},
  {Shkolnik}, {Jardine}, {Cameron}, \& {Deleuil}}]{Fares2013}
{Fares}, R., {Moutou}, C., {Donati}, J.~F., {et~al.} 2013, \mnras, 435, 1451

\bibitem[{{Farrell} {et~al.}(1999){Farrell}, {Desch}, \& {Zarka}}]{Farrell1999}
{Farrell}, W.~M., {Desch}, M.~D., \& {Zarka}, P. 1999, \jgr, 104, 14025

\bibitem[{{Franklin} \& {Burke}(1958)}]{Franklin1958}
{Franklin}, K.~L. \& {Burke}, B.~F. 1958, \jgr, 63, 807

\bibitem[{{Gray} {et~al.}(2001){Gray}, {Napier}, \& {Winkler}}]{Gray2001}
{Gray}, R.~O., {Napier}, M.~G., \& {Winkler}, L.~I. 2001, \aj, 121, 2148

\bibitem[{{Grie{\ss}meier}(2015)}]{Griessmeier2015}
{Grie{\ss}meier}, J.-M. 2015, in Astrophysics and Space Science Library, Vol.
  411, Characterizing Stellar and Exoplanetary Environments, ed. H.~{Lammer} \&
  M.~{Khodachenko}, 213

\bibitem[{{Griessmeier}(2017)}]{Griessmeier2017}
{Griessmeier}, J.~M. 2017, in Planetary Radio Emissions VIII, ed. G.~{Fischer},
  G.~{Mann}, M.~{Panchenko}, \& P.~{Zarka}, 285--299

\bibitem[{{Grie{\ss}meier} {et~al.}(2005){Grie{\ss}meier}, {Motschmann},
  {Mann}, \& {Rucker}}]{Griessmeier2005}
{Grie{\ss}meier}, J.~M., {Motschmann}, U., {Mann}, G., \& {Rucker}, H.~O. 2005,
  \aap, 437, 717

\bibitem[{{Grie{\ss}meier} {et~al.}(2007{\natexlab{a}}){Grie{\ss}meier},
  {Preusse}, {Khodachenko}, {Motschmann}, {Mann}, \&
  {Rucker}}]{Griessmeier2007b}
{Grie{\ss}meier}, J.~M., {Preusse}, S., {Khodachenko}, M., {et~al.}
  2007{\natexlab{a}}, \planss, 55, 618

\bibitem[{{Grie{\ss}meier} {et~al.}(2011){Grie{\ss}meier}, {Zarka}, \&
  {Girard}}]{Griessmeier2011}
{Grie{\ss}meier}, J.~M., {Zarka}, P., \& {Girard}, J.~N. 2011, Radio Science,
  46, RS0F09

\bibitem[{{Grie{\ss}meier} {et~al.}(2007{\natexlab{b}}){Grie{\ss}meier},
  {Zarka}, \& {Spreeuw}}]{Griessmeier2007a}
{Grie{\ss}meier}, J.~M., {Zarka}, P., \& {Spreeuw}, H. 2007{\natexlab{b}},
  \aap, 475, 359

\bibitem[{{Gurumath} {et~al.}(2018){Gurumath}, {Hiremath}, \&
  {Ramasubramanian}}]{Gurumath2018}
{Gurumath}, S.~R., {Hiremath}, K.~M., \& {Ramasubramanian}, V. 2018, IAU
  Symposium, 340, 242

\bibitem[{{Hess} \& {Zarka}(2011)}]{Hess2011}
{Hess}, S.~L.~G. \& {Zarka}, P. 2011, \aap, 531, A29

\bibitem[{{Hill}(2001)}]{Hill2001}
{Hill}, T.~W. 2001, \jgr, 106, 8101

\bibitem[{{Hinson} {et~al.}(1998){Hinson}, {Twicken}, \&
  {Karayel}}]{Hinson1998}
{Hinson}, D.~P., {Twicken}, J.~D., \& {Karayel}, E.~T. 1998, \jgr, 103, 9505

\bibitem[{{Ip} {et~al.}(2004){Ip}, {Kopp}, \& {Hu}}]{Ip2004}
{Ip}, W.-H., {Kopp}, A., \& {Hu}, J.-H. 2004, \apjl, 602, L53

\bibitem[{{Jardine} \& {Collier Cameron}(2008)}]{Jardine2008}
{Jardine}, M. \& {Collier Cameron}, A. 2008, \aap, 490, 843

\bibitem[{{Jeffers} {et~al.}(2018){Jeffers}, {Mengel}, {Moutou}, {Marsden},
  {Barnes}, {Jardine}, {Petit}, {Schmitt}, {See}, {Vidotto}, \& {BCool
  Collaboration}}]{Jeffers2018}
{Jeffers}, S.~V., {Mengel}, M., {Moutou}, C., {et~al.} 2018, \mnras, 479, 5266

\bibitem[{{Kavanagh} {et~al.}(2021){Kavanagh}, {Vidotto}, {Klein}, {Jardine},
  {Donati}, \& {{\'O} Fionnag{\'a}in}}]{Kavanagh2021}
{Kavanagh}, R.~D., {Vidotto}, A.~A., {Klein}, B., {et~al.} 2021, \mnras, 504,
  1511

\bibitem[{{Kavanagh} {et~al.}(2020){Kavanagh}, {Vidotto}, {{\'O}
  Fionnag{\'a}in}, {Bourrier}, {Fares}, {Jardine}, {Helling}, {Moutou},
  {Llama}, \& {Wheatley}}]{Kavanagh2020}
{Kavanagh}, R.~D., {Vidotto}, A.~A., {{\'O} Fionnag{\'a}in}, D., {et~al.} 2020,
  in Solar and Stellar Magnetic Fields: Origins and Manifestations, ed.
  A.~{Kosovichev}, S.~{Strassmeier}, \& M.~{Jardine}, Vol. 354, 305--309

\bibitem[{{Kavanagh} {et~al.}(2022){Kavanagh}, {Vidotto}, {Vedantham},
  {Jardine}, {Callingham}, \& {Morin}}]{Kavanagh2022}
{Kavanagh}, R.~D., {Vidotto}, A.~A., {Vedantham}, H.~K., {et~al.} 2022, \mnras,
  514, 675

\bibitem[{{Kislyakova} {et~al.}(2014){Kislyakova}, {Holmstr{\"o}m}, {Lammer},
  {Odert}, \& {Khodachenko}}]{Kislyakova2014}
{Kislyakova}, K.~G., {Holmstr{\"o}m}, M., {Lammer}, H., {Odert}, P., \&
  {Khodachenko}, M.~L. 2014, Science, 346, 981

\bibitem[{{Kislyakova} {et~al.}(2016){Kislyakova}, {Pilat-Lohinger}, {Funk},
  {Lammer}, {Fossati}, {Eggl}, {Schwarz}, {Boudjada}, \&
  {Erkaev}}]{Kislyakova2016}
{Kislyakova}, K.~G., {Pilat-Lohinger}, E., {Funk}, B., {et~al.} 2016, \mnras,
  461, 988

\bibitem[{{Kopp} {et~al.}(2011){Kopp}, {Schilp}, \& {Preusse}}]{Kopp2011}
{Kopp}, A., {Schilp}, S., \& {Preusse}, S. 2011, \apj, 729, 116

\bibitem[{{Koskinen} {et~al.}(2010){Koskinen}, {Cho}, {Achilleos}, \&
  {Aylward}}]{Koskinen2010}
{Koskinen}, T.~T., {Cho}, J. Y.~K., {Achilleos}, N., \& {Aylward}, A.~D. 2010,
  \apj, 722, 178

\bibitem[{{Lai} {et~al.}(2010){Lai}, {Helling}, \& {van den Heuvel}}]{Lai2010}
{Lai}, D., {Helling}, C., \& {van den Heuvel}, E.~P.~J. 2010, \apj, 721, 923

\bibitem[{{Llama} {et~al.}(2011){Llama}, {Wood}, {Jardine}, {Vidotto},
  {Helling}, {Fossati}, \& {Haswell}}]{Llama2011}
{Llama}, J., {Wood}, K., {Jardine}, M., {et~al.} 2011, \mnras, 416, L41

\bibitem[{{Maggio} {et~al.}(2011){Maggio}, {Sanz-Forcada}, \&
  {Scelsi}}]{Maggio2011}
{Maggio}, A., {Sanz-Forcada}, J., \& {Scelsi}, L. 2011, \aap, 527, A144

\bibitem[{{Marsden} {et~al.}(2014){Marsden}, {Petit}, {Jeffers}, {Morin},
  {Fares}, {Reiners}, {do Nascimento}, {Auri{\`e}re}, {Bouvier}, {Carter},
  {Catala}, {Dintrans}, {Donati}, {Gastine}, {Jardine}, {Konstantinova-Antova},
  {Lanoux}, {Ligni{\`e}res}, {Morgenthaler}, {Ram{\`\i}rez-V{\`e}lez},
  {Th{\'e}ado}, {Van Grootel}, \& {BCool Collaboration}}]{Marsden2014}
{Marsden}, S.~C., {Petit}, P., {Jeffers}, S.~V., {et~al.} 2014, \mnras, 444,
  3517

\bibitem[{{Mauk} {et~al.}(2020){Mauk}, {Clark}, {Gladstone}, {Kotsiaros},
  {Adriani}, {Allegrini}, {Bagenal}, {Bolton}, {Bonfond}, {Connerney}, {Ebert},
  {Haggerty}, {Kollmann}, {Kurth}, {Levin}, {Paranicas}, \& {Rymer}}]{Mauk2020}
{Mauk}, B.~H., {Clark}, G., {Gladstone}, G.~R., {et~al.} 2020, Journal of
  Geophysical Research (Space Physics), 125, e27699

\bibitem[{{Mauk} {et~al.}(2017){Mauk}, {Haggerty}, {Paranicas}, {Clark},
  {Kollmann}, {Rymer}, {Mitchell}, {Bolton}, {Levin}, {Adriani}, {Allegrini},
  {Bagenal}, {Connerney}, {Gladstone}, {Kurth}, {McComas}, {Ranquist},
  {Szalay}, \& {Valek}}]{Mauk2017}
{Mauk}, B.~H., {Haggerty}, D.~K., {Paranicas}, C., {et~al.} 2017, \grl, 44,
  4410

\bibitem[{{Mengel} {et~al.}(2016){Mengel}, {Fares}, {Marsden}, {Carter},
  {Jeffers}, {Petit}, {Donati}, {Folsom}, \& {BCool
  Collaboration}}]{Mengel2016}
{Mengel}, M.~W., {Fares}, R., {Marsden}, S.~C., {et~al.} 2016, \mnras, 459,
  4325

\bibitem[{{Mignone} {et~al.}(2007){Mignone}, {Bodo}, {Massaglia}, {Matsakos},
  {Tesileanu}, {Zanni}, \& {Ferrari}}]{Mignone2007}
{Mignone}, A., {Bodo}, G., {Massaglia}, S., {et~al.} 2007, in JENAM-2007, ``Our
  Non-Stable Universe'', 96--96

\bibitem[{{Mignone} {et~al.}(2010){Mignone}, {Tzeferacos}, \&
  {Bodo}}]{Mignone2010}
{Mignone}, A., {Tzeferacos}, P., \& {Bodo}, G. 2010, Journal of Computational
  Physics, 229, 5896

\bibitem[{{Miller} {et~al.}(2012){Miller}, {Gallo}, {Wright}, \&
  {Dupree}}]{Miller2012}
{Miller}, B.~P., {Gallo}, E., {Wright}, J.~T., \& {Dupree}, A.~K. 2012, \apj,
  754, 137

\bibitem[{{Miller} {et~al.}(2015){Miller}, {Gallo}, {Wright}, \&
  {Pearson}}]{Miller2015}
{Miller}, B.~P., {Gallo}, E., {Wright}, J.~T., \& {Pearson}, E.~G. 2015, \apj,
  799, 163

\bibitem[{{Mittag} {et~al.}(2017){Mittag}, {Robrade}, {Schmitt}, {Hempelmann},
  {Gonz{\'a}lez-P{\'e}rez}, \& {Schr{\"o}der}}]{Mittag2017}
{Mittag}, M., {Robrade}, J., {Schmitt}, J.~H.~M.~M., {et~al.} 2017, \aap, 600,
  A119

\bibitem[{Murray \& Dermott(2000)}]{Murray2000}
Murray, C.~D. \& Dermott, S.~F. 2000, Solar System Dynamics (Cambridge
  University Press)

\bibitem[{{Nichols} \& {Milan}(2016)}]{Nichols2016}
{Nichols}, J.~D. \& {Milan}, S.~E. 2016, \mnras, 461, 2353

\bibitem[{{Nicholson} {et~al.}(2016){Nicholson}, {Vidotto}, {Mengel},
  {Brookshaw}, {Carter}, {Petit}, {Marsden}, {Jeffers}, {Fares}, \& {BCool
  Collaboration}}]{Nicholson2016}
{Nicholson}, B.~A., {Vidotto}, A.~A., {Mengel}, M., {et~al.} 2016, \mnras, 459,
  1907

\bibitem[{{Parker}(1958)}]{Parker1958}
{Parker}, E.~N. 1958, \apj, 128, 664

\bibitem[{{Preusse} {et~al.}(2006){Preusse}, {Kopp}, {B{\"u}chner}, \&
  {Motschmann}}]{Preusse2006}
{Preusse}, S., {Kopp}, A., {B{\"u}chner}, J., \& {Motschmann}, U. 2006, \aap,
  460, 317

\bibitem[{{Preusse} {et~al.}(2007){Preusse}, {Kopp}, {B{\"u}chner}, \&
  {Motschmann}}]{Preusse2007}
{Preusse}, S., {Kopp}, A., {B{\"u}chner}, J., \& {Motschmann}, U. 2007,
  \planss, 55, 589

\bibitem[{{Route}(2019)}]{Route2019}
{Route}, M. 2019, \apj, 872, 79

\bibitem[{{Saur} {et~al.}(2013){Saur}, {Grambusch}, {Duling}, {Neubauer}, \&
  {Simon}}]{Saur2013}
{Saur}, J., {Grambusch}, T., {Duling}, S., {Neubauer}, F.~M., \& {Simon}, S.
  2013, \aap, 552, A119

\bibitem[{{Saur} {et~al.}(2018){Saur}, {Janser}, {Schreiner}, {Clark}, {Mauk},
  {Kollmann}, {Ebert}, {Allegrini}, {Szalay}, \& {Kotsiaros}}]{Saur2018}
{Saur}, J., {Janser}, S., {Schreiner}, A., {et~al.} 2018, Journal of
  Geophysical Research (Space Physics), 123, 9560

\bibitem[{{Saur} {et~al.}(2021){Saur}, {Willmes}, {Fischer}, {Wennmacher},
  {Roth}, {Youngblood}, {Strobel}, \& {Reiners}}]{Saur2021}
{Saur}, J., {Willmes}, C., {Fischer}, C., {et~al.} 2021, \aap, 655, A75

\bibitem[{{See} {et~al.}(2015){See}, {Jardine}, {Fares}, {Donati}, \&
  {Moutou}}]{See2015}
{See}, V., {Jardine}, M., {Fares}, R., {Donati}, J.~F., \& {Moutou}, C. 2015,
  \mnras, 450, 4323

\bibitem[{{Shkolnik} {et~al.}(2008){Shkolnik}, {Bohlender}, {Walker}, \&
  {Collier Cameron}}]{Shkolnik2008}
{Shkolnik}, E., {Bohlender}, D.~A., {Walker}, G. A.~H., \& {Collier Cameron},
  A. 2008, \apj, 676, 628

\bibitem[{{Shkolnik} {et~al.}(2003){Shkolnik}, {Walker}, \&
  {Bohlender}}]{Shkolnik2003}
{Shkolnik}, E., {Walker}, G.~A.~H., \& {Bohlender}, D.~A. 2003, \apj, 597, 1092

\bibitem[{{Shkolnik} {et~al.}(2005){Shkolnik}, {Walker}, {Bohlender}, {Gu}, \&
  {K{\"u}rster}}]{Shkolnik2005}
{Shkolnik}, E., {Walker}, G.~A.~H., {Bohlender}, D.~A., {Gu}, P.~G., \&
  {K{\"u}rster}, M. 2005, \apj, 622, 1075

\bibitem[{{Storey} \& {Hummer}(1995)}]{Storey1995}
{Storey}, P.~J. \& {Hummer}, D.~G. 1995, \mnras, 272, 41

\bibitem[{{Strobel} \& {Atreya}(1983)}]{Strobel1983}
{Strobel}, D.~F. \& {Atreya}, S.~K. 1983, in Physics of the Jovian
  Magnetosphere, 51--67

\bibitem[{{Strugarek} {et~al.}(2019{\natexlab{a}}){Strugarek}, {Ahuir}, {Brun},
  {Donati}, {Moutou}, \& {R{\'e}ville}}]{Strugarek2019a}
{Strugarek}, A., {Ahuir}, J., {Brun}, A.~S., {et~al.} 2019{\natexlab{a}}, in
  SF2A-2019: Proceedings of the Annual meeting of the French Society of
  Astronomy and Astrophysics, ed. P.~{Di Matteo}, O.~{Creevey}, A.~{Crida},
  G.~{Kordopatis}, J.~{Malzac}, J.~B. {Marquette}, M.~{N'Diaye}, \& O.~{Venot},
  Di

\bibitem[{{Strugarek} {et~al.}(2019{\natexlab{b}}){Strugarek}, {Brun},
  {Donati}, {Moutou}, \& {R{\'e}ville}}]{Strugarek2019b}
{Strugarek}, A., {Brun}, A.~S., {Donati}, J.~F., {Moutou}, C., \&
  {R{\'e}ville}, V. 2019{\natexlab{b}}, \apj, 881, 136

\bibitem[{{Strugarek} {et~al.}(2014){Strugarek}, {Brun}, {Matt}, \&
  {R{\'e}ville}}]{Strugarek2014}
{Strugarek}, A., {Brun}, A.~S., {Matt}, S.~P., \& {R{\'e}ville}, V. 2014, \apj,
  795, 86

\bibitem[{{Strugarek} {et~al.}(2015){Strugarek}, {Brun}, {Matt}, \&
  {R{\'e}ville}}]{Strugarek2015}
{Strugarek}, A., {Brun}, A.~S., {Matt}, S.~P., \& {R{\'e}ville}, V. 2015, \apj,
  815, 111

\bibitem[{{Tabata} \& {Shirai}(2000)}]{Tabata2000}
{Tabata}, T. \& {Shirai}, T. 2000, Atomic Data and Nuclear Data Tables, 76, 1

\bibitem[{{Treumann}(2006)}]{Treumann2006}
{Treumann}, R.~A. 2006, \aapr, 13, 229

\bibitem[{{Turner} {et~al.}(2016){Turner}, {Christie}, {Arras}, {Johnson}, \&
  {Schmidt}}]{Turner2016}
{Turner}, J.~D., {Christie}, D., {Arras}, P., {Johnson}, R.~E., \& {Schmidt},
  C. 2016, \mnras, 458, 3880

\bibitem[{{Turner} {et~al.}(2019){Turner}, {Grie{\ss}meier}, {Zarka}, \&
  {Vasylieva}}]{Turner2019}
{Turner}, J.~D., {Grie{\ss}meier}, J.-M., {Zarka}, P., \& {Vasylieva}, I. 2019,
  \aap, 624, A40

\bibitem[{{Turner} {et~al.}(2021){Turner}, {Zarka}, {Grie{\ss}meier}, {Lazio},
  {Cecconi}, {Emilio Enriquez}, {Girard}, {Jayawardhana}, {Lamy}, {Nichols}, \&
  {de Pater}}]{Turner2021}
{Turner}, J.~D., {Zarka}, P., {Grie{\ss}meier}, J.-M., {et~al.} 2021, \aap,
  645, A59

\bibitem[{{Turnpenney} {et~al.}(2018){Turnpenney}, {Nichols}, {Wynn}, \&
  {Burleigh}}]{Turnpenney2018}
{Turnpenney}, S., {Nichols}, J.~D., {Wynn}, G.~A., \& {Burleigh}, M.~R. 2018,
  \apj, 854, 72

\bibitem[{{Turnpenney} {et~al.}(2020){Turnpenney}, {Nichols}, {Wynn}, \&
  {Jia}}]{Turnpenney2020}
{Turnpenney}, S., {Nichols}, J.~D., {Wynn}, G.~A., \& {Jia}, X. 2020, \mnras,
  494, 5044

\bibitem[{{Van Doorsselaere} {et~al.}(2011){Van Doorsselaere}, {Wardle}, {Del
  Zanna}, {Jansari}, {Verwichte}, \& {Nakariakov}}]{Doorsselaere2011}
{Van Doorsselaere}, T., {Wardle}, N., {Del Zanna}, G., {et~al.} 2011, \apjl,
  727, L32

\bibitem[{{Varela} {et~al.}(2022){Varela}, {Brun}, {Strugarek}, {R{\'e}ville},
  {Zarka}, \& {Pantellini}}]{Varela2022}
{Varela}, J., {Brun}, A.~S., {Strugarek}, A., {et~al.} 2022, \aap, 659, A10

\bibitem[{{Varela} {et~al.}(2016){Varela}, {Reville}, {Brun}, {Pantellini}, \&
  {Zarka}}]{Varela2016}
{Varela}, J., {Reville}, V., {Brun}, A.~S., {Pantellini}, F., \& {Zarka}, P.
  2016, \aap, 595, A69

\bibitem[{{Varela} {et~al.}(2018){Varela}, {R{\'e}ville}, {Brun}, {Zarka}, \&
  {Pantellini}}]{Varela2018}
{Varela}, J., {R{\'e}ville}, V., {Brun}, A.~S., {Zarka}, P., \& {Pantellini},
  F. 2018, \aap, 616, A182

\bibitem[{{Vidal-Madjar} {et~al.}(2003){Vidal-Madjar}, {Lecavelier des Etangs},
  {D{\'e}sert}, {Ballester}, {Ferlet}, {H{\'e}brard}, \&
  {Mayor}}]{VidalMadjar2003}
{Vidal-Madjar}, A., {Lecavelier des Etangs}, A., {D{\'e}sert}, J.-M., {et~al.}
  2003, \nat, 422, 143

\bibitem[{{Vidotto} \& {Donati}(2017)}]{Vidotto2017}
{Vidotto}, A.~A. \& {Donati}, J.~F. 2017, \aap, 602, A39

\bibitem[{{Vidotto} {et~al.}(2012){Vidotto}, {Fares}, {Jardine}, {Donati},
  {Opher}, {Moutou}, {Catala}, \& {Gombosi}}]{Vidotto2012}
{Vidotto}, A.~A., {Fares}, R., {Jardine}, M., {et~al.} 2012, \mnras, 423, 3285

\bibitem[{{Vidotto} {et~al.}(2015){Vidotto}, {Fares}, {Jardine}, {Moutou}, \&
  {Donati}}]{Vidotto2015}
{Vidotto}, A.~A., {Fares}, R., {Jardine}, M., {Moutou}, C., \& {Donati}, J.~F.
  2015, \mnras, 449, 4117

\bibitem[{{Vidotto} {et~al.}(2010){Vidotto}, {Jardine}, \&
  {Helling}}]{Vidotto2010}
{Vidotto}, A.~A., {Jardine}, M., \& {Helling}, C. 2010, \apjl, 722, L168

\bibitem[{{Vidotto} {et~al.}(2011){Vidotto}, {Jardine}, \&
  {Helling}}]{Vidotto2011}
{Vidotto}, A.~A., {Jardine}, M., \& {Helling}, C. 2011, \mnras, 411, L46

\bibitem[{{Wang} \& {Ford}(2011)}]{Wang2011}
{Wang}, J. \& {Ford}, E.~B. 2011, \mnras, 418, 1822

\bibitem[{{Weber} {et~al.}(2018){Weber}, {Erkaev}, {Ivanov}, {Odert},
  {Grie{\ss}meier}, {Fossati}, {Lammer}, \& {Rucker}}]{Weber2018}
{Weber}, C., {Erkaev}, N.~V., {Ivanov}, V.~A., {et~al.} 2018, \mnras, 480, 3680

\bibitem[{{Weber} {et~al.}(2017){Weber}, {Lammer}, {Shaikhislamov}, {Chadney},
  {Khodachenko}, {Grie{\ss}meier}, {Rucker}, {Vocks}, {Macher}, {Odert}, \&
  {Kislyakova}}]{Weber2017}
{Weber}, C., {Lammer}, H., {Shaikhislamov}, I.~F., {et~al.} 2017, \mnras, 469,
  3505

\bibitem[{{Wood} {et~al.}(2021){Wood}, {M{\"u}ller}, {Redfield}, {Konow},
  {Vannier}, {Linsky}, {Youngblood}, {Vidotto}, {Jardine},
  {Alvarado-G{\'o}mez}, \& {Drake}}]{Wood2021}
{Wood}, B.~E., {M{\"u}ller}, H.-R., {Redfield}, S., {et~al.} 2021, \apj, 915,
  37

\bibitem[{{Zarka}(1998)}]{Zarka1998}
{Zarka}, P. 1998, \jgr, 103, 20159

\bibitem[{{Zarka}(2007)}]{Zarka2007}
{Zarka}, P. 2007, \planss, 55, 598

\bibitem[{Zarka {et~al.}(2004)Zarka, Cecconi, \& Kurth}]{Zarka2004}
Zarka, P., Cecconi, B., \& Kurth, W.~S. 2004, Journal of Geophysical Research:
  Space Physics, 109

\bibitem[{{Zarka} {et~al.}(2021){Zarka}, {Magalh{\~a}es}, {Marques}, {Louis},
  {Echer}, {Lamy}, {Cecconi}, \& {Prang{\'e}}}]{Zarka2021}
{Zarka}, P., {Magalh{\~a}es}, F.~P., {Marques}, M.~S., {et~al.} 2021, Journal
  of Geophysical Research (Space Physics), 126, e29780

\bibitem[{{Zarka} {et~al.}(2018){Zarka}, {Marques}, {Louis}, {Ryabov}, {Lamy},
  {Echer}, \& {Cecconi}}]{Zarka2018}
{Zarka}, P., {Marques}, M.~S., {Louis}, C., {et~al.} 2018, \aap, 618, A84

\bibitem[{{Zarka} {et~al.}(2001){Zarka}, {Treumann}, {Ryabov}, \&
  {Ryabov}}]{Zarka2001}
{Zarka}, P., {Treumann}, R.~A., {Ryabov}, B.~P., \& {Ryabov}, V.~B. 2001,
  \apss, 277, 293

\bibitem[{{Zhilkin} \& {Bisikalo}(2019)}]{Zhilkin2019}
{Zhilkin}, A.~G. \& {Bisikalo}, D.~V. 2019, Astronomy Reports, 63, 550

\bibitem[{{Zhilkin} \& {Bisikalo}(2020)}]{Zhilkin2020}
{Zhilkin}, A.~G. \& {Bisikalo}, D.~V. 2020, Astronomy Reports, 64, 563

\end{thebibliography}

%===========================================================================
%===========================================================================
%======================= APPENDIX ==========================================
%===========================================================================
%===========================================================================
%===========================================================================
\clearpage
\appendix
\section{Poynting fluxes as function of radial distance from the planet}\label{section:appendix:S_vs_r}
\begin{figure}[h]
	\centering  
	\includegraphics[width=0.99\linewidth]{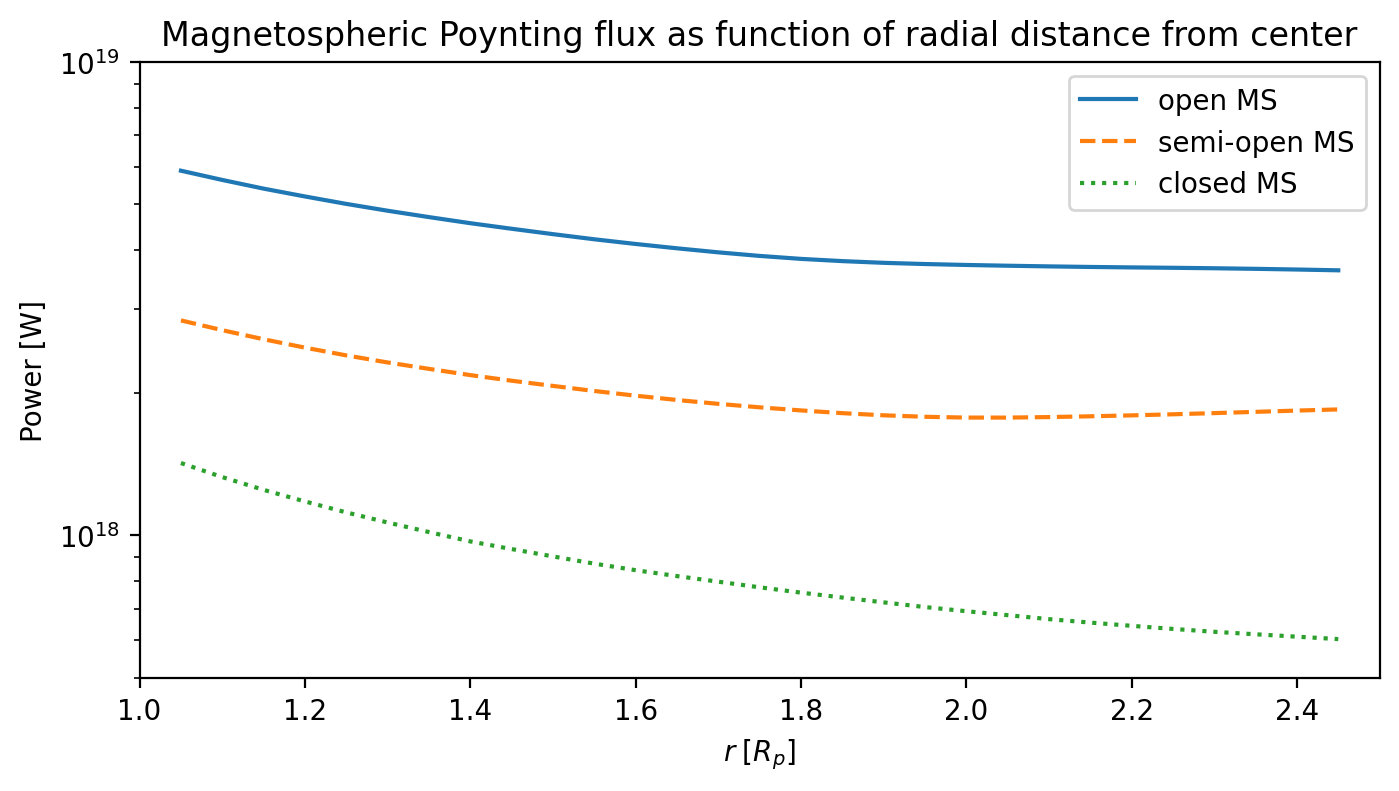}
	\caption{Magnetospheric Poynting flux as function of radial distance from the planet's center calculated from Eq. \ref{eq:S-integral-mag} for the open (blue), semi--open (orange) and closed (green) magnetosphere model.}
	\label{fig:S_vs_r}
\end{figure}
Here we study the dependence of magnetospheric Poynting fluxes as a function of radial distance from the planet. The particular choice of integrating Poynting fluxes over a spherical shell with radius $r=2R_p$ is based on the fact that radio emission around Jupiter and most solar system planets with intrinsic magnetic fields originates from altitudes of about 1 $R_p$ or higher above the planets \citep[see e.g.][]{Zarka1998} rather than from within the ionosphere or even further below. We do not have evidence if this translates to extrasolar planets but simply assume so. For the ECMI mechanism to work efficiently the ratio between electron gyro frequency and plasma frequency, $\nu_{g,e}/\nu_{p,e}$, must be significantly larger than 1 which turns the magnetospheric region at high altitudes to favored radio source regions where plasma densities are strongly reduced compared to other regions of the magnetosphere. \citep{Zarka1998,Treumann2006, Weber2017}. Considering the plasma and neutral particle density within the modeled magnetosphere we also chose a location (i.e. the radius) where ion--neutral collisions are significantly reduced (e.g. above the ionosphere). The favored radius according to this constraint lies between 1.3 and 3 $R_p$ (see Sect. \ref{section:appendix:atmosphere} and Fig. \ref{fig:neutralAtmDensity}).\\
In order to show that the choice of $r=2$ $R_p$ or any other radial distance within the magnetosphere does not influence the results and the derived conclusions significantly (despite controlling the emission frequency since stronger magnetic fields cause higher gyro frequencies) we integrated the Poynting flux (Eq. \ref{eq:S-integral-mag}) for shells with radii between 1 and 2.5 $R_p$ above the planet. The results are displayed in Fig. \ref{fig:S_vs_r} for the open (blue), semi-open (orange) and closed (green) magnetosphere model. We note that the Poynting flux is not a conserved quantity in this system since several possibilities for conversion from or to electromagnetic energy exist within the magnetosphere (e.g. deceleration due to magnetic stresses, ion--neutral collisions, conversion between electromagnetic and thermal energy). As it can be seen in Fig. \ref{fig:S_vs_r} the Poynting flux variation as function of $r$ amounts to a factor of $\sim$1.5--2 for the open MS, $\sim$ 1 -- 1.5 for the semi--open MS and $\sim$2 for the closed MS. Given the uncertainties with which our study is anyway afflicted (such as the uncertainty of the stellar wind density, magnetic field strength, etc.), the uncertainty by the choice of the shell radius upon the Poynting fluxes is negligible. The choice of magnetic field strength, for instance, has much larger influence on the Poynting fluxes due to its $B^2$ dependence.
\section{On the effect of magnetic diffusion on the results}\label{section:diffusion}
\begin{figure*}
	\centering  
	\includegraphics[width=0.49\linewidth]{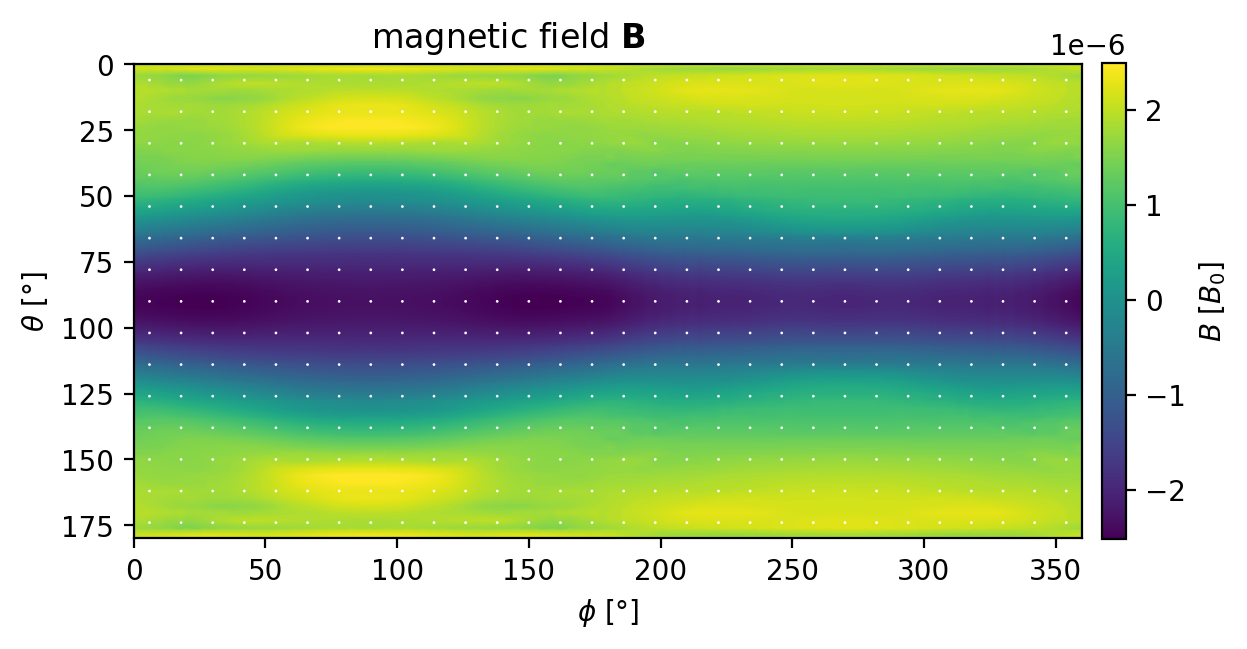}
	\includegraphics[width=0.49\linewidth]{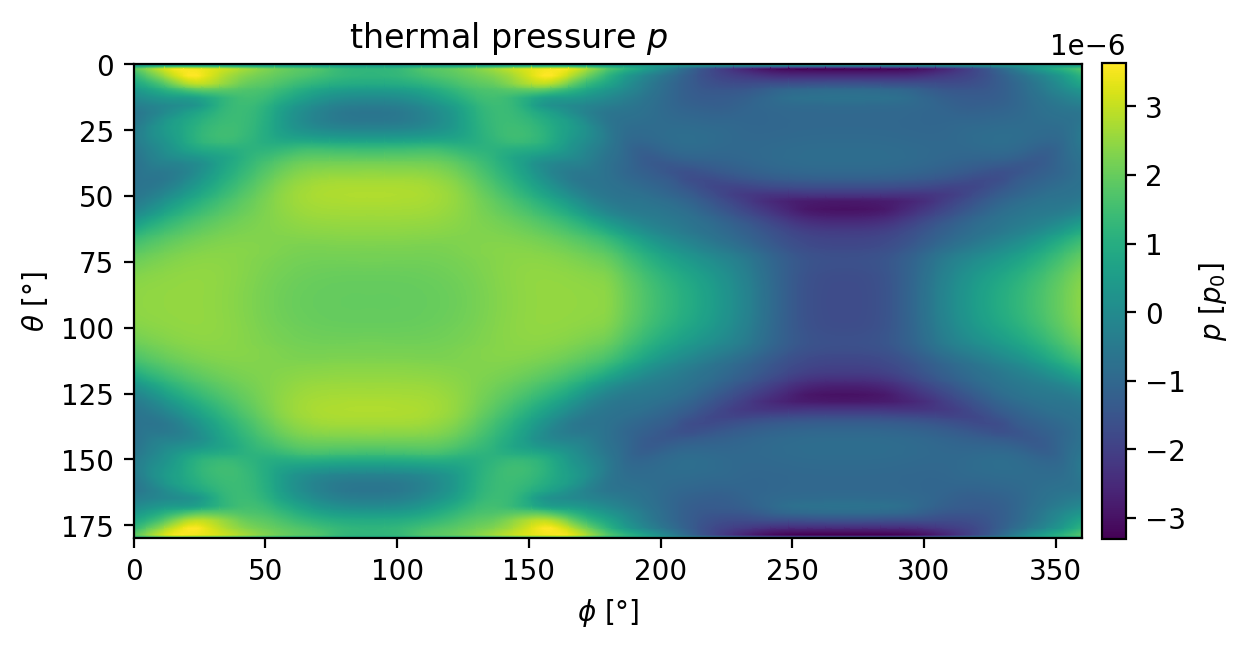}\\
	\includegraphics[width=0.49\linewidth]{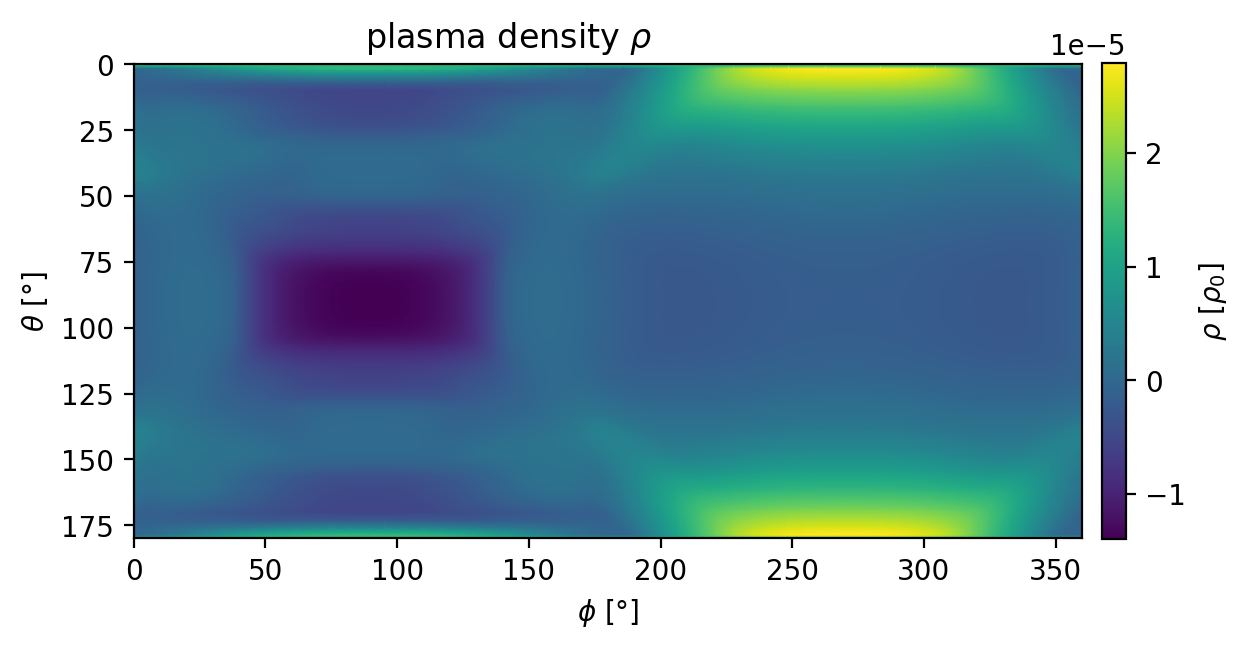}
	\includegraphics[width=0.49\linewidth]{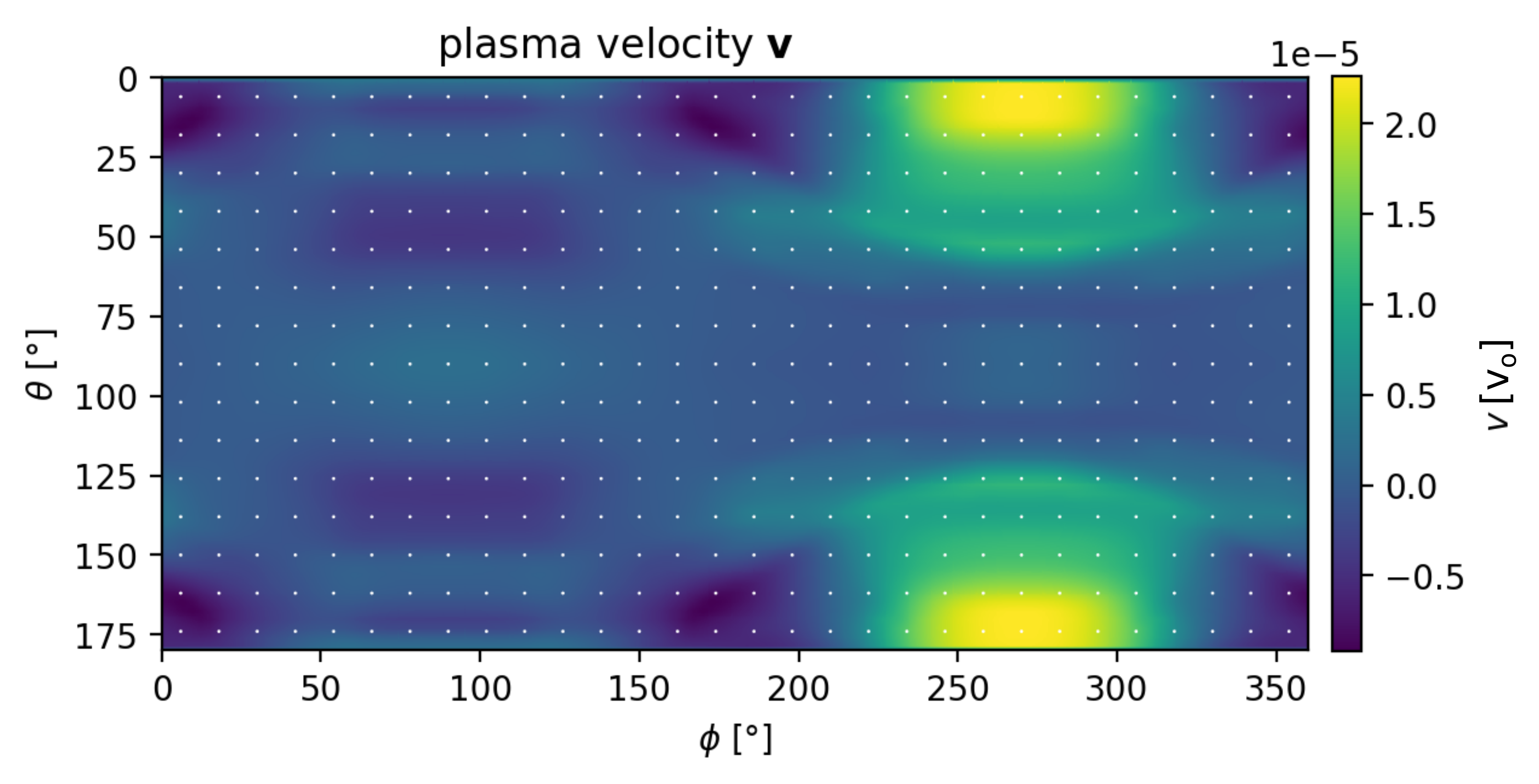}
	\caption{Deviations of resistive MHD simulations from our simulations discussed in the main text using the basic model (Table \ref{table:parameters}). The results correspond to simulations with Pedersen conductance of 100S. The four panels display magnetic field (upper left), pressure (upper right), density (bottom left) and velocity (bottom right) maps at $r=2R_p$.}
	\label{fig:diffusion_compare_100S}
\end{figure*}
\begin{figure*}
	\centering  
	\includegraphics[width=0.49\linewidth]{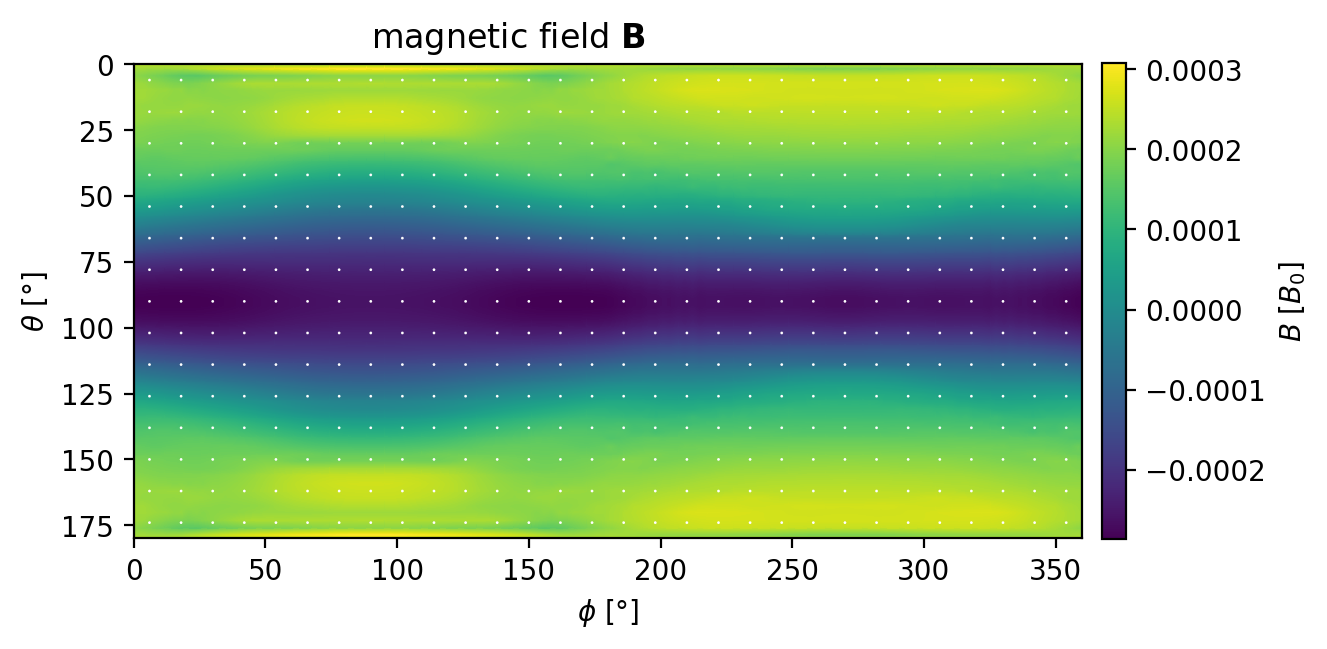}
	\includegraphics[width=0.49\linewidth]{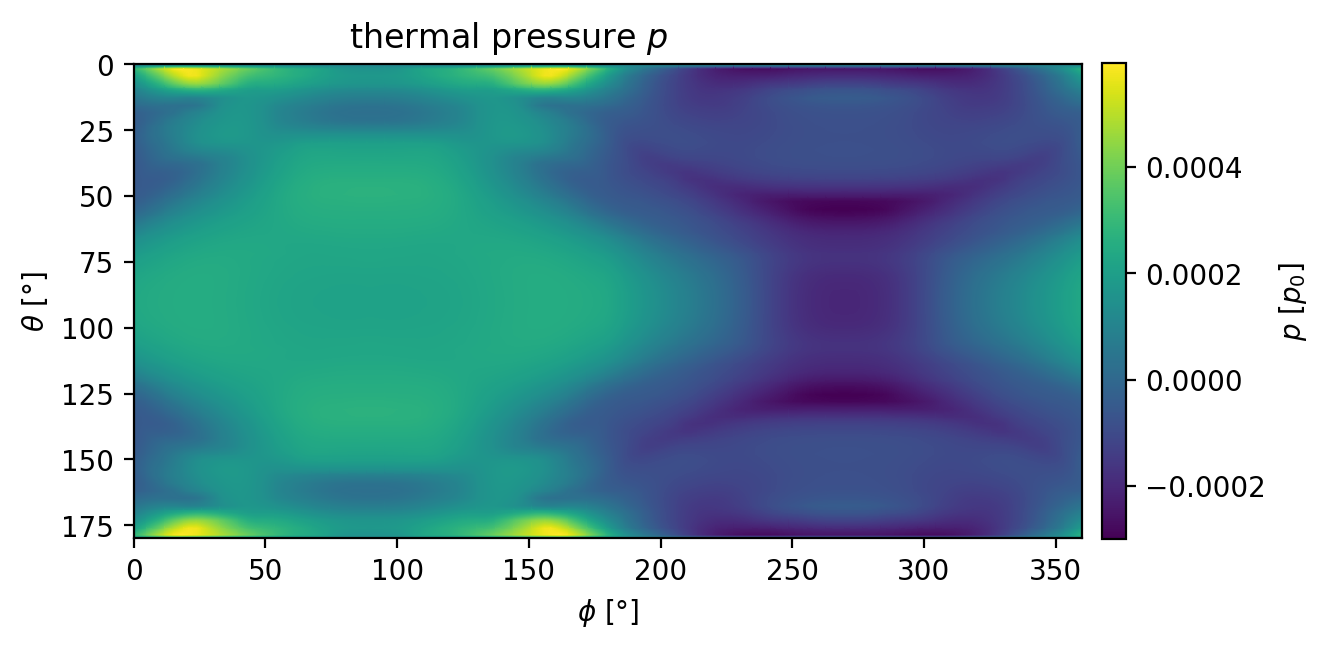}\\
	\includegraphics[width=0.49\linewidth]{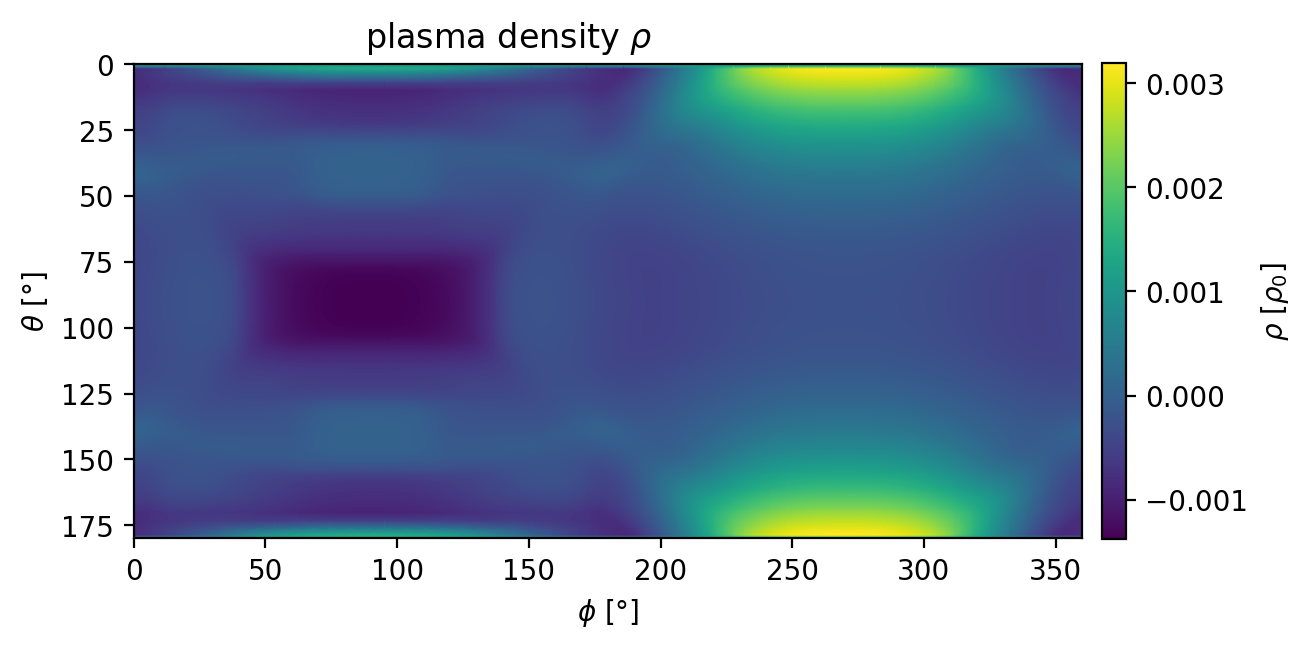}
	\includegraphics[width=0.49\linewidth]{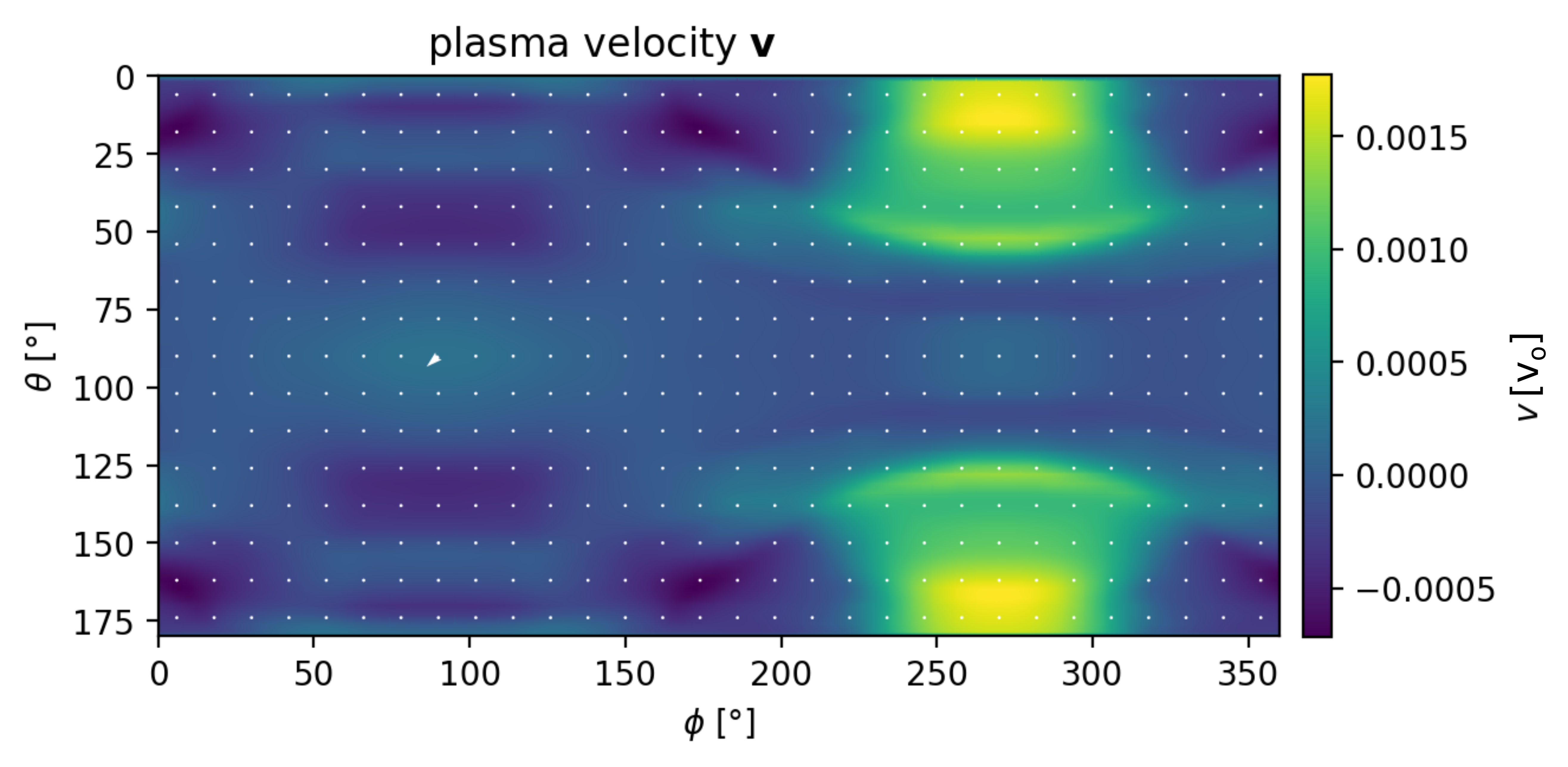}
	\caption{Deviations of resistive MHD simulations from our simulations discussed in the main text using the basic model (Table \ref{table:parameters}). The results correspond to simulations with Pedersen conductance of 1S. The four panels display magnetic field (upper left), pressure (upper right), density (bottom left) and velocity (bottom right) maps at $r=2R_p$.}
	\label{fig:diffusion_compare_1S}
\end{figure*}
In our MHD model we include the ionosphere and atmosphere of $\tau$ Boötis b through collisions with the plasma. We do not include the effects of collisions in the induction equation for computational time reasons. In this section we will justify this simplification by comparing simulations with different physical diffusion rates (i.e. different ionospheric conductivities) compared to our basic model without diffusion.\\ \\
Therefore we add a diffusion term to the induction equation (Eq. \ref{induction-equation}) which then reads
\begin{equation}
	\frac{\partial \vec{B}}{\partial t} = \nabla \times \left[\vec{v} \times \vec{B} - \eta \nabla \times \vec{B}\right] \; .
\end{equation} 
We scale the magnetic diffusivity $\eta=\eta(r)$ with the atmospheric scale height so that it correlates with the neutral particle density,
\begin{equation}
	\eta(r) = \eta_0 \exp\left(\frac{R_p - r}{H}\right)\;,
\end{equation}
where $\eta_0$ is the surface magnetic diffusivity, $R_p=72875$ km the planetary radius \citep{Wang2011}, $r$ the radial distance from the planet center and $H=4373$ km the scale height (see Table \ref{table:parameters} and Sect. \ref{section:TauBoötis_model}).\\
We assume a range of height integrated Pedersen conductivities, $\Sigma_P$ between 1 and 100 S. Together with the scale height we can calculate the surface magnetic diffusivity with
\begin{equation}
	\eta_0 = H \mu_0^{-1} \Sigma_P^{-1}\,,
\end{equation}
where $\mu_0$ is the vacuum permeability. The expected range of surface diffusivities is therefore $\eta_0^{(1)}\approx 3.5\times10^{10}$ and $\eta_0^{(2)} = 3.5\times10^{12}$ m$^2$ s$^{-1}$ for conductances of 100 S and 1 S, respectively. \\
In order to study the effect of magnetic diffusion on our simulation results, we compare the simulated plasma variables $\vec{B}$, $p$, $\rho$ and $\vec{v}$ within the magnetosphere of $\tau$ Boötis b with and without magnetic diffusion near the planet at $r=2$ $R_p$. We restrict ourselves to simulations with open magnetospheres using the basic stellar wind model.
\\ 
The deviations of the 4 plasma variables simulated with diffusion from those simulated without are displayed in Figs. \ref{fig:diffusion_compare_100S} ($\eta_0^{(1)}\approx 3.5\times10^{10}$ m$^2$ s$^{-1}$, $\Sigma_P=100$ S) and \ref{fig:diffusion_compare_1S} ($\eta_0^{(2)}\approx 3.5\times10^{12}$ m$^2$ s$^{-1}$, $\Sigma_P=1$ S) given in arbitrary normalized units. The finite conductivity causes only very small deviations from our basic model at $r=2R_p$ on the order of $10^{-3}$ for a diffusive model with $1$ S Pedersen conductance. Maximum deviations at $r=2R_p$ for the diffusive model with $\Sigma_P= 100$ S amount to factor of $10^{-5}$ (Fig. \ref{fig:diffusion_compare_100S}) compared with the non--diffusive model.
\\
We thus conclude that the effect of magnetic diffusion on our results (i.e. the magnetospheric Poynting flux) at 2 Rp above the ionosphere/atmosphere of the planet is negligible.

\section{Structure of the interactions: Parameter study}\label{section:appendix:structure_of_interaction}
\begin{figure*}
	\centering  
	\includegraphics[width=0.49\linewidth]{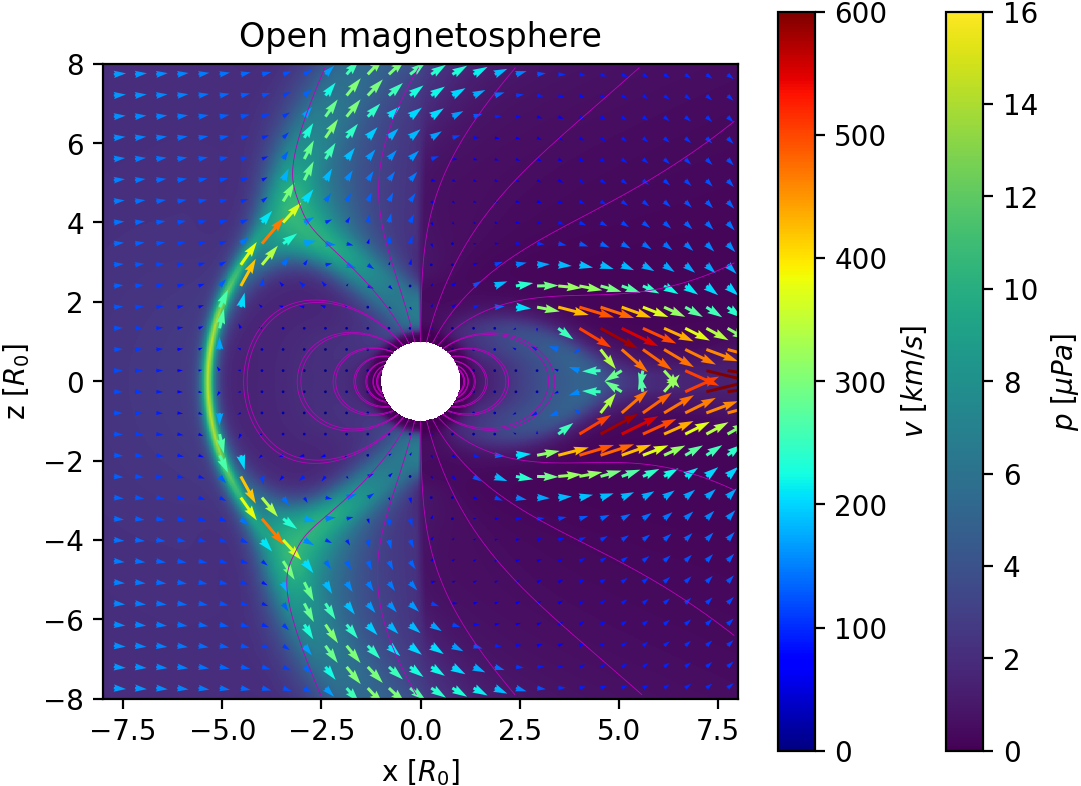}
	\includegraphics[width=0.49\linewidth]{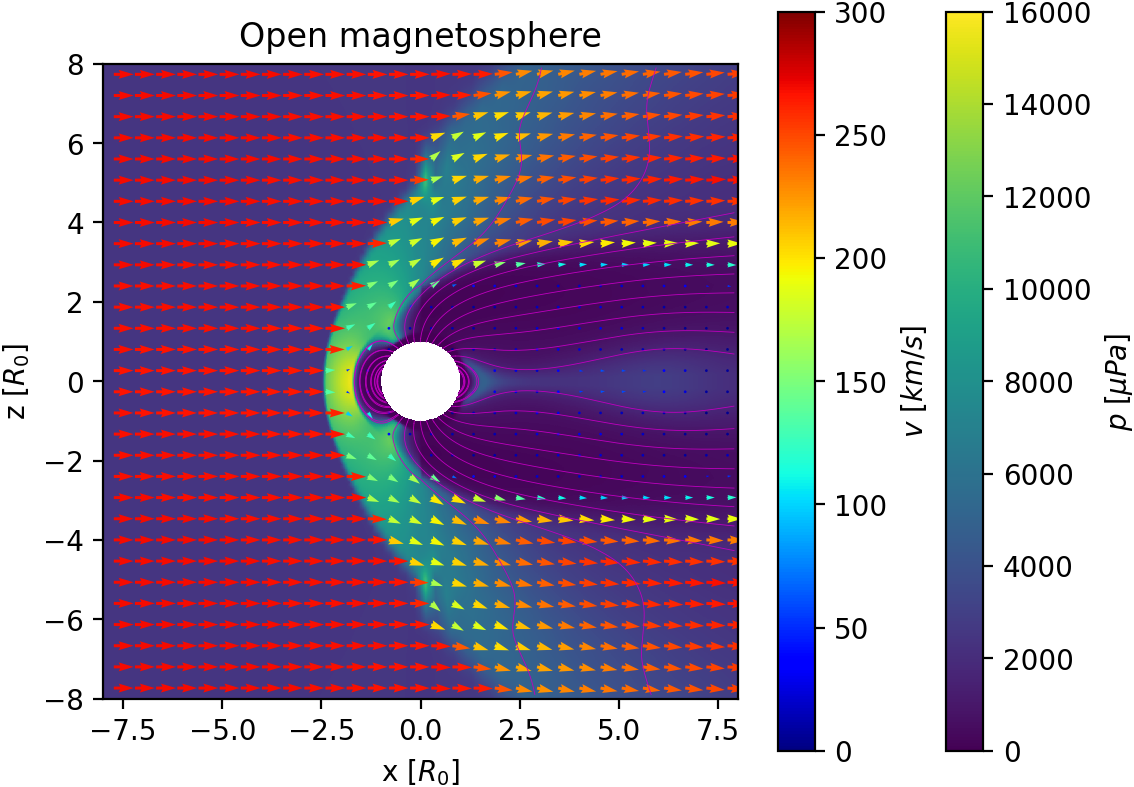}\\
	\includegraphics[width=0.49\linewidth]{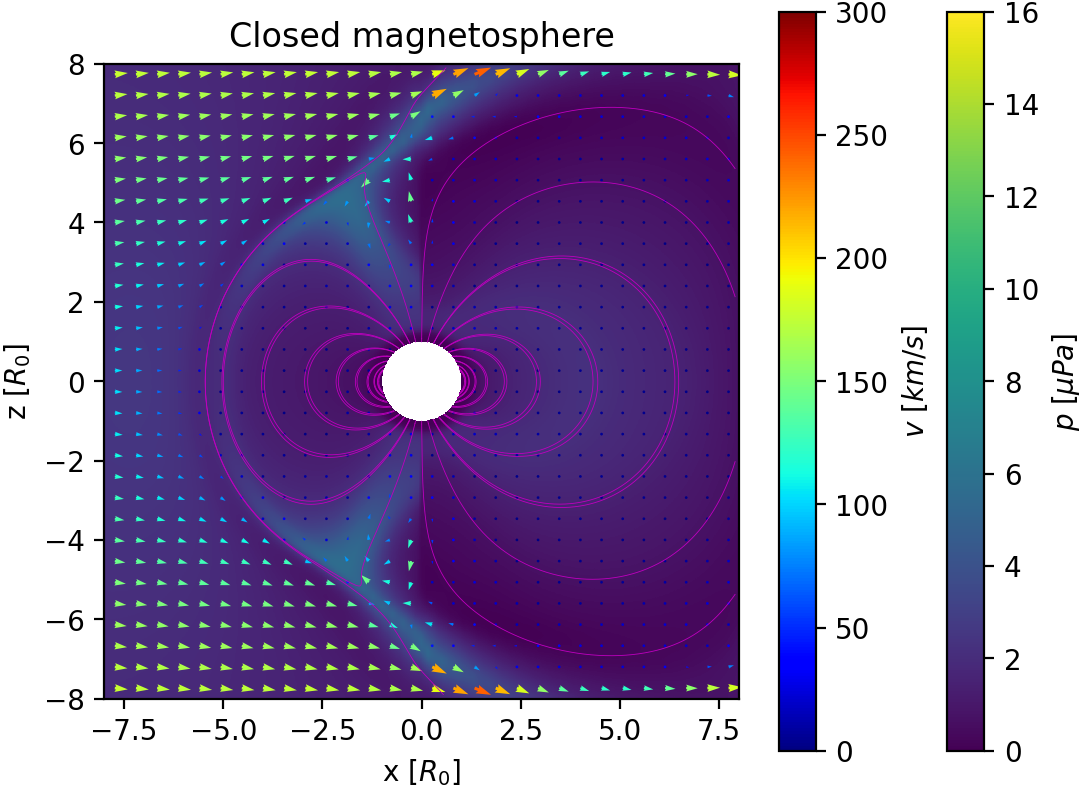}
	\includegraphics[width=0.49\linewidth]{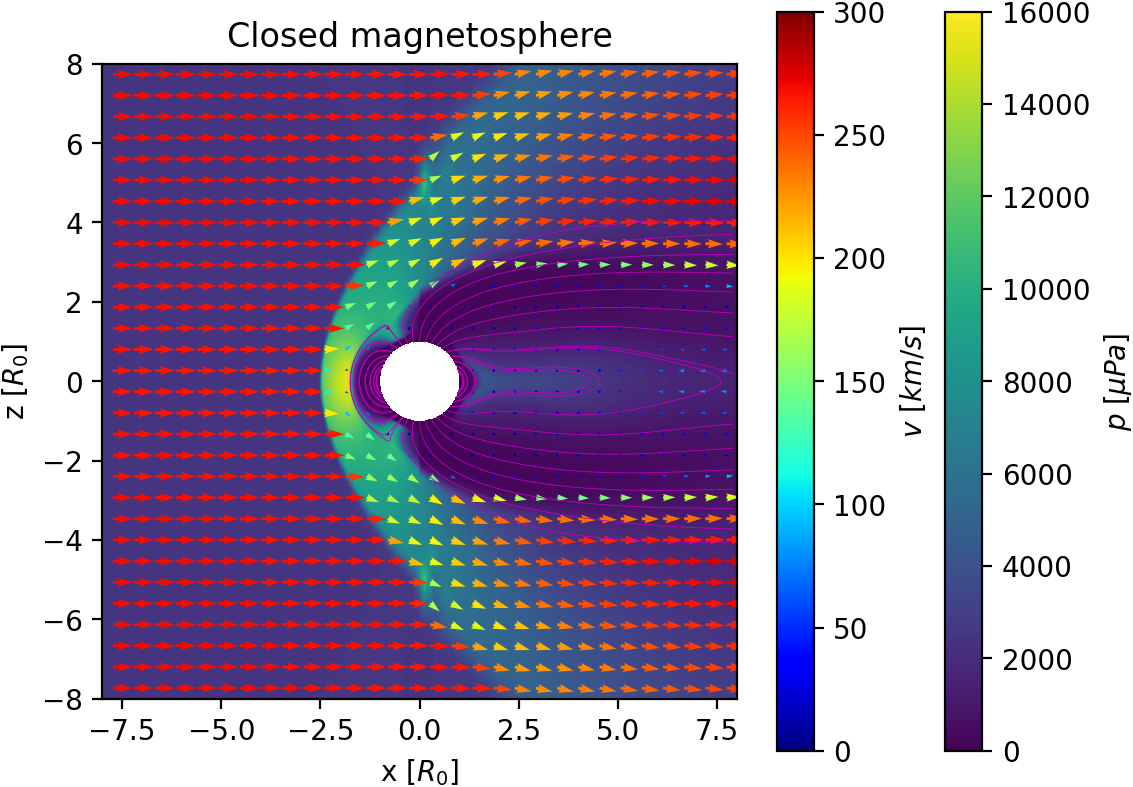}
	\caption{Plasma interaction for density $\rho_{sw} = 0.05$ $\rho_0$ (left panels) and density $\rho_{sw} = 100$ $\rho_0$ (right panels). Displayed are velocity fields (colored arrows, left colorbars) and plasma pressure (color contours, right colorbars) in the xz-plane for the open MS ($\theta_B=0^\circ$,top) and closed ($\theta_B=180^\circ$, bottom) MS case. Closed and open magnetospheric field lines are colored in magenta.}
	\label{fig:vel_plots_extremeCases}
\end{figure*}
In order to illustrate the effect of stellar wind density and pressure on the magnetospheric structure, we show xz--plane slices (Figs. \ref{fig:vel_plots_extremeCases}) similar to Fig. \ref{fig:vel_plots} for the extreme cases ($0.05$ $\rho_{sw}$ and $100$ $\rho_{sw}$) of our parameter study (Sects. \ref{section:results:stellar-wind-variability} and \ref{section:discussion_stellarwind_variability}). The plots show plasma velocity and thermal pressure in the near space environment of $\tau$ Boötis b. 
We use spherical coordinates to numerically describe the space environment around $\tau$ Boötis b. These coordinates have mathematical singularities along the pole axis. For exceptionally small upstream plasma densities, this can lead to numerical artifacts (i.e. jumps for the scalar variables along the pole axis in the PLUTO code). This is visible in Figure \ref{fig:vel_plots_extremeCases} (left). In the vector fields and thus the Poynting fluxes this discontinuity is negligible. This effect occurring at the extremely low upstream conditions thus does not have an effect on the conclusions of this work.
\\
The effect of stellar wind pressure and density on the size of the magnetosphere is clearly visible. The day side magnetopause location is $\sim 5$ $R_p$ for the lowest density case ($0.05$ $\rho_{sw}$, left panels) and $\sim 1.2$ $R_p$ for the highest density case ($100$ $\rho_{sw}$, right panels). Also visible is the sub--fast nature of the interaction in the lowest density case, where no bow shock forms upstream and thus incident plasma flow is perturbed prior to intersecting with the magnetosphere.

\section{On the neutral atmosphere model assumptions and its interaction with the plasma} \label{section:appendix:atmosphere}
\begin{figure*}
	\centering  
	\includegraphics[width=0.49\linewidth]{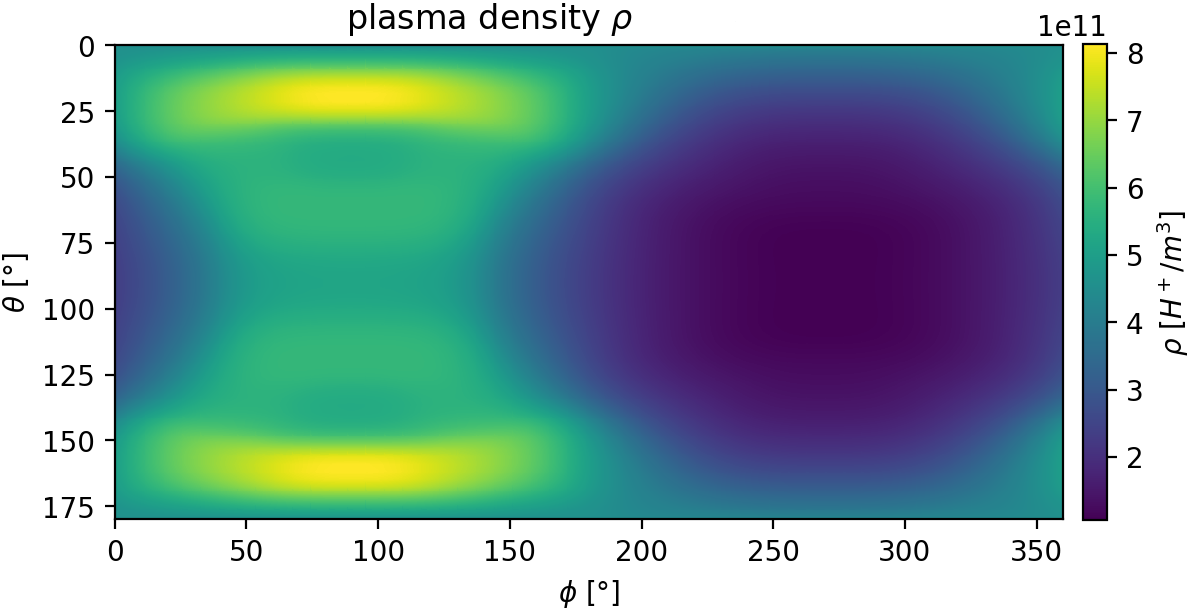}
	\includegraphics[width=0.49\linewidth]{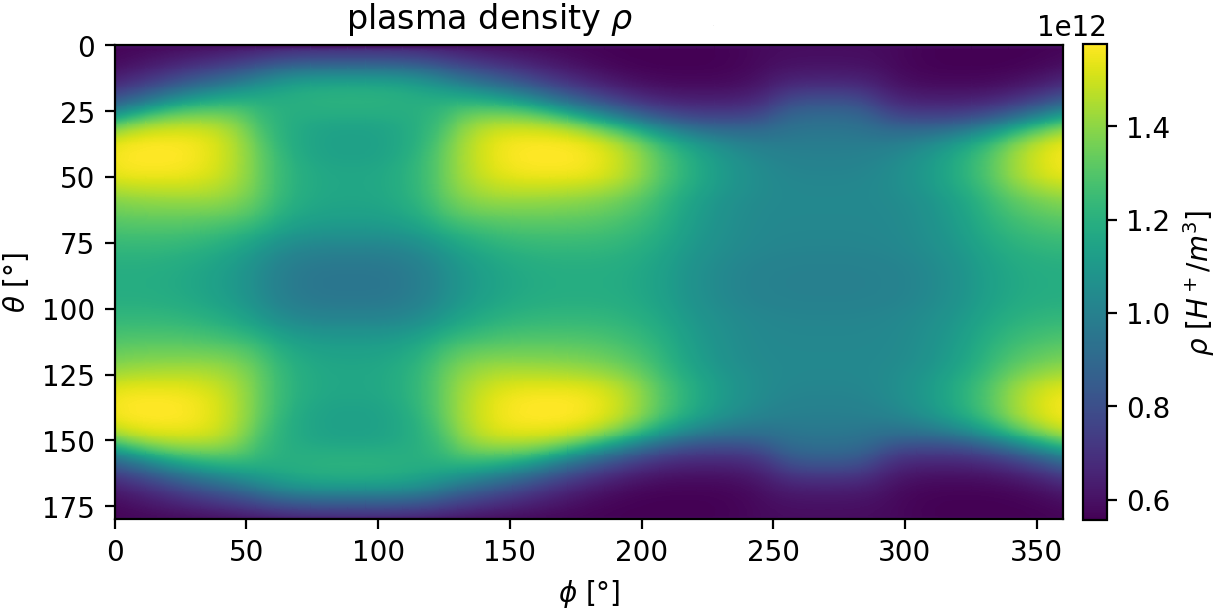}
	\caption{Plasma density maps over a sphere with $r=2$ $R_p$ for the open (left panel) and closed (right panel) MS case.}
	\label{fig:neutralAtmInteraction}
\end{figure*}
\begin{figure}
	\centering  
	\includegraphics[width=0.99\linewidth]{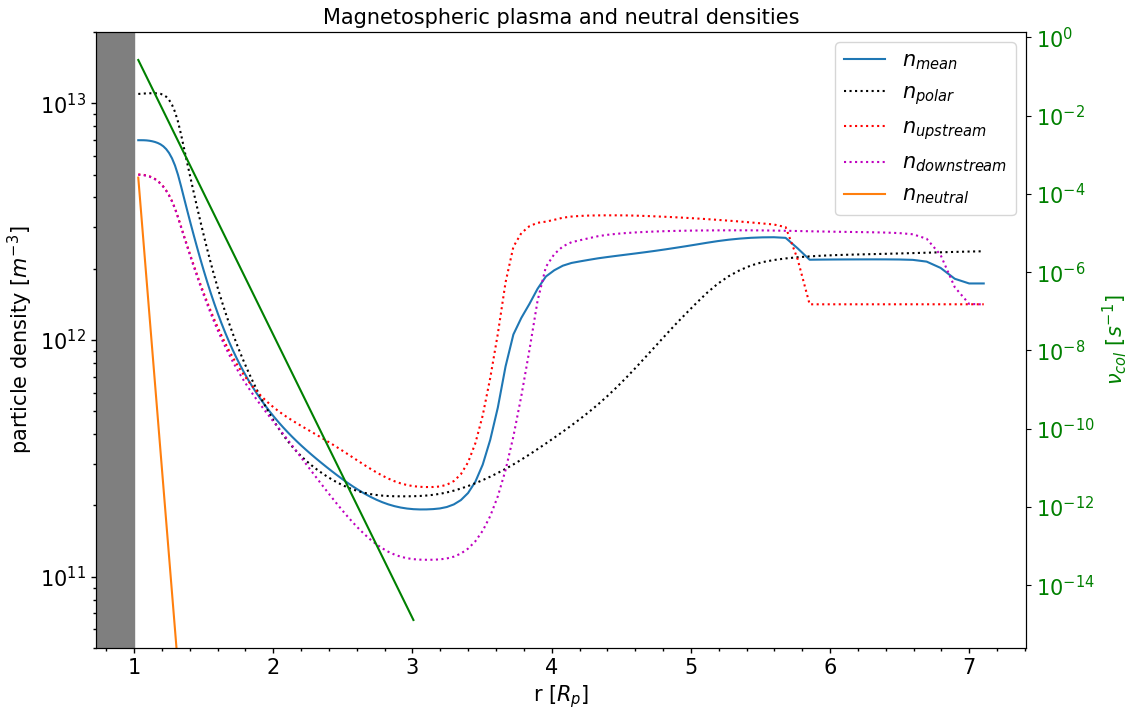}
	\caption{Plasma density profiles along the polar axis (black dotted line), upstream (red) and downstream x--axis (magenta). The blue solid line denotes the mean plasma density profile. The orange solid line shows the neutral particle density according to our atmosphere model (Eq. \ref{eq:atmosphere}). The green solid line (right y--axis) denotes ion--neutral collision frequencies.}
	\label{fig:neutralAtmDensity}
\end{figure}

In this section we discuss some properties and assumptions on our neutral atmosphere model presented in Sect. \ref{section:TauBoötis_model} as well as how the atmosphere affects the plasma focusing on our basic model (Table \ref{table:parameters}).\\
The aim of this work is not a detailed description of the ionosphere of the planet, but its magnetosphere and larger space environment. Our simplistic atmosphere model only acts through collisions with the plasma and affects the photo--ionization rate (i.e. plasma production) which both directly scale with the neutral particle density (Eqs. \ref{eq:atmosphere}, \ref{eq:ionisation}). Due to the exponential decrease of neutral particle density (Eq. \ref{eq:atmosphere}) with radial distance from the planet the atmosphere's effect on the plasma population drastically decreases with increasing altitude. The denser the neutral atmosphere the more the magnetospheric plasma is decelerated which leads to plasma pile up around the planet mimicking an ionosphere. Figure \ref{fig:neutralAtmDensity} shows plasma density profiles within the magnetosphere as function of radial distance from the center. The black dotted line represents the plasma density along the polar axes. Red and magenta lines represent equatorial upstream and downstream profiles respectively. The orange solid line denotes the neutral particle density according to Eq. \ref{eq:atmosphere}, the green dotted line shows the corresponding ion--neutral collision frequency. As visible in Fig. \ref{fig:neutralAtmDensity} the effect of ion--neutral collisions and thus the amount of plasma pile up is drastically reduced above an altitude near $1.3$ $R_p$. There the neutral particle density is $n(r=1.3R_p) \approx 5\times
10^{10}$ m$^{-3}$. The ion--neutral collision frequency is $\sim 0.5$ s$^{-1}$ at the surface and drastically decreases with altitude. In our studies we focus on the region 1$R_p$ above the planets surface and thus above the ionospheric shell. At $r=2$ $R_p$ the collisions are negligible because the collision frequency has decreased to about $\sim 10^{-7}$ s$^{-1}$.\\
The large scale height of 4373 km increases the extent of the atmosphere but is needed in order to sufficiently resolve the atmosphere in our model. However, the atmospheres of Hot Jupiter exoplanets are expected to be strongly inflated due to intense stellar irradiation \citep[e.g.][]{VidalMadjar2003} which is partially mimicked by the large scale height. The surface neutral particle density $n_0$ corresponds to an atmosphere pressure near $10^{-3}$ nbar assuming a temperature between 1000 and 2000 K. Here the mixing ratio of plasma (hydrogen ions) and neutral particles is roughly 50 \%. The neutral atmosphere consists of molecular hydrogen which is the most abundant molecule in all solar system gas giants.\\
In order to demonstrate the minor role of the neutral atmosphere at $r=2$ $R_p$ we show plasma density maps over a shell at this altitude in Fig. \ref{fig:neutralAtmInteraction}. The left and right panels display the density map of the open and closed MS case, respectively. There the neutral particle density is $n(r=2$ $R_p) \approx 4.6\times 10^{5}$ m$^{-3}$ according to Eq. \ref{eq:atmosphere} and Fig. \ref{fig:neutralAtmDensity} leading to a neutral--plasma mixing ratio of about $10^{-6}$. Regions of high plasma density (i.e. at the day side) indicate regions where the interaction between the neutral atmosphere and plasma is strongest (at $r=2$ $R_p$) in terms of ion--neutral collisions and recombination. However, the effect of the neutral atmosphere on the plasma is an order of magnitude lower than in regions within the ionosphere below $\sim 1.3$ $R_p$.
\end{document}